%% file: gc_var_part1.tex
\def\hst{{\sl HST}}

\def\wfc3{WFC3}
\def\vlt{{\sl VLT}}

\documentclass[12pt,preprint]{aastex}
\usepackage{appendix}
\usepackage{epsf}
\usepackage[final]{pdfpages}
\usepackage{color}
\usepackage{graphicx}
\usepackage{graphics}
\usepackage{epsfig}
\usepackage{longtable}
\usepackage{multirow}
\usepackage{booktabs}
\begin{document}
\title{Near-Infrared Variability Study of the Central 2.3\arcmin$\times$2.3\arcmin\ of the Galactic Centre I. Catalog of Variable Sources}
\author{Hui Dong$^{1}$, Rainer Sch{\"o}del$^1$, Benjamin F. Williams$^2$, Francisco Nogueras-Lara$^1$, Eulalia Gallego-Cano$^1$,Teresa Gallego-Calvente$^1$, Q. Daniel Wang$^3$, Mark R.  Morris$^4$, Tuan Do$^4$, Andrea Ghez$^4$}

\affil{$^1$ Instituto de Astrof\'{i}sica de Andaluc\'{i}a (CSIC), Glorieta de la Astronom\'{i}a S/N, E-18008 Granada, Spain}\affil{$^2$ Department of Astronomy, Box 351580, University of Washington, Seattle, WA 98195, USA}\affil{$^3$ Department of Astronomy, University of Massachusetts,
Amherst, MA, 01003, USA}\affil{$^4$Department of Physics and Astronomy, University of California, Los
Angeles, CA, 90095, USA}\affil{E-mail: hdong@iaa.es}

\begin{abstract}
We used four-year baseline \hst/WFC3 IR observations of the 
Galactic Centre in the F153M band (1.53 $\mu$m) to identify variable stars 
in the central $\sim$2.3\arcmin$\times$2.3\arcmin\ field. We classified 
3845 long-term (periods from months to years) and 76 short-term 
(periods of a few days or less) variables among a total sample of 33070 stars. 
For 36 of the latter ones, we also 
derived their periods  ($<$3 days). Our catalog 
not only confirms bright long period variables and 
massive eclipsing binaries identified in previous works, but 
also contains many newly recognized dim variable stars. For 
example,  we found $\delta$ Scuti and RR Lyrae stars towards the Galactic Centre for the first time, 
as well as one BL Her star (period $<$ 1.3 d). We cross-correlated our catalog   
with previous spectroscopic studies and found that 
319 variables have well-defined stellar types, such as 
Wolf-Rayet, OB main sequence, supergiants and 
asymptotic giant branch stars. We used colours 
and magnitudes to infer the probable variable types 
 for those stars without accurately measured 
periods or spectroscopic information.
We conclude that  
the majority of unclassified variables could potentially be 
eclipsing/ellipsoidal binaries 
and Type II Cepheids. 
Our source catalog will be 
valuable for future studies aimed at constraining 
the distance, star formation history and massive 
binary fraction of  the 
Milky Way nuclear star cluster.  
 \end{abstract}

\section{Introduction}
The Galactic Centre (GC) contains a massive black 
hole (MBH), Sgr A* (4$\times 10^6~M_{\odot}$) and 
 is located at a distance of only $\sim$8 
 kpc~\citep{ghe08,gil09,cha15,boe16}, 100 times closer than the nearest nucleus of 
 a comparable system, the Andromeda galaxy.
 Because of 
its proximity, the GC provides us with a unique lab to 
 study the interaction between a MBH 
 and its environment. 
 Since both star formation and AGN activities require the accretion of matter, knowledge 
 of the star formation history in the GC 
 could give us a hint about its activity history. 
  
  There are already many works on the star formation history in the GC.  
Three young massive star clusters (the Arches, Quintuplet and Central clusters, 
2-6 Myrs old, $\sim10^4$ M$_{\odot}$, 
\citealt{fig99,fig02,gen03}) 
were previously thought to be the major contributor of 
the star formation over the past 10 Myr. However, 
\citet{don12} identify a similar number of 
evolved massive stars outside those three clusters, which indicate that 
significant star formation might also have 
taken place beyond the clusters. 
\citet{nis16} find 
13 intermediate-age stars that likely formed 50-500 Myr 
ago. 
\citet{fig04} study the stellar population in 
a number of GC fields and conclude that a  
continuous star formation history could best explain the data. 
\citet{pfu11} (see also~\citealt{blu03}) investigate the star formation 
history in the central parsec of the GC and propose that
the star formation rate 
(SFR) 
experienced a maximum at 10 Gyr 
ago, then decreased until 1-2 Gyr ago and  
increased again during the last several hundred million years.  
In a word, there is still considerable debate as to the 
precise star formation history of 
the GC. 
  
A detailed study of stellar populations in the 
GC is seriously hampered by foreground 
 extinction, which is not only large, with typical 
 values A$_{Ks}\sim$2.5~\citep{sch10}, but also 
 changes on scales of arcseconds~\citep{sco03,sch07,sch10}. 
 The GC is totally obscured in the ultraviolet and 
 optical bands, which would be optimal to 
 distinguish between blue, young massive stars and 
 red, old low-mass stars. In the near-infrared (IR), 
intrinsic colour indices are small ($J$$-$$K$ $\leq$1 mag and 
$H$$-$$K$ $\leq$0.3 mag), and the observed colours of stars at the GC 
are dominated by the extreme reddening. Without highly 
precise and accurate multi-band photometry, 
young massive stars can therefore be easily confused 
with less extinguished old, low-mass stars. Spectroscopy 
 can identify late-type stars via their CO bandhead absorption. 
 However, the need for high angular resolution  
 limits observations to small 
regions~\citep{man07,do09,lu13,do13, do15,sto15}. 
 
 \subsection{Variable Stars in the GC}
Variable stars provide us with a tool to study the 
 star formation history in the GC. Unlike 
 apparent magnitudes, the 
 periods and amplitudes of stars are unrelated to their foreground 
 extinction.  Stars 
 with different initial masses evolve to certain stages and 
 cross the instability strip in the 
Hertzprung-Russell diagram to become variables 
 with different periods, amplitudes and luminosities  
 (see more on this in \S\ref{ss:type}). Therefore, 
such variables trace 
 stellar populations of different ages, while 
 their number densities could be used to infer the corresponding SFR at a 
 given epoch. For 
 example,~\citet{mat11} find three classical Cepheids 
 in the central 40 pc of the GC, from which they derive a SFR of 
0.075 M$_{\odot}$/yr about 20-30 Myr ago.

 Monitoring stellar variability in the GC started in 
 the 1980s~\citep{hal89}. 
Early works can be roughly divided into two groups: 
1) wide field-of-view (FoV) imaging with small telescopes  
($<$4 meter) and 2) images and integral field units 
assisted by adaptive optics (AO) on 8-10 meter-class telescopes. 
The advantage of the former method is that the photometry
is relatively stable for bright stars. 
The disadvantage is that the seeing-limited resolution 
is very poor, $\geq$1\arcsec\ and the confusion is severe for faint stars. 
The confusion limit problem can be largely alleviated by AO. 
However, AO has a small FoV  
(at least for the current almost exclusively used single-conjugated AO systems) 
and the point spread function (PSF) is potentially large at off-axis angles, 
larger than the so-called isoplanatic angle ($\sim$$15\arcsec$ at $2~\mu$m). 
In this latter case, the variability of the PSF 
can lead to systematic photometric errors of up to a few 0.1 mag 
(e.g.,~\citealt{sch10a}). Additionally, the absolute zero point 
can be difficult to determine in AO observations of extremely 
crowded fields, when the PSF cannot be measured with 
high accuracy. Although this could be compensated by cross-calibrating 
different epochs with the help of presumably non-variable stars (e.g.,~\citealt{raf07}), 
zeropoint variations between epochs may remain an issue.

Instead, we will here use four-year baseline Hubble Space Telescope (\hst ) observations to 
identify variable stars 
in the Milky Way nuclear star cluster (MWNSC hereafter). 
These observations were obtained with the Wide Field Camera 3 (WFC3) IR camera.  
The large FoV (123\arcsec$\times$136\arcsec ), high angular resolution 
(full width half maximum, FWHM, $\sim$0.15\arcsec\ at 1.5 $\mu$m), high sensitivity,  
stable PSF and accurate zeropoint (0.01 \rm{mag}) 
of the observations allow us to accurately 
detect many dim variable stars with 
small amplitudes, except in the central few arcseconds around Sgr A*, 
where the source confusion remains too problematic. 
In this paper, we will use this dataset to produce a catalog 
of variable stars. In two follow-up papers (Dong et al. 2017b,c in preparation), we will 
 use the RR Lyrae (hereafter RRL) stars to constrain the distance and 
 old stellar population and to derive the binary fraction of massive stars 
 in the MWNSC. 

 \subsection{Types of Variable Stars}\label{ss:type}
Periodic variable stars are classified according to their  
 periods, luminosities, shapes of light curves and 
 peak-to-peak amplitudes (PPAs). A complete summary of various types 
 is beyond the scope of this paper. However, we briefly 
 summarize the properties of some frequently encountered types: 
 Long Period Variables (hereafter LPVs), Classical Cepheids (hereafter CCEPs), 
Type II Cepheids (hereafter T2Cs), RRLs and $\delta$ Scuti 
 stars. 

 LPVs are red giant stars which vary on time scales from 
 weeks to years and can be divided into three groups: 
regular variables (Miras), semi-regular variables (SRVs) and 
OGLE small 
amplitude red giants (OSARGs)~\citep{sos13}. 
These stars evolve on the first-ascent red giant branch (RGB) or  
asymptotic giant branch (AGB). They occupy three out 
of five sequences (`A', `B', `C') identified in the period-magnitude 
diagram of Large Magellanic Cloud by~\citet{woo99} and~\citet{woo00}~\citep{sos04}.  Among these 
three types, Miras are the best studied 
ones due to their large amplitudes:   
Their $K$ band amplitudes are $\sim$0.6 mag 
on average~\citep{woo00}. Their  
period-luminosity (PL) 
relationship was first established by~\citet{gla81} and~\citet{fea89}. 
They have been widely used as a distance indicator for remote galaxies. 
Frequently, SiO maser emission is observed from Miras, arising in the shells 
produced by outward propagating shocks~\citep{gra09}. The colours of Miras are 
typically reddened by circumstellar dust produced in their slow, 
dense stellar winds. Compared to Miras, OSARGs and SRVs are dimmer by up 
to 1 mag at the $K$ band~\citep{ita04}. 
 
CCEPs (1$\leq$P$\leq$100 d) are also well-studied and frequently used 
to measure the distance of 
remote galaxies, because of their brightness and regular variation. Their $V$ 
band amplitude are $\sim$0.7 mag on average~\citep{kla09}. 
They are yellow 
supergiants with initial masses of 4-20 $M_{\odot}$~\citep{tur96}. Their ages are $\sim$ 100 Myr 
\citep{bec77} and their periods are anti-correlated with their ages and masses. 
Therefore, they can also be 
used as an age tracer for stellar populations~\citep{alc99}.

T2Cs are population II, old, low-mass and typically metal-poor 
stars often found in globular clusters, thick disk, bulge and halo of our Galaxy. 
Such variables are 
the immediate progeny of horizontal branch (HB) stars with little envelope 
mass~\citep{cat09}. They 
have a similar period range as CCEPs, but are 1.5-2 mag dimmer~\citep{san06}. 
Their amplitudes  
can reach up to 1 mag in the $J$ and $H$ bands. They can be divided into three groups 
according to their periods: BL Her stars (1$<$$P$$<$4 d), W Vir stars 
(4$<$$P$$<$20 d) and RV Tau (RV) stars (20$<$$P$$<$100 d)~\citep{sos08}. 
These different types of variables represent the different evolutionary phases of 
low-mass stars after they leave the HB.~\citet{mat06} show that in the 
near-IR band, T2Cs follow a narrow PL relationship, 
which is not sensitive to the metallicity and can be used as a  
distance indicator. 

RRLs (0.2$\leq$$P$$\leq$1.0 d,~\citealt{cat09}) are old low-mass 
core-helium-burning giants ($>$10 Gyr,~\citealt{wal89,lee92}) in the mass range of 
0.55 to 0.80 
M$_{\odot}$~\citep{mar15}. According to the shape of the light curves, 
they can be divided into fundamental-mode (Type ab, hereafter RRab), 
first-overtone (type c, RRc hereafter) and rare double-mode (type d). 
RRab stars are on average intrinsically brighter and have 
higher amplitudes and longer period than RRc stars. 
Normally, the period of 0.4 d  
distinguishes RRab and RRc stars.

$\delta$ Scuti stars are main-sequence (MS) or early post-MS stars 
near the lower part of the instability trip~\citep{mac07,pol10}. Their periods are less 
than 0.2 d. Compared to the four types above, they are fainter, 
have small PPA and are therefore very hard to be detected at the GC. 

Besides the five types above, there exist other, but much rarer, 
variables, such as  
1) red supergiant stars, evolved helium burning 
stars with initial mass between 10 and 30 
$M_{\odot}$, with surface temperature 3000-4000 K~\citep{ver12};  
2) luminous blue variables (LBVs), evolved massive 
stars with initial mass $>$30 $M_{\odot}$~\citep{cro07}; 3) OB 
main sequence stars~\citep{lef09}. 

Finally, another frequently encountered class of variables 
consist of eclipsing/ellipsoidal binaries. 
For example,~\citet{sos14} and~\citet{paw14} suggest that 
binary scenario could explain the other two sequences (`D' and `E') 
in the period-magnitude 
diagram of~\citet{woo99} and~\citet{woo00}, which are roughly 0.5 mag 
and 2 mag dimmer than Miras, respectively.

\section{Observations and Data Reduction}\label{s:data}
\subsection{HST Dataset}\label{ss:data_hst}
We focus on a field of $\sim$2.3\arcmin$\times$2.3\arcmin\ ($\sim$ 5 pc $\times$ 5 pc) centered 
approximately on Sgr A*. This field was repeatedly observed by the 
 WFC3 IR camera with the F153M 
filter from Jun, 2010 to April, 2014. The effective wavelength of the 
filter is 1.53 $\mu$m, which is close to the 
traditional Johnson/Glass H 
band ($\sim$1.63 $\mu$m). These observations came 
from two groups of \hst\ programs: 1) 
 Programs GO-11671, GO-12318, GO-12667 (PI, Andrea Ghez) and GO-13049 (PI, Tuan Do;~\citealt{hos15,sto15}) observed 
 Sgr A* with one pointing on each of 
 Jun, 2010, Sep, 2011, Aug, 2012 and Feb, 2014, to study 
stellar proper motions and present-day mass function 
of the MWNSC;  
2) Programs GO-13316 (PI, Howard Bushouse) and 
GO-13403 (PI, Nicolas Grosso) monitored 
Sgr A* from Feb, 2014 to April, 2014~\citep{mos16}, a period corresponding to the peri-center passage of the G2 object around Sgr A*~\citep{gil12}. These 
two programs included 56 pointings. The pointings and 
rotation angles of the observations from these six programs were different by less than 0.3\arcsec\ and 3.1 degrees. 

The time coverage of the data set can be divided into eight periods. Table~\ref{t:obs} lists the observing dates, \hst\ program IDs, durations, numbers of pointings and dithered exposures, 
as well as total effective exposure times. In summary, the 60 pointings consist of 290 dithered exposures; 
most of the pointings have four exposures, while the three 
pointings in 2010, 2011, 2012 have 21 exposures each and the one pointing in March 11, 2014 
has three exposures. All these pointings have used a sub-pixel dithering pattern. 
The total exposure time is $\sim$90 ks. The exposure times for individual dithered 
exposures range between 250 s to 350 s. 
The durations of the eight observational blocks are from 0.3 hours to 
15 hours. Because of the four-year time baseline of the total observations, as well as  
the duration of each period, this dataset can help us to 
identify variable stars with not only periods of years, 
but also periods of days.  

We also used the \hst\ WFC3/IR observations of the MWNSC in the F127M band (1.27 $\mu$m, an 
analog of the Johnson/Glass J band, 1.22 $\mu$m) 
from Programs GO-11671 and GO-12182 (PI, Tuan Do,~\citealt{sto15}, see also Table 1) 
to measure the F127M-F153M colours  
of variable candidates. This information was valuable to distinguish foreground/background 
stars from those belonging to the MWNSC. Program GO-11671 included one pointing consisting of 
12 dithered exposures for a total exposure time of 7.2 ks. Its pointing and rotation angle were the same as the F153M observations in 
the same \hst\ program. Program GO-12182 included nine pointings, each 
of which had four dithered exposures, totaling 1.8 ks exposure time. These pointings mapped the outer regions 
of the MWNSC, which were not covered by the Program GO-11671. They were used for variable 
candidates in the F153M band outside the FoV of Program GO-11671, but detected in the other programs, especially 
GO-13316 and 13403. Sub-pixel dithering patterns were employed. 
The observations in the F127M band were taken on Aug 17, 2010 
and May 20, 2011, respectively.

\subsection{Data Reduction}\label{ss:reduction}
\begin{figure*}[!thb]
  \centerline{    
       \epsfig{figure=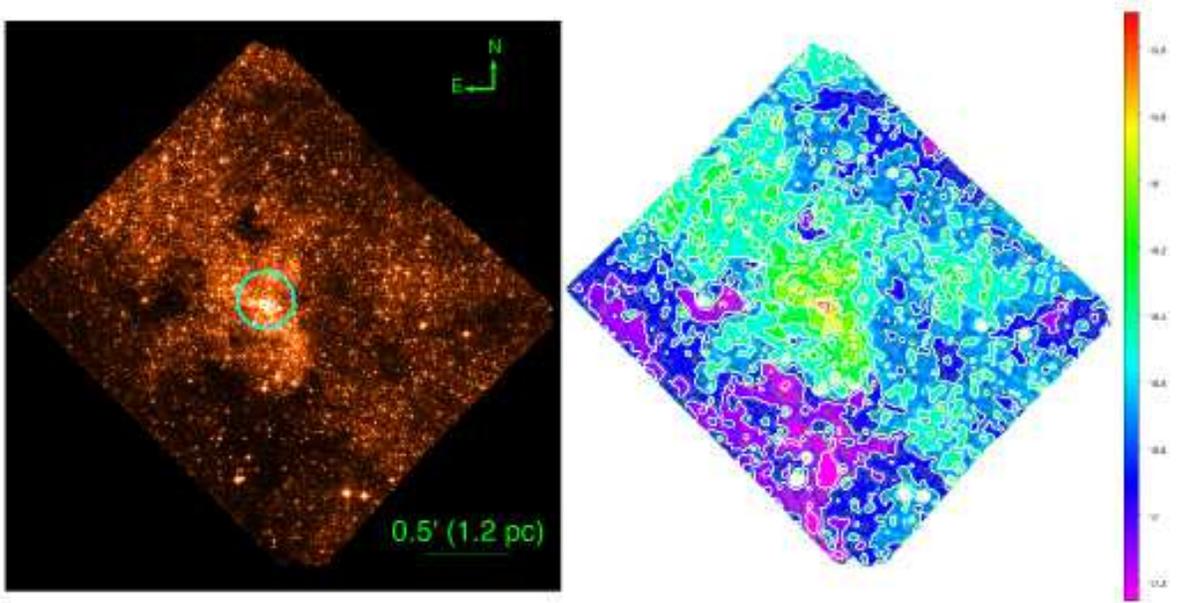,width=1.0\textwidth,angle=0}
       }
 \caption{Left: \hst/WFC3 F153M observations of the MWNSC. 
The cyan circle indicates a radius of 10 arcseconds around Sgr A*. High extinction regions 
 with low stellar number densities can easily be recognized. Right: the spatial distribution of the 
 logarithm of the smoothed surface brightness (in units of {\rm ergs s$^{-1}$ cm$^{-2}$ \AA$^{-1}$ 
 arcsec$^{-2}$}). The white contours divide the image into ten regions with the 
 surface densities defined in Table~\ref{t:dl}.}
 \label{f:sgra_f153m}
 \end{figure*}

We first performed the basic reduction on the dithered exposures and combined them into mosaics. 
We downloaded \hst\ raw data and calibration files from the Multimission Archive at STScI (MAST). 
We then used the \hst\ pipeline, OPUS version and CALWFC3 version 2.1 to perform 
the basic calibration steps on individual dithered exposures, such as identifying bad pixels, 
bias correction, dark subtraction and flat fielding. In PyRAF, the `Tweakreg' task aligned 
individual dithered exposures and the `Astrodrizzle' task corrected for the distortion, 
masked out defects (including cosmic rays), and combined the dithered exposures into a 
mosaic image. Considering the large proper motions of stars in the MWNSC\footnote{
\citet{sch09}  study the proper motions of 6124 stars within 
1.0 pc ($\sim$ 25\arcsec ) of Sgr A*. These stars could move on average 0.1 
and even 
reach 1.28 WFC3/IR pixel (0.13\arcsec/pixel for the WFC3/IR camera) 
during the four-year time baseline of the F153M observations. 
These values decrease to 0.003 and 0.05 WFC3/IR pixels, if we consider only the observations from 2014.}, 
we used `Tweakreg' and `Astrodrizzle' tasks to produce one mosaic image per each year 
for the F153M observations.  
For demonstration purpose, we also constructed a mosaic image using all the 290 dithered 
F153M exposures, which is shown in the left panel of 
Fig.~\ref{f:sgra_f153m}. 
Because the F127M observations were pointed toward different lines-of-sight, we 
produced mosaic images for each pointing, respectively. The method described in the appendix of ~\citet{don11} 
was used to align the four F153M mosaic images and ten F127M mosaic images 
with an accuracy of 0.013'', i.e., one tenth of the pixel size of the \hst\ WFC3/IR 
camera. 

The right panel of Fig.~\ref{f:sgra_f153m} shows a surface brightness image, 
produced by smoothing the image in the left panel, using a 
median filter with size of 2\arcsec, more than ten times the 
FWHM of the WFC3/IR 
camera. This image is used to derive the 
photometric uncertainty discussed below. We divided the image  
into ten regions based on the logarithm of the surface 
brightness (in units of ergs s$^{-1}$ cm$^{-2}$ \AA$^{-1}$ arcsec$^{-2}$) from 
-17.29 to -15.51 with a step size of 0.2 dex. The region with the lowest surface 
density has the smallest stellar density 
(see the left panel) and suffers extreme extinction because of the 
presence of a dark cloud. More information about these ten 
regions is given in Table~\ref{t:dl}.

The `DOLPHOT' package\footnote{http://americano.dolphinsim.com/dolphot/}~\citep{dol00} was then 
used to detect individual sources, extract photometry and perform artificial star tests. 
`DOLPHOT' is a mature stellar photometry package, specifically developed for the 
analysis of the data taken by \hst\ WFC3 and ACS and has 
been widely used in the \hst\ treasury and large programs, such as the Panchromatic Andromeda 
Hubble Treasury ~\citep{dal12a} and the ACS Nearby Galaxy Survey~\citep{dal12b}. 
The `DOLPHOT' package first aligns individual dithered exposures with their common stars. 
Then, the routine merges dithered exposures as a temporary merged image, on which 
the source detection is performed. For each dithered exposure, bright sources 
are used to determine the PSFs and 
aperture corrections at different locations in the WFC3/IR camera. 
Finally the routine measures stellar magnitudes not only in the temporary merged image, 
but also in the individual dithered exposures through the PSF fitting method with the centroid fixed to 
the spatial locations determined from the temporary merged image. `DOLPHOT' also gives   
the photometric uncertainty, which only considers the Poisson fluctuations produced by the electrons 
in the camera. The `DOLPHOT' parameters given in table 2 of~\citet{wil14} were adopted, because they were 
well designed to detect faint stars in crowded regions. Due to the large proper motions,  
we performed the source detection for the F153M mosaics for each of the four years separately. 
On the other hand, at the F127M band, the source detection was performed on 
individual pointings. 

The quality control parameters provided by `DOLPHOT' were used to cull the output source catalog. 
These parameters are signal-to-noise ratio (S/N), sharpness, 
crowding, flag and `CHI'. The `sharpness' determines whether one object is a star or a cosmic-ray/extended 
background galaxy. The `crowding' parameter expresses how significantly 
nearby bright sources affect photometry of dim sources. The `flag' parameter indicates  
various problems, such as too many bad and saturated pixels or photometry aperture extending off the camera. 
`CHI' is the reduced chi-square of the PSF fitting. 
The criteria given in~\citet{dal12a} were used: S/N $>$ 4, 
sharpness$^2<$ 0.1 and crowd $<$ 0.48, as well as 
flag $\leq$ 4, according to the `DOLPHOT' menu. `CHI' is normally not 
used to cull sources~\citep{dal12a,dal12b,wil14}. 
However, this quality parameter seems to be related to some potential artifacts in the F153M light 
curves (see more discussion in \S\ref{s:method} and Appendix B). We 
also removed the sources that were 
less than 5 pixels away from the CCD edge. 

We merged the source catalogs. First, we used the astrometry 
information determined above to combine the four source catalogs in the F153M band. 
We searched among catalogs for the counterparts within 0.2\arcsec\ radius and $\pm$2 mag 
difference. The 0.2\arcsec\ searching radius is large  
enough to link high-velocity stars detected at different epochs. After that, we 
obtained the magnitudes of individual sources in the 290 dithered exposures. We kept only the 
sources detected in all four catalogs\footnote{Because of the differences in the pointings and 
rotation angles, $\sim$10\% of the FoV shown in the left panel of 
Fig.~\ref{f:sgra_f153m} has been observed by only the dithered exposures in 2014. We also  
kept the sources detected in this region.}. Finally, there 
were 33,070 sources left. Second, the F127M catalogs were corrected for the astrometry and cross-correlated 
with the F153M catalog. 29,526 F153M sources have available F127M magnitudes. 
Table~\ref{t:sou} gives the ID, celestial coordinate in 2012\footnote{For the sources 
detected only 
in the FoV of the Programs GO-13316 and GO-13403, their celestial coordinates are those 
measured in 2014}, 
the mean F153M magnitude of the four catalogs and their corresponding 
uncertainty\footnote{The F153M photometric 
uncertainty ($\sigma_{F153M}$, hereafter) is the largest value among the photometric errors 
from the four catalogs derived from the artificial star tests and 
the standard deviation of the four magnitude.}, as well as the F127M magnitude and uncertainty 
($\sigma_{F127M}$ hereafter), if available. Table~\ref{t:dithered} gives the Julian dates, F153M 
magnitude ($m_i$, `i' means the i$^{th}$ exposures), S/N, sharpness$^2$, crowd, flag and CHI of sources at individual dithered exposures. 

\begin{figure*}[!thb]
  \centerline{    
       \epsfig{figure=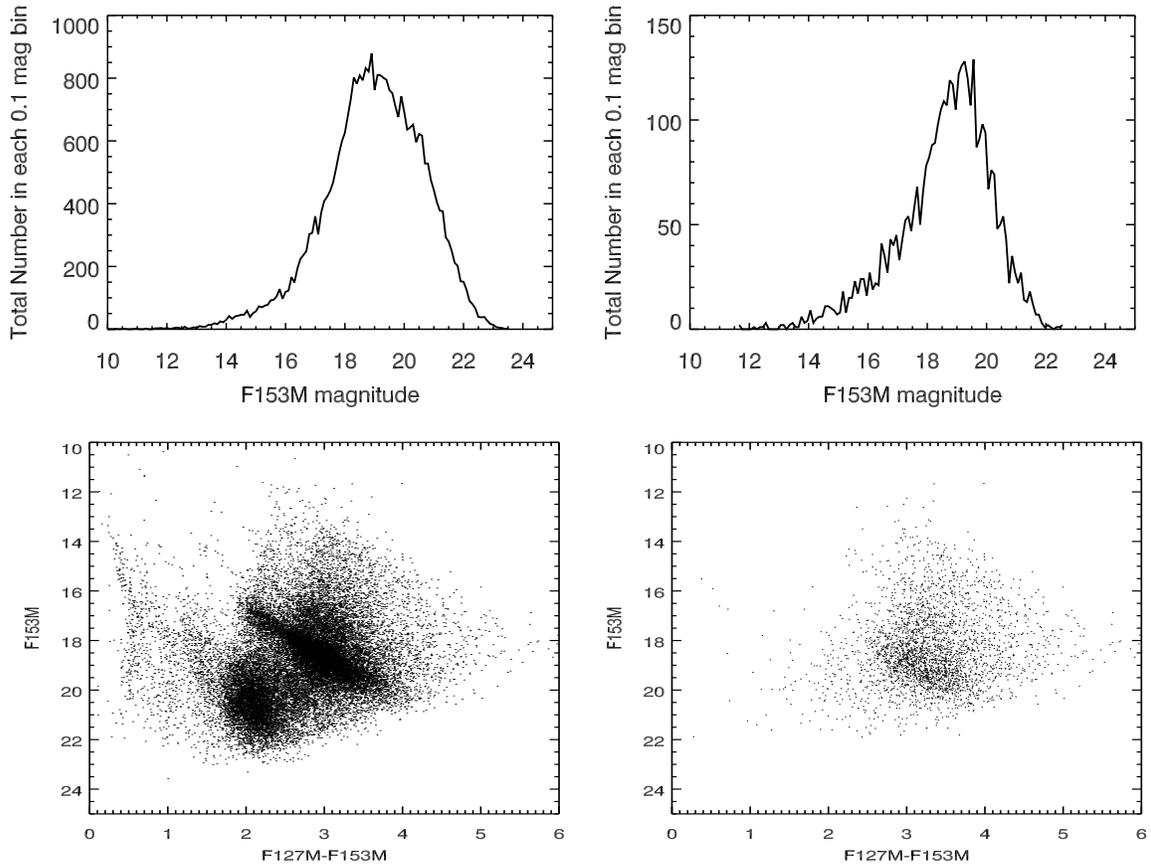,width=1.0\textwidth,angle=0}
       }
 \caption{Top panels: the F153M magnitude distribution of all detected sources (left) and variable 
 candidates (right). Bottom panels: the colour-magnitude diagram (F127M-F153M vs. F153M) of 
 the detected sources (left) and variable candidates (right).}
\label{f:mag_cmd}
 \end{figure*}

The F153M magnitude distribution and the colour magnitude diagram (CMD, F127M-F153M VS F153M) are given in left panels of Fig.~\ref{f:mag_cmd}. The diagonal structure from 
[F127M-F153M, F153M]=[2, 17] to [3.6, 20] 
is caused by the Red Clump (RC) stars. The concentration of stars at 
[F127M-F153M, F153M]$\sim$[2, 21] is the RGB or 
MS stars in the foreground Galactic bulge. Three F127M$-$F153M colours, 
1.7, 2.2 and 3.8 mag can be used to roughly divide our stars into the foreground, the less extinguished 
foreground Galactic bulge 
population, the GC population and the background (which corresponds $A_K$=1.6, 2.0, 3.5 by 
assuming the extinction law given in~\citet{sch10}). The sources with 
F127M$-$F153M$>$3.8 mag could also be red supergiants  
(such as IRS 7, $F127M$-$F153M$=4.0 mag) and AGB stars in the GC embedded in their  
circumstellar dust.

Artificial star tests were performed to determine the photometric uncertainty 
($\sigma_{F127M}$ and $\sigma_{F153M}$), magnitude variation among dithered 
exposures ($\sigma_v$, hereafter) and our detection limit. 
An artificial star catalog was produced for each source catalog, 
the luminosity, color and spatial distribution of which were similar to the observed sources. 
We used  `DOLPHOT'  to repeatedly insert independent 
batches of artificial stars into the original dithered exposures at 
both the F127M and F153M bands. 
The number of artificial stars in each batch was not allowed to 
exceed 5\% of the number observed, so that the surface density would not be 
significantly changed. The source detection of `DOLPHOT' was run on these simulated 
images again. 
Thereafter, we divided the recovered artificial stars in each of the ten regions 
defined in the right panel of Fig.~\ref{f:sgra_f153m} by their magnitudes with a 0.5 mag stepsize. 
For each magnitude and surface brightness bin, we calculated 
1) the median absolute value and 68\% percentile of the difference 
between the output and input magnitudes in the mosaic images; 2) 
the standard deviation of the output magnitudes 
of the 290 dithered exposures. 
The values in 1) then characterize the absolute photometric uncertainties,  
$\sigma_{F127M}$/$\sigma_{F153M}$ and 2)  
determines the magnitude variation, $\sigma_v$,  
among the 290 dithered exposures. 
In 1), the former value represents the systematic error introduced by 
the confusion and the latter value is the 
statistic error.  
Finally, for each source, from their mean F153M magnitudes, or $m_{i}$,  
and local surface brightness, we
interpolated the relationship determined above to obtain their 
$\sigma_{F127M}$, $\sigma_{F153M}$ and $\sigma_{v, i}$. 
More discussion about the origin and meaning of $\sigma_{F153M}$ and $\sigma_{v}$ are 
given in Appendix A. Here, we just emphasize that since $\sigma_{F153M}$ includes the systematic uncertainty 
introduced by source confusion, it overestimates the photometric variation among 
dithered exposures, especially for dim stars, while $\sigma_{v,i}$ does not show such a bias. 
Therefore, $\sigma_{v,i}$ will be used in \S\ref{s:method} to 
identify variable stars. 

\begin{figure*}[!thb]
  \centerline{    
       \epsfig{figure=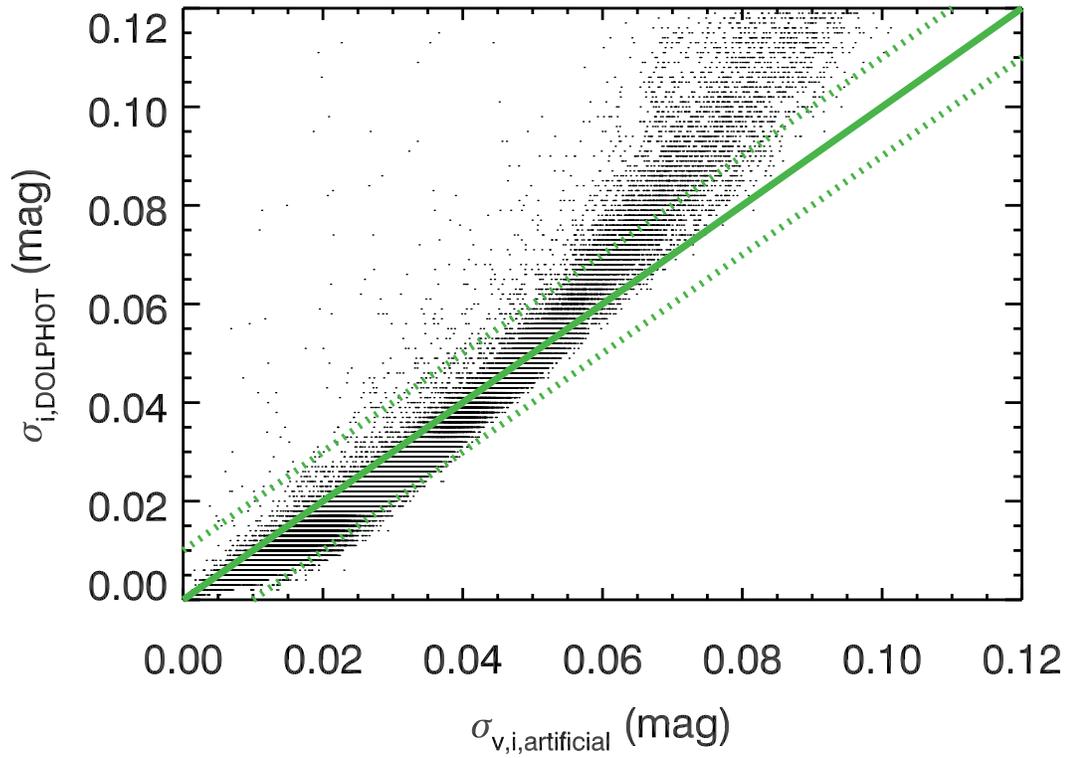,width=1.0\textwidth,angle=0}
      }
 \caption{Comparison of the $\sigma_{v,i}$ derived by the artificial star
 tests ($\sigma_{v,i,artificial}$) and the photometric uncertainties from the `DOLPHOT' ($\sigma_{i,DOLPHOT}$).
  The solid and dashed lines represent $\sigma_{i,DOLPHOT}$=$\sigma_{v,i,artificial}$ and 
  $\sigma_{i,DOLPHOT}$=$\sigma_{v,i,artificial}\pm0.01$, respectively.}
\label{f:sigma_com}
 \end{figure*}

\begin{figure*}[!thb]
  \centerline{    
       \epsfig{figure=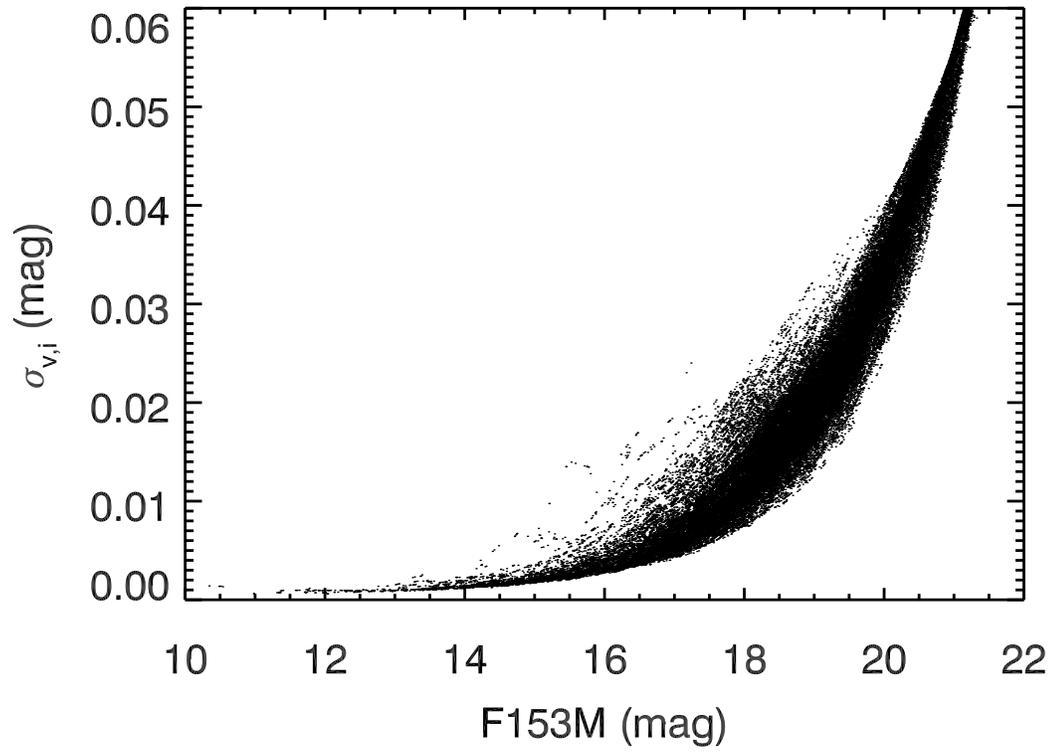,width=1.0\textwidth,angle=0}
       }
 \caption{The $\sigma_{v,i}$ as a function of the input F153M 
 magnitude.}
\label{f:mag_sigma}
 \end{figure*}

Fig.~\ref{f:sigma_com} shows the comparison of $\sigma_{v,i}$ derived from 
the artificial star tests and the magnitude uncertainty given by the `DOLPHOT' routine ($\sigma_{i, DOLPHOT}$). 
The former one is larger by up to 0.01 mag for stars with F153M $<$ 21, but smaller for 
dimmer stars. In Appendix A, we found that the latter one could explain
 the magnitude variation among dithered exposures for the stars with F153M $>$ 21 better. 
Therefore, for each star, we used the larger one of the two uncertainty measurements (artificial star test or formal uncertainty given by `DOLPHOT') 
as the final $\sigma_{v,i}$, which 
is also given in Table \ref{t:dithered}. Fig.~\ref{f:mag_sigma} presents the relationship between the 
F153M magnitude, $m_{i}$, and $\sigma_{v,i}$. 

\begin{figure*}[!thb]
  \centerline{    
       \epsfig{figure=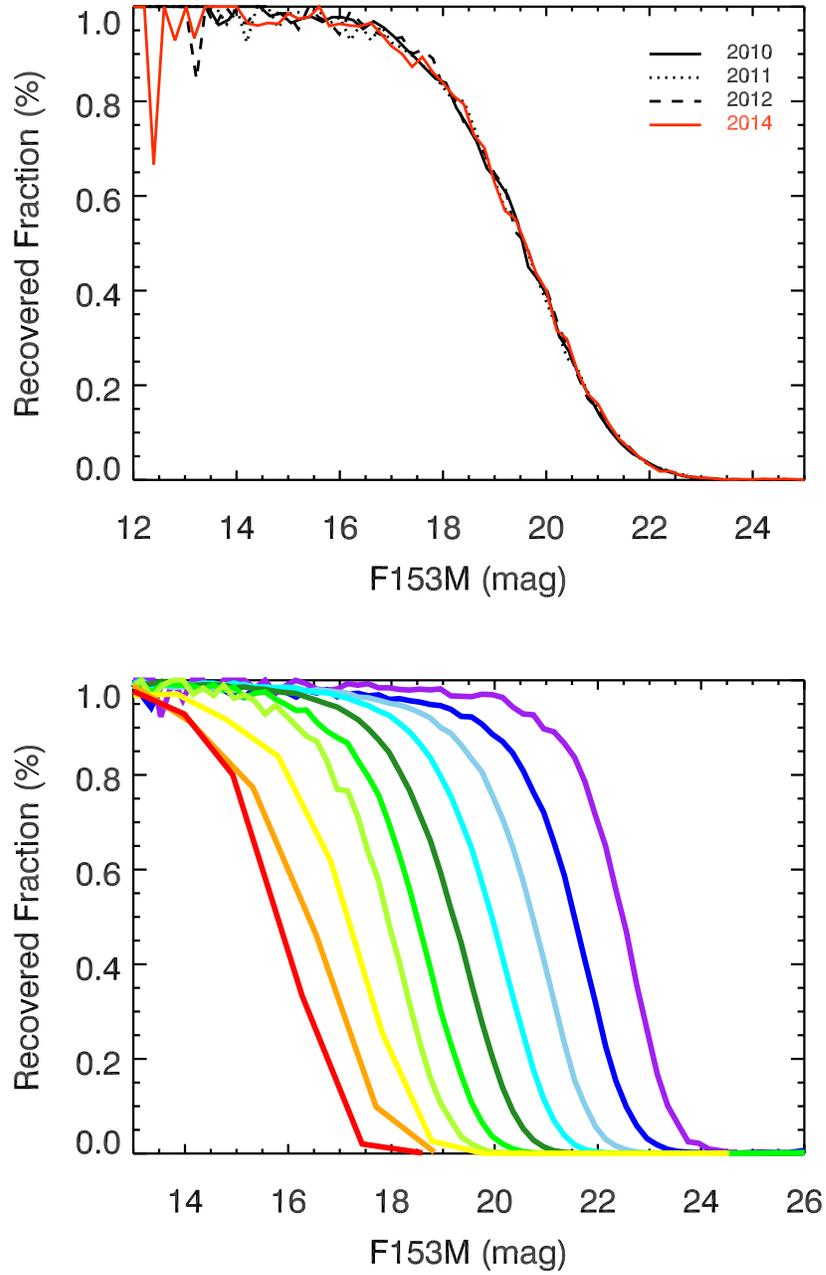,width=0.8\textwidth,angle=0}
       }
 \caption{The recovered fraction as a function of input F153M magnitude from 
 the artificial star tests for the four different years (top panel) and 
 regions with different surface brightness given in Fig.~\ref{f:sgra_f153m} (bottom panel). 
 In the bottom panel, from left to right, the regions indicated by the solid lines with 
 different colours have decreasing surface densities.}
\label{f:det_limit}
 \end{figure*}

Fig.~\ref{f:det_limit} shows the completeness as a function of the input F153M 
magnitudes for each of the four catalogs (the top panel) and regions 
with different surface brightness for the one 
in 2014 (the bottom panel). In the top panel, we find  
that even though the total exposure time in 2014 (227 dithered exposures, $\sim$68 ks) is 
more than nine times longer than those in 2010-2012 (21 dithered exposures and 7.3 ks for each year), the 
90\% (50\%) completeness detection limits of the four years are similar: $\sim$17.4 mag (19 mag). This illustrates that 
the observations are limited by confusion, and increasing the exposure time 
does not lead to the detection of more 
faint stars. The 90\% (50\%) completeness detection limits 
of different regions are given in Table~\ref{t:dl}. In the F153M band, we can 
detect sources in the lowest surface brightness region that are 6 mag fainter
 than in the highest surface brightness region. 
 
\section{Methods}\label{s:method}
We used the least $\chi^2$ method to select variable stars. For each source, 
\begin{eqnarray}
weight_i=\frac{1/\sigma_{v,i}^2}{\sum 1/\sigma_{v,i}^2}\\
\bar{m}=\sum m_i \times weight_i\\
\chi^2/d.o.f = \sum_i \frac{(m_i - \bar{m})^2}{\sigma_{v,i}^2}/(n-1)
\end{eqnarray}
where `n' is 
the total number of dithered exposures used and the sums are over the `n' exposures. 	
We only accepted the photometry from the dithered exposures with 
S/N$_{i}>$4, sharpness$^2_i<$0.1, crowd$_i<$0.48 and flag$_i<=$4. At least three data points ($n\geq3$) are 
needed to calculate $\chi^2/d.o.f$. 

\begin{figure*}[!thb]
  \centerline{    
       \epsfig{figure=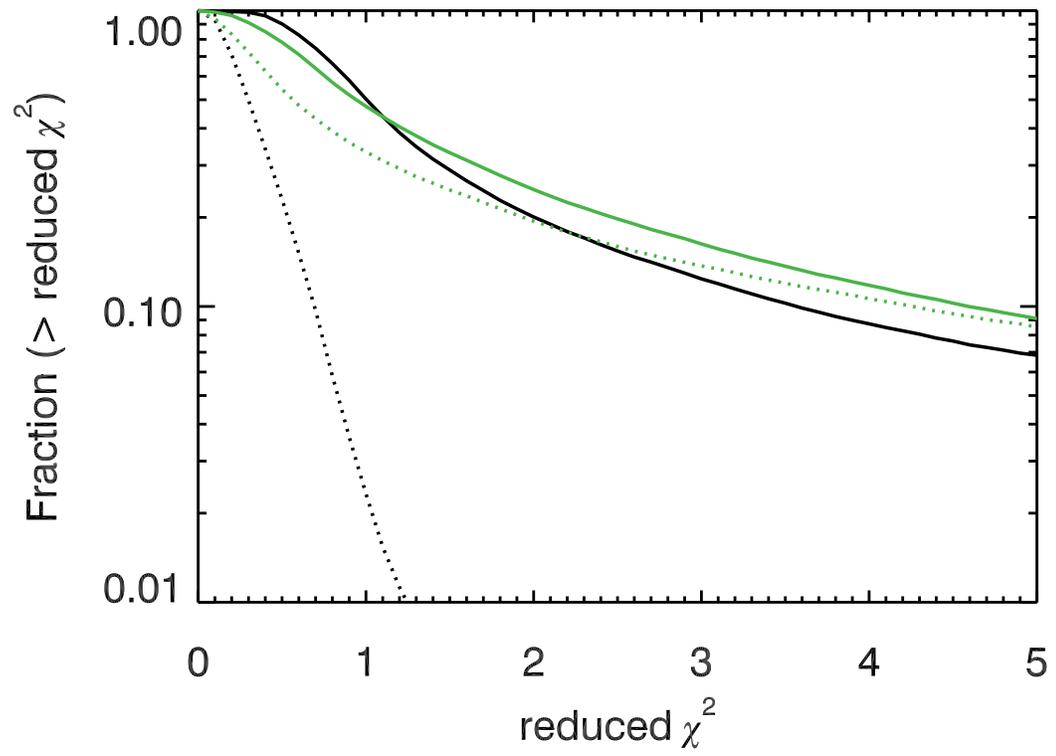,width=1.0\textwidth,angle=0}
       }
 \caption{The cumulative distribution 
 function ($>$ reduced $\chi^2$) for $\chi^2_{y}$ (green solid line), 
$\chi^2_{d}$ (green dotted line), $\chi^2_{y,b}$ (black solid line), 
$\chi^2_{d,b}$ (black dotted line), see \S\ref{s:method} for a description of these values.}
\label{f:chi_dis}
 \end{figure*}

We analyzed the variability in two categories: among years (long-term) or days (short-term). 
For the former case, all 290 dithered exposures are used to calculate $\chi^2/$d.o.f 
($\chi^2_y$ hereafter, the subscript `y' means `year'). For the latter case, 
we used only the dithered exposures in three periods: groups 5, 6 and 7 in Table~\ref{t:obs}. 
These groups are characterized by long-duration observations ($\geq$10 hours). We  
derived $\chi^2/$d.o.f for each of the three periods and 
used the maximum value ($\chi^2_d$ hereafter, the subscript `d' means `day') 
as the criterion for whether a star is variable or not.  We added 
 0.025 and 0.01 mag systematic uncertainties to $\sigma_{v,i}$, when 
calculating $\chi^2_y$ and $\chi^2_d$, respectively (see Appendix A). 
Table~\ref{t:sou} gives the $\chi^2_y$ and $\chi^2_d$ for individual sources. 
Table~\ref{t:sv} gives the 68\% and 90\% dispersions, as well as the 
median of the $\chi^2_y$ and 
$\chi^2_d$ values. 
The median values are around one, which means 
that $\sigma_{v,i}$ is reasonably estimated. 
Fig.~\ref{f:chi_dis} gives the 
cumulative distribution functions 
of $\chi^2_y$ (the green lines) 
and $\chi^2_d$ (the black lines). 
As expected, $\chi^2_y$ extends to larger values than $\chi^2_d$ because 
the latter is not sensitive to long-term variables.

\begin{figure*}[!thb]
  \centerline{    
       \epsfig{figure=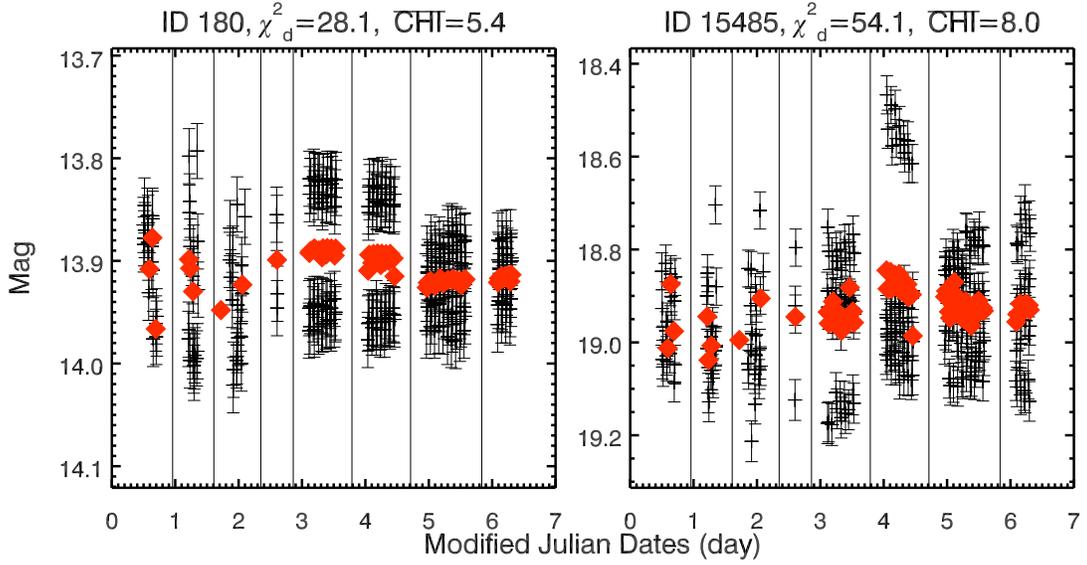,width=1.0\textwidth,angle=0}
       }
 \caption{The \hst/WFC3 F153M light curves of two stars 
 with different magnitudes, the variability of which is not considered to be real. 
 The black data points are from the 290 dithered exposures, while the red filled 
 diamonds are the magnitudes after the binning procedures described in \S\ref{s:method}.
 The seven vertical lines divide the data points into eight groups as shown in Table~\ref{t:obs}: 
 The first three groups are observations in 2010, 2011 and 2012, while the last five groups are 
 observations from February to April, 2014. 
 For easy demonstration, a constant has been subtracted from the Julian 
 dates of dithered exposures in each group, so that the modified Julian date of the first dithered 
 exposure in each group is 0.5 day later than the last 
 dithered exposure in the previous group. In the title of each figure, we give the source ID, 
 $\chi^2_{d}$ as defined 
 in \S\ref{s:method}, and $\overline{CHI}$ defined in Appendix B.}
\label{f:artifact}
 \end{figure*}

When visually examining light curves of individual sources, we found that 
the majority of sources with large $\chi^2_d$ showed rapid, alternating 
intensity variations among 
dithered exposures (e.g., Fig.~\ref{f:artifact}). These sources widely distribute in the detector, but 
preferred the regions with high surface brightness. Although we could not exclude 
the possibility that this variability was real for some individual stars, 
we suspected that this phenomenon was typically due to 
instrumental effects (see more discussion in Appendix B). Therefore, 
we binned the data for individual sources  
to reduce this effect on the derived $\chi^2_d$ and $\chi^2_y$. For the 57 pointings in 
2014, we binned the dithered exposures in individual pointings. 
For each of 
the three pointings from 2010 to 2012, we divided the 21 dithered exposures into five bins (the first four bins  
include four dithered exposures, while the last one includes five exposures), 
so that each bin has similar total  exposure time   
to those in 2014.
Therefore, in Eqns.(1)-(3), m$_i$ is the mean magnitudes 
of these dithered exposures in each bin and the uncertainty is the maximum of 
their standard deviation and their $\sigma_{v,i}$, then divided by the number of 
dithered exposures in each bin. Using the standard deviation of magnitudes of
individual dithered exposures can cause the pointings with outliers in magnitude to  
have larger uncertainty and lower weight in the final reduced $\chi^2/d.o.f$, while using 
the $\sigma_{v,i}$ can prevent the bins which have very similar magnitudes among 
dithered exposures in these bins 
to have overly small uncertainty and too much weight in the reduced $\chi^2/d.o.f$. 
The next step was to use  
Eqns. (1)-(3) to derive $\chi^2_{y,b}$ and $\chi^2_{d,b}$ (the extra subscript `b' means 
`bin'), which are also given in Table~\ref{t:sou}. 
Table~\ref{t:sv} gives the statistic distributions of $\chi^2_{y,b}$ and 
$\chi^2_{d,b}$ and Fig.~\ref{f:chi_dis} gives their 
cumulative distribution functions (green and black dotted lines). 
From the cumulative distribution, we conclude that 
the difference between $\chi^2_{y}$ and $\chi^2_{y,b}$ is smaller than that 
between $\chi^2_{d}$ and $\chi^2_{d,b}$. 
 
By using the procedure given in \S\ref{s:method}, we found 3879 and 77 
sources in our two categories with variability over the four-year 
baseline and within one day, respectively. 
The former sources have $\chi^2_y$$>$3, $\chi^2_{y,b}$$>$3, 
while the latter have $\chi^2_d>2$, $\chi^2_{d,b}$$>$2. 42 stars  
fall in both categories.

Further visual examination was carried out individually for all 
these variables by six people (HD, RS, FN, EG, TG and Siro Benvenutto Gallego, 
the son of EG). We only considered a source as a real variable star, if it was selected 
by more than half of us. 
This examination was 
useful to remove sources with apparently large $\chi^2$, 
caused by potential artifacts. In total, 
 34 and 4 stars are removed from the two categories as non-variable stars. 
On the other hand, we also included 3 additional sources: IDs 2495, 11656 and 12097,
 with $\chi^2_d>2$, $\chi^2_{d,b}<2$ 
due to large photometric uncertainty, as potential candidates having periods of days. 
In the final tally, there are 3894 variable 
 candidates: 3845 and 76 sources are in the two categories, respectively, 
 and 27 candidates are in both. 
 These sources are listed in Table~\ref{t:sou}.

 \section{Results}\label{s:result}
\begin{figure*}[!thb]
  \centerline{    
       \epsfig{figure=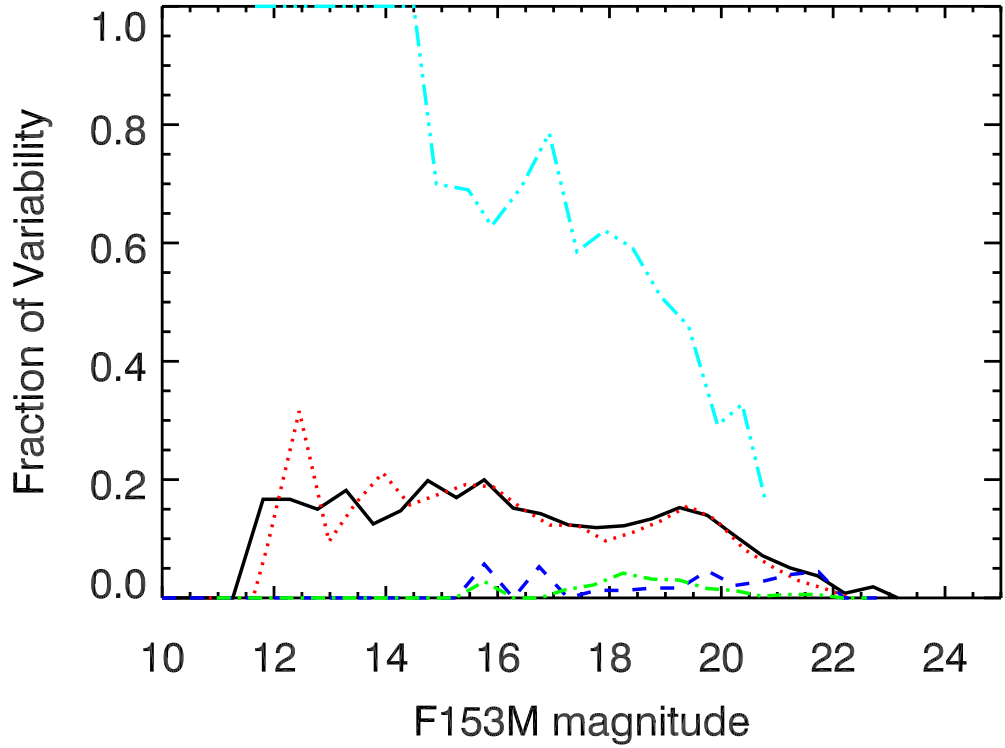,width=1.0\textwidth,angle=0}
       }
 \caption{The fraction of variable stars as a function of the F153M magnitude. The black solid 
 line: all stars, blue dashed: stars with F127M-F153M$<$1.7 (presumably foreground stars), 
 green dash-dot: 1.7$<$F127M-F153M$<$2.2 (foreground Galactic bulge), red dotted: 
 2.2$<$F127M-F153M$<$3.8 (GC) and cyan dash dot dot: 
 background or AGB stars in the GC embedded in 
 the interstellar dust.}
 \label{f:var_frac}
 \end{figure*}

\begin{figure*}[!thb]
  \centerline{   
       \epsfig{figure=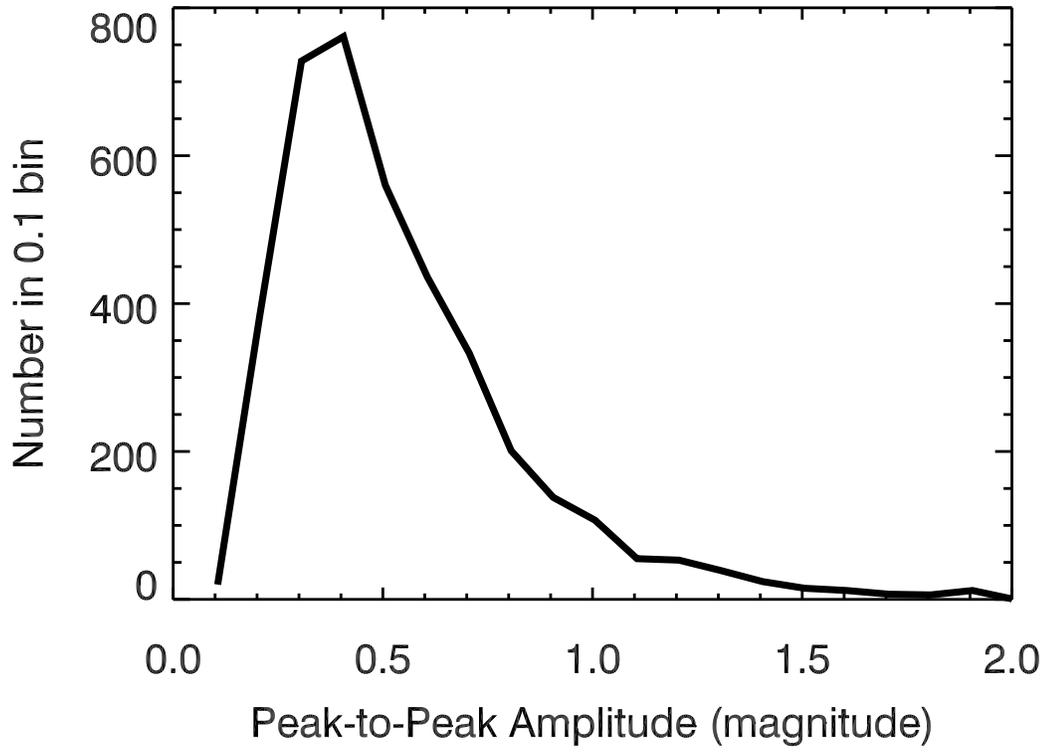,width=1.0\textwidth,angle=0}
       }
 \caption{The peak-to-peak amplitude distribution of the variable stars.}
 \label{f:amplitude_var}
 \end{figure*}

The right panel of Fig.~\ref{f:mag_cmd} and Figs.~\ref{f:var_frac} 
and~\ref{f:amplitude_var} show 
the magnitude distribution, CMD, the fraction and 
PPA distributions of variable stars. 
The variable candidates are widely distributed throughout the CMD. 
indicating their strong differential reddening and 
various origins.
Except for a few 
foreground stars, the majority of variable 
candidates should be in the Galactic bulge or MWNSC, according to their 
F127M-F153M colour. 
Fig.~\ref{f:var_frac} indicates that the fraction of variable stars 
is roughly 15\% for all the magnitude ranges and 
increases with the F127M-F153M color. Especially, 51\% of stars 
with F127M-F153M$>$3.8 are variables. The PPA distribution 
of variable stars has 
a peak around 0.5 mag. 

\begin{figure*}[!thb]
  \centerline{    
       \epsfig{figure=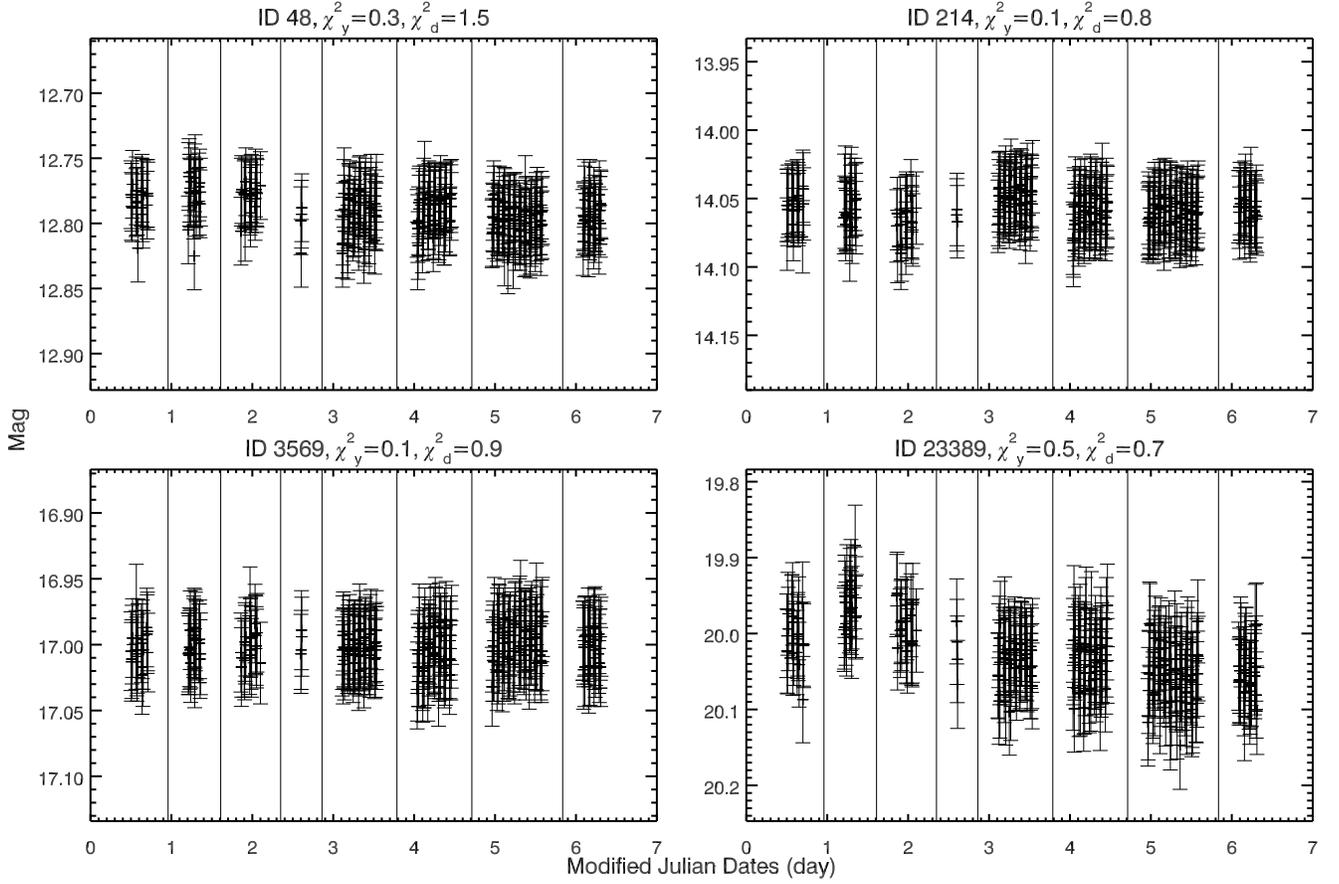,width=1.2\textwidth,angle=0}
       }
 \caption{The \hst/WFC3 F153M light curves of four non-variable stars 
 with different magnitudes. The symbols (`pluses' and vertical lines) are the 
 same as those in Fig.~\ref{f:artifact}. In the title of each figure, we give the source ID, 
 $\chi^2_{y}$ and  $\chi^2_{d}$ defined in \S\ref{s:method}.}
\label{f:noval}
 \end{figure*}

\begin{figure*}[!thb]
  \centerline{    
       \epsfig{figure=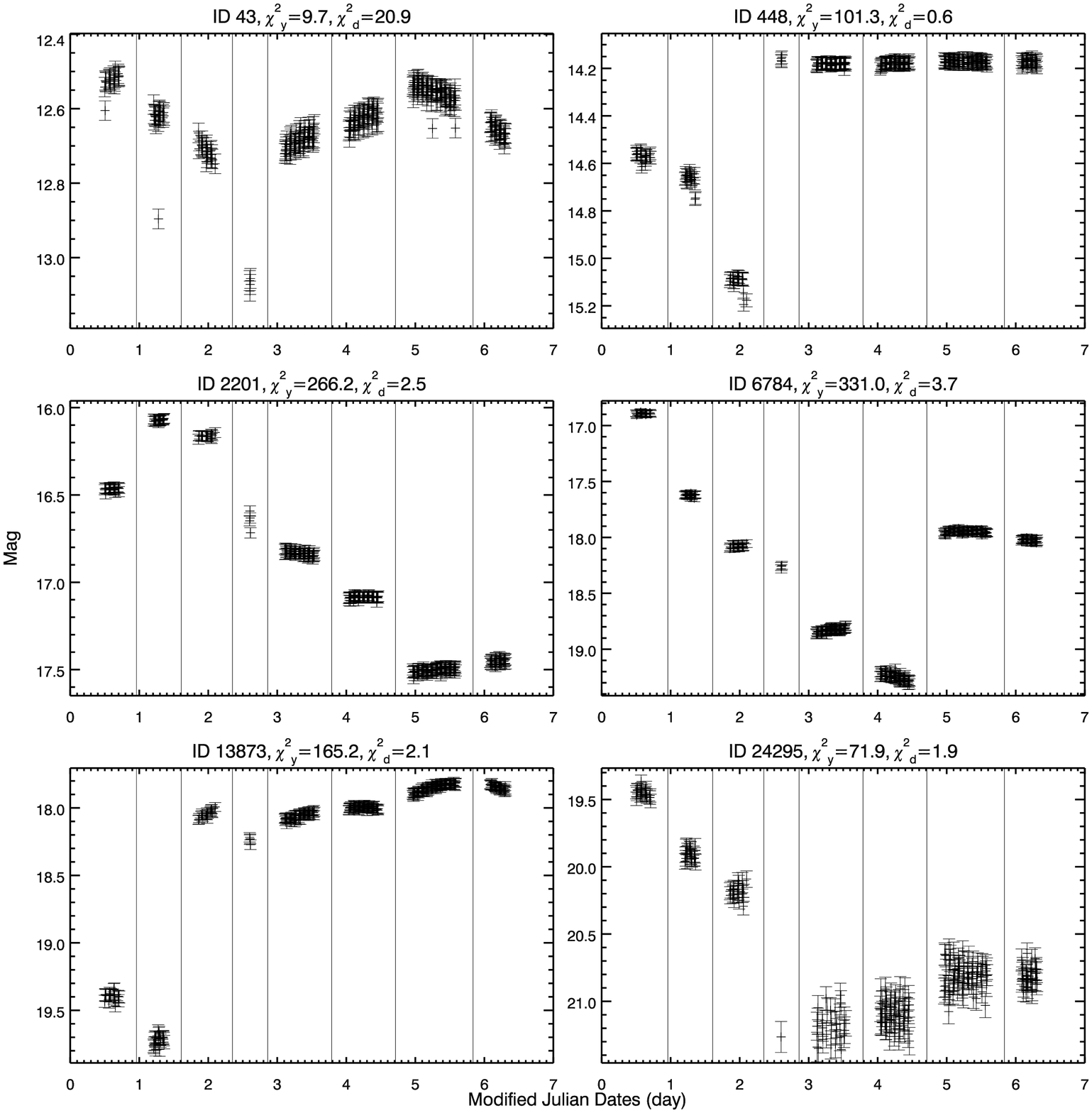,width=1.2\textwidth,angle=0}
       }
 \caption{The \hst/WFC3 F153M light curves of four stars 
 with different magnitudes, which are identified to vary on yearly time scales. The symbols 
 (`pluses' and vertical lines) are the 
 same as those in Fig.~\ref{f:artifact}. In the title of each figure, we give the source ID, 
 $\chi^2_{y}$ and  $\chi^2_{d}$, as defined in \S\ref{s:method}.}
\label{f:longval}
 \end{figure*}

Fig.~\ref{f:noval} shows examples of light curves for 
non-variable stars, while Fig.~\ref{f:longval} illustrates 
light curves for stars which vary among years. 
 This sample of stars covers a 
 large F153M magnitude range from $\sim$ 12 to 20 mag. The first three sources 
 in Fig.~\ref{f:noval} exhibit very little variation, if any; The standard deviations of their 
 magnitudes are smaller than 0.014 mag. While the deviation 
 increases to 0.04 mag for ID 23389, the large 
 $\sigma_{v,i}\sim$0.1 mag excludes it as a variable star. In contrast, 
 the six stars in Fig.~\ref{f:longval} all show  
 large variations 
with PPAs from 1 (ID 448) to 2 mag (ID 6784). 
The magnitudes of IDs 43, 2201, 6784 and 13873 with $\chi^2_d>2$ 
also vary on daily timescale.

From the light curves, we divided our 76 stars with intraday variability 
 into three subgroups, based on the difference in their observed light curve characteristics: 
 \begin{itemize}
 
 \item Subgroup 1: Stars with a significant fraction of their periods covered 
 by our dataset, especially 
by the observation block on April 2-3, 2014. The periods of these stars can be accurately 
determined (see more discussion in \S\ref{sss:rrl}).  
 Figs.~\ref{f:rrlyrae_1} to~\ref{f:rrlyrae_3} show the light curves for the 28 
stars in this subgroup. 

\item Subgroup 2: Eclipsing binary candidates, with the dips being a small fraction of their periods, 
fortunately covered 
by our dataset 
(Fig. \ref{f:eclipse}). ID 1365, a known contact massive eclipsing 
binary system~\citep[E60 in][]{pfu14}, shows a sinusoidal light curve. 
The other stars in this subgroup show similar features in their light curves: 
the magnitudes are roughly constant for awhile, become dimmer during a short period 
and quickly return to their constant value. Except for IDs 2980 and 21717, 
the periods of the other sources seem to be 
longer than those in subgroup 1 (see more discussion in \S\ref{sss:eclipse}). 

\item Subgroup 3: The remaining sources 
cannot be unambiguously assigned to any variability subgroup, because our dataset 
only includes a small fraction of their periods, due to the poor temporal sampling. 
Fig. \ref{f:shortval} gives 
nine examples. 
 \end{itemize}
  
As a next step, we tried to derive the periods of some short-term variables. Because large gaps, 
from 10 to 547 days, exist  
between the observing runs in 
Table~\ref{t:obs} and the longest duration of any individual observing run 
is only 15 hours, 
it is not possible to unambiguously determine the periods for most of our variable 
candidates. We focused on stars in Subgroups 1 and 2. We 
used the Lomb-Scargle periodogram~\citep{lom76,sca82} technique to 
calculate the periods of 36 stars given in Figs.~\ref{f:rrlyrae_1} to~\ref{f:eclipse} 
(except for ID 30400 in Fig.~\ref{f:eclipse}, because its light curve includes only one dip). 
Specifically, we used the 
IDL routine `lomb.pro' to perform the calculation. 
The routine calculates the power spectral density of the Fourier 
components of the light curves. 
The routine does not account for the binary 
nature of eclipsing systems, so treats the two dips 
in the light curves of eclipsing binary candidates 
 as the same origin. Therefore, we 
 doubled the period derived from the periodogram analysis for these sources. 
The IDs, periods, F153M magnitudes and colours of the 
36 sources are given in Table~\ref{t:period}.

 \section{Discussion}\label{s:discussion}
In this section, we first compare our results with previous works 
in \S\ref{ss:previous} to check the efficiency of our method 
and dataset to identify variables. Then, in \S\ref{ss:cross} 
we cross-correlate our source catalog with existing 
catalogs having spectroscopic identifications. We discuss 
the variable types for the stars with determined period in 
\S\ref{s:result} and then statistically classify the other variables 
without period measurements in \S\ref{ss:ec_sc} and 
\S\ref{ss:origin}, respectively.  

\subsection{Comparison with Previous Work}\label{ss:previous}
The reliability of our variability analysis can be assessed by comparing our results with 
those of prior studies. Table~\ref{t:references} summarizes existing variability studies in the 
GC, the instruments that were used, the FoV and covered time range, as well as 
the total number of 
variables found and the number that we recovered. 

Variability studies of the GC began in the 1980s~\citep{hal89}. 
Small telescopes ($<$ 4 m) were used to observe the GC in a 
relatively small number of epochs. Therefore, these studies could only identify variable 
stars, but could not give their periods. 
\citet{tam96} and~\citet{blu96} reported  
IRS 7, 9, 12N, 28, 14SW and 10E (IRS 10E* or IRS 10EE) 
to be variable stars, all of which were also identified as variables by our method.

Some long-term monitoring studies of the GC with large FoV camera 
were published later~\citep{gla01,mat09,mat11,mat13,pee07}. 
\citet{gla01} reported    
409 LPV candidates in the central 24\arcmin$\times$24\arcmin\ field, 
 10 of which fall into our FoV. Three of these 10 stars failed to appear as variables 
in our study for good reasons: two with periods of $\sim350$ d  
are near the edge of our FoV, detected only 
in March and April 2014, while the  
other, `3-2389' in Table 2 of~\citet{gla01}, has a very long period of 408 
d and a $K$ band amplitude of 1.4 mag.  
`3-2389' has been detected in a SiO maser survey~\citep{deg04} and 
should be a Mira. In our dataset, its magnitude increased 0.1 mag 
from 2010 to 2012, and then appeared constant from 
February to April, 2014. Therefore, our dataset could 
miss the part of light curve in which the magnitude changes significantly. 

\citet{mat09,mat11,mat13} identified 549 Miras, three CCEPs, 18 T2Cs, 
24 eclipsing binaries, one pulsating star and one Cepheid-like 
variable from the SIRIUS survey\footnote{Simultaneous 
3-colour InfraRed Imager for Unbiased Surveys (SIRIUS) was 
taken by the Infrared Survey Facility (IRSF) 
in South Africa, with a pixel scale of 0.45\arcsec~\citep{nag03}. 
The survey includes the region $|l|$ $<$ 2 degree and $|b|$ $<$ 1
degree. }, which monitored the central 20\arcmin$\times30$\arcmin\ field from 2001 to 2008.
Only 17 of their Miras 
fall into 
the FoV of our dataset, two of which are not classified 
as variables by us. One is near the edge of 
our FoV, only detected in April, 2014. The other one is `604' in Table 6 
of~\citet{mat09} with H=12.48 and PPA in the H band = 0.38 mag. 
According to its celestial 
coordinate, its counterpart should be ID 105 (0.06\arcsec\ away) in our catalog, 
with F153M = 13.5 mag, 
$\chi^2_y$=1.6 and PPA = 0.17 mag, a non-variable source in our catalog. 
Instead, about 4.4\arcsec\ away, ID 87 is variable and has F153M=13.4 mag
, $\chi^2_y$=9.4 and PPA=0.32 mag. We suspect that maybe ID 86 is the real 
counterpart of `604' in~\citet{mat09}.

\citet{pee07} reported  
 98 variable stars and 14 stars with well defined periods separately in their Tables 1 and 2. 94 and 
 14 of them fall into our field-of-view and 37 and 11 are classified by us as variables.  
Because Peeples et al. used a crude root-mean-squared cut method to identify variable stars 
 (log (rms) $>$ a$\times m_o$ +b, where `rms' and `m$_o$' are the standard deviation and mean 
 of the observed magnitudes, while `a' and `b' are coefficients defined by Peeples et al.), 
 we suspect that 
 their catalog contains many false  variables. Therefore, we concentrate 
 on their stars with well defined periods. Among the three of the 14 stars, which are not 
 classified as variable in our dataset, PSDJ174540.25-290027.2 (IRS 16NE) has been 
 considered to be photometric non-variable in many 
 references, such as~\citet{raf07} and~\citet{pfu14}. Its light curve in Fig. 12 
 of~\citet{pee07} is very noisy. We suspect that the period claimed by~\citet{pee07} is due to 
 some instrumental effects. 
The other two stars are  PSDJ174535.60-290035.4 and PSDJ174540.16-290055.7. We 
checked sources near them in Table~\ref{t:sou}, and found that, while the magnitudes 
of some of them seem to change from 2010 to 2012, they do not pass our 
selection criteria. Therefore, 
 these two sources need further investigation. 

\citet{ott99} performed the first high-angular-resolution variability study of 
the GC (FWHM $\sim$0.15\arcsec ), using the speckle camera 
SHARP on the ESO NTT. 
They also used the least $\chi^2$ method and 
found that 122 of the 218 stars in their sample with 
$m_K<$13 showed variability larger than 3 sigma. In particular, they 
 first reported that IRS 16SW (ID 43, in Fig.~\ref{f:longval}) is an eclipsing binary. 
 In our data set, 195 of their 218 stars have counterparts, 
 115 of which are variables, according to Ott et al. However, only 32 
 of the 195 stars are identified as variables
 by our method. 
Fig.~\ref{f:ott_compare} compares 
 the reduced $\chi^2$ of the 195 stars from Ott et al. and our dataset. There is 
 no clear correlation. We suggest that this inconsistency could be due to two problems 
 in the method used in Ott et al.: 
 1) Unsuitable choice of flux calibrators. Ott et al. anchored the photometry to 
eight stars, including IRS 16C, IRS 16NW and IRS 29N, 
which they assumed to be non-variable, even though they found that the 
reduced chi-squared of IRS16C reaches 4.1. Instead,~\citet{raf07} 
identified IRS 16NW and IRS 29N (see also~\citealt{hor02}) as variable stars with 
$\sim$0.2 mag variation in the $K$ band.
Our dataset identified IRS 29N as a variable star as well, with PPA equal 
 to 0.66 mag in the F153M band. IRS 16C has 
 $\chi^2_y$=4.7, $\chi^2_{y,b}$=2.0 and PPA=0.3 mag. The systematic 
 uncertainty introduced by the unsuitable flux calibrators could have led to the 
 misidentification of  some bright 
 non-variable sources with small photometric uncertainties as variable stars. 
 Instead, 
 for our \hst\ dataset, the systematic uncertainty of 
 its photometric zeropoints is only 1\%. 
2) Ott et al. determined the photometry using three methods 
  with the same empirical PSF extracted from the images. The standard deviation of the 
  photometry derived from these methods is used to 
 represent the real fluctuation variation among exposures. We suspect that this method 
 could underestimate the real uncertainty, especially for the dim stars.
  For example, Fig.~\ref{f:ott_compare_dmk} shows the $K$ band magnitude distribution
  ($m_K$, from Table 2 in Ott et al.) for the variable stars defined by 
 Ott et al., but classified as non-variable (solid line) and variable (dashed line) stars 
 by our dataset. 30\% in the former case have $m_K>$12.5, while this value decreases 
 to 12.5\% for the latter cases. 
  
 Since 2005, more long term observations of the GC 
 taken by 8m-class telescopes have become available.~\citet{raf07} performed a  
variability study in the central 5\arcsec\ with a 10 year baseline observation with 
Keck/NIRC camera. They found 15 variables out of 131 stars with $m_k$$<$16. 
Compared to their study, which reaches an angular resolution of about 0.06\arcsec , 
in this region, our dataset suffers from confusion due to the poor resolution. 
Therefore, only 10 of their 15 variables are detected and only 3 of those ten stars, 
IRS 16SW, IRS 29N and S0-34 have
been identified as variables. Both IRS 16NW and IRS 16CC\footnote{~\citet{raf07} classify 
IRS 16CC as variable at the `L' band, not the `K' band. } are 
classified as variable by Rafelski et al., but 
not by us (PPA=0.22 mag and 0.16 mag), or by~\citet{ott99} or~\citet{pfu14}. 
\citet{pfu14} used \vlt/SINFONI spectroscopy and NACO imaging data 
from 2003 to 2013 and identified IRS 16NE as a non-variable star 
in photometry\footnote{The radial velocity of IRS 16NE change with a 
period of 224 days, which suggests that it is a binary system. 
The non-variability in its photometry could 
probably be due to the inclination angle. }
and E60 as a close 
WR eclipsing binary 
with a period of 2.276 days, which  
are consistent with our dataset (IRS 16NE: ID 14, E60: ID 1365). 
Thanks to the stable 
performance of the \hst/WFC3 IR camera, our light curve for E60 (the top left figure in Fig.~\ref{f:eclipse}) 
has significantly smaller photometric uncertainties than that taken by 
the NACO camera on the 8 m \vlt\ telescope (Figure 3 in~\citealt{pfu14}). 

Besides near-IR variability, \citet{rei07} found 15 SiO 
masers within 50\arcsec\ of Sgr A*, 11 of which are not 
associated with any LPV candidates given in \citet{gla01}. 
We find 
near-IR counterparts for all of these maser stars, except for IRS 19NW, 
SiO12 and SiO 17. 
The \hst\ WFC3/IR image did not allow us to disentangle
IRS 19NW from the bright non-variable IRS 19 (ID 26, F153M=12.2 mag, 
$\chi^2_y$=0.55, $\chi^2_{y,b}$=0.26), while the giant 
counterparts of SiO 12 and SiO 17 could be 
background stars that are not detected 
because of large foreground extinction. 
11 of the 12 masers that we detect are classified as variable stars;  
SiO11 does not fully pass our criteria ($\chi^2_y$=5.3, $\chi^2_{y,b}$=2.9, PPA=0.28), but 
its variability is apparent in its light curve. 

Table~\ref{t:reference_var} summarizes the cross-correlation between 
our source catalogs and those in the previous variable studies. 
In summary, the four-year baseline \hst/WFC3 observations used here form a unique dataset for  
identifying variable stars in the GC, except in the central few arc seconds, due to 
the strong source confusion there. Thanks to the stable photometric zeropoint of 
these observations and the method which we used 
to determine $\sigma_{v,i}$, we are able to confirm most of the well-measured  
variable stars found in previous work. 

\subsection{Cross-correlation with Spectroscopic Catalogs}\label{ss:cross}
In this subsection, we cross-correlate our variable source catalog with existing 
spectroscopic observations, which will help us understand their stellar types.   
Specifically, we compare with 
1)~\citet{pau06},~\citet{bar09} and~\citet{fel15}, which concentrate on massive stars; 2)
~\citet{man07} and~\citet{fel17}, which focus on late-type stars; 3)~\citet{blu03},
~\citet{do13},~\citet{sto15} and~\citet{nis16}, which include both 
early-type and late-type stars. 
Most of the targets 
of the spectroscopic observations are brighter than 16 mag in the $K$ band. 

Table~\ref{t:reference} summarizes our variable star candidates 
with spectroscopically studied counterparts. 
In total, \citet{pau06}, \citet{bar09} and~\citet{fel15} identify (or tentatively classify) 153 WR 
stars, OB supergiants and 
main-sequence stars. We detected all of them, 
except for 22 near Sgr A* 
due to confusion. 
38 of these 131 stars  are identified as variables. The cause of variability 
of these early-type stars is probably binarity or confusion with other sources. 
We will give a more thorough discussion 
of the binary fraction 
of massive stars in the MWNSC in Dong et al. 2017c (in preparation).  
\citet{man07} and~\citet{fel17} list 329 and 990 late-type stars (i.e. cool stars with 
molecular CO absorption lines\footnote{Late-type stars (F-type or later,~\citealt{do15}) are not equivalent to 
evolved, low mass stars, such as AGB stars or RGB stars. Late-type stars  
could also be red supergiants (luminosity class I), related to recent 
star formation, for example, IRS 7~\citep{car00,pfu11,fel17}.}), 
21 and 223 of which, respectively are variables, according to our detections. 
\citet{blu03} derive surface temperatures  
and bolometric luminosities for 79 
stars in the GC, which includes AGB and supergiant stars. 78 of these 79 stars fall into our 
FoV and are divided into three groups (according to~\citealt{blu03}): 20 as 
Miras, 43 as giants (III), 15 as supergiants (I). We identify 
17 (85\%), 17 (40\%) and 4 (27\%) of the 
stars in these three groups respectively as variables. 
\citet{do13}, \citet{sto15} and \citet{nis16} report 286, 371 and 20 stars, respectively, 
including 86 early-type stars (i.e. hot stars without CO absorption lines, 
definitely related to recent star formation), 
536 late-type stars and 55 
without spectral identifications. 29, 16 and 7 of the stars 
listed by~\citet{do13}, \citet{sto15} and \citet{nis16} show 
variability. Except for 7 and 5 variable stars in~\citet{do13} and~\citet{nis16}, 
all the other variable stars are defined as late-type stars.

In summary, 319 variable stars in our catalog have available 
spectroscopic identifications, which includes 13 WR stars, 19 OB stars, 4 supergiants, 13 early-type stars, 18 Miras, 17 giants and 235 late-type stars.

\subsection{Classifying Variables with Well-Determined Periods}\label{ss:ec_sc}
In this subsection, we attempt to assign variability types to our identified 
variables having well-determined light curves and measured periods.   
Those without period measurements are considered in the next subsection.

\subsubsection{Eclipsing Binaries}\label{sss:eclipse}
Depending on the distances between stars 
and their sizes, eclipsing binaries can be divided into detached, semidetached 
and contact binaries. Since the stars in a detached binary are still within their Roche lobe and 
evolve separately, the light curve stays constant except when one star 
eclipses the other. The depth of the dip caused by the eclipsing depends 
on the relative sizes of the two stars. 
 In a contact binary, the two stars are so close 
to each other that their filled Roche lobes interact. As a result, there is no plateau  
in the light curves. The light curve of a semidetached binary lies between those of the 
detached 
and contact binaries. 

There are ten eclipsing binaries in our variable catalog (ID 43 in Fig.~\ref{f:longval} and 
nine in Fig.~\ref{f:eclipse}). Most of them 
should be detached binaries, for example, 
most of those in Fig.~\ref{f:eclipse}. Among them, 
ID 2980 shows quick variation with 
a period of $\sim$0.5 d. The difference in the depth of the dips excludes 
the possibility that the variation is due to a pulsating single star (see top right panel in 
Fig.~\ref{f:eclipse}, especially 
in Group 5 with modified Julian dates $\sim$ 3.2 d). The 
other three, ID 43 (IRS 16SW,~\citealt{raf07}), 
 ID 1365 (E60,~\citealt{pfu14}) and ID 217171 
 are 
 contact binaries, the light curves of which have a sinusoidal shape. 
 According to the F127M-F153M colours of the ten eclipsing binaries 
 given in Table~\ref{t:period}, two of them, ID 1404 and 
 ID 2980, are in the foreground, three are in the Galactic bulge and the other 
 six, including ID 43 and ID 1365, are within the GC. 

Our variable star catalog may contain unclassified eclipsing binaries. 
The periods as well as the duration and amplitude of 
eclipses are determined by physical parameters of the 
binary systems. Due to inadequate sampling, however, our data 
may not cover the light curves of many potential 
binaries adequately. Therefore, a significant fraction 
of the 76 stars with intraday variability 
could be eclipsing binaries.

In addition, sources that 
have not been detected as variables 
may also be eclipsing binaries. 
The depth of the dip caused by an eclipse in 
the light curve of a binary is a strong function of the relative 
sizes of the stars and their orbital inclination angle. Such a dip, 
especially in a long-period binary, may be missed in an 
observation. Future observations 
taken by the \hst/WFC3 with optimal cadences are highly desirable to confirm 
our binary candidates and find new ones. Of particular 
interest are massive binaries (see Dong et al. 2017c, in preparation), which 
may provide useful insights into 
star formation processes in the unique environment of the MWNSC.

\subsubsection{$\delta$ Scuti, RRL and T2C}\label{sss:rrl}
Among the 28 stars shown in 
Figs~\ref{f:rrlyrae_1} to~\ref{f:rrlyrae_3}, 
4, 8, 13 and 3 of them have periods of $<$0.2, 0.2-0.4, 0.4-1 and $>$1 d,  
the expected ranges for $\delta$ Scuti, RRc, RRab and 
Cepheids, respectively. Here we discuss these classes of 
variable candidates individually:

\textbf{$\delta$ Scuti}: 
The four $\delta$ Scuti candidates are 
IDs 6164, 12215, 20785
 and 29285, with PPA=0.33, 0.25, 0.40 and 0.75 mag. 
 The OGLE survey
\citep{pol10} shows that the $I$ band amplitude of a sample of 
the  $\delta$ Scuti stars is 0.21$\pm$0.1 mag, although the maximum 
amplitude at the $I$ band could reach 0.9 mag. Therefore 
ID 20785 may have a too large amplitude in the F153M band to 
be $\delta$ Scuti 
and may instead be an eclipsing binary system. According to the 
F127M-F153M colours, ID 29285 is likely 
 in the foreground, while the other three sources 
 are in the Galactic bulge. 
 
\textbf{RRLs}: a detailed analysis 
of the 21 stars with periods between 0.2 and 1 d  will 
be presented in Dong et al. 2017b (in preparation). 
They report discoveries of 3 RRcs and 4 RRabs, plus three RRab candidates. 
In particular, IDs 8735, 10520, 22197 and 22312 show the typical sawtooth 
light curves of RRabs.

\textbf{T2C}: The three sources with periods $>$ 1 d could be CCEPs, T2Cs or eclipsing binaries.
The key differences between the first two types are their intrinsic luminosities, while the last one does not follow the PL relationship of CCEPs or T2Cs.  We adopted Equations (6)-(11)
in ~\citet{mat13} as the PL relationships for CCEPs and T2Cs. 
Their $J$ and $H$ are for the SIRIUS filters 
($J$: 1.25 $\mu$m and $H$: 1.63 $\mu$m). 
Because of the different effective wavelengths, 
the absolute magnitudes at F127M/F153M bands 
are not the same to those at the $J$/$H$ bands, especially 
for cool stars with strong water vapor absorption 
at the $J$ and $H$ bands~\citep{lan00}. 
For example, the Mira-type LPV S Car, the near-IR spectrum of which 
is given in Figure 8 of~\citet{lan00} has F153M - $H$=0.23. 
Therefore, we used the method given in 
the Appendix of Dong et al. 2017b to calculate the transformation from 
the SIRIUS to WFC3/IR absolute magnitudes through the stellar atmosphere model 
of~\citet{cas04} with surface temperatures between 3500 $K$ and 10$^4$ $K$: 
\begin{eqnarray}
F127M=J_{SIRIUS}-0.006-0.109\times( J_{SIRIUS} - H_{SIRIUS} )\\
F153M=H_{SIRIUS}-0.004+0.184\times( J_{SIRIUS} - H_{SIRIUS} )
\end{eqnarray}
With the extinction law of~\citet{sch10}\footnote{The extinction law 
given in~\citet{nis09} with a slope of -2.0 is flatter than that given in~\citet{sch10}. 
\citet{sch10} obtain the law from RC stars detected by \vlt/NACO, while  
\citet{nis09} constrain the slope from the observations taken by the 1.4 m IRSF telescope  
on the central 3$\times$1 degree, which suffer from strong confusion in the 
central few arc minutes. Therefore, we 
prefer the extinction law given in~\citet{sch10} because it was obtained 
with high angular resolution observations very close to the field that we are studying.}, 
the same method is used to derive the relative extinction, 
$A_{F127M}$=1.49$A_{F153M}$ and $A_{F153M}$=2.08$A_{Ks}$ 
(\vlt/NACO $K_s$, 2.18 $\mu$m). 
Then from the F127M and F153M apparent  
magnitudes we derived their extinctions and distances, as listed in Table~\ref{t:t2c}. 
We can see that if the three sources were CCEPs, they would be too far away to 
be in the Milky Way. ID 23605 is 
more likely a short-period T2C, i.e. BL Her star, probably only slightly beyond the GC. If 
ID 23523 and ID 26112 were T2C stars on the far-side of the Galactic disk or halo, then 
they would suffer less extinction than the average of the MWNSC, 
which is extremely unlikely. They might therefore be mis-classified eclipsing binaries. 
Fig.~\ref{f:t2c_can_cmd} gives 
the CMDs of detected sources within 2\arcsec\ of the three T2C candidates. ID 23605 
is redder than the majority of nearby sources, while ID 23523 and ID 26112 are not. This 
supports our typing of these three sources, i.e. ID 23605 is a T2C star, while IDs 23523 
and 26112 are eclipsing binaries. 

\subsection{Classifying variables with no measured period}\label{ss:origin}
Only a tiny portion of our 3894 variable stars have spectroscopic 
observations (8.2\%) or 
determined periods (1\%), 
which are important for the classification. 
We therefore only statistically study our unclassified variable stars, based on their locations in the CMD, compared with those variable stars having  
existing spectroscopic identification in the CMD 
(F127M-F153M vs F153M, Fig.~\ref{f:cmd_demon}), 
and compare with the magnitudes and colours of four well-known types of 
variable stars (Fig.~\ref{f:cmd_demon_vari}). 

Fig.~\ref{f:cmd_demon} shows all our variable candidates in the CMD. 
Stars classified through spectroscopy are overplotted in red open symbols, while  
stars classified via their light curves and periods are shown 
as yellow filled symbols. Most of the stars with 
spectroscopic identifications are brighter than 18 mag at the F153M band, 
because of the sensitivity limits of such observations. On the other hand, 
the stars with variable types determined from only their periods concentrate in 
fainter parts of the CMD. That is partly due to the origins of these variable stars: 1) 
$\delta$ Scuti and RRL stars are known to be dim; 2) T2Cs with periods of $\sim$1 d are dimmer than 
those of 10 and 100 days by $\sim$ 2.4 and 4.8 mag at the F153M band, according 
to their PL relationship. 

The PL relationships of Miras, CCEP, T2C and RRL 
provide us with intrinsic luminosities and 
colours. Applying the distance modulus of the 
GC (8.0 kpc) and using foreground extinction 
($A_K$=2, 2.5, 3, 3.5 mag), 
we overlay the potential locations of these stars 
on the CMDs shown in Fig.~\ref{f:cmd_demon_vari}. 
The PL relationships for CCEPs and 
T2Cs are given in \S\ref{sss:rrl}. For Miras, we used 
Equations (1) and (2) of~\citet{mat09} derived 
from Miras in the LMC.~\citet{whi08} suggest that 
the relationships in the LMC and Galactic bulge are similar. The PL relationships  
for RRL are given in Equations (4) and (5) 
of~\citet{cat04} from theoretical calculation, assuming the metallicity, 
[$Fe$/$H$]=-1.0~\citep{wal91}. The $J$ and $H$ in~\citet{mat09} and
 \citet{cat04}
are for the SIRIUS filters and 
Johnson-Cousins-Glass system (J: 1.22 $\mu$m and H: 1.63 $\mu$m), respectively. 
As in \S\ref{sss:rrl}, we used the same method given in the 
Appendix of Dong et al. 2017b 
to calculate the transformation from 
the Johnson-Cousins-Glass (JCG) system to WFC3/IR magnitudes:
\begin{eqnarray}
F127M=J_{JCG}-0.007-0.142\times( J_{JCG} - H_{JCG} )\\
F153M=H_{JCG}-0.005+0.198\times( J_{JCG} - H_{JCG} )
\end{eqnarray} 
The relative extinction $A_{F127M}$/$A_{F153M}$ given 
in \S\ref{sss:rrl} is also used. 
We can see that the brightness 
decreases from Miras, CCEP, T2Cs to RRL. Miras 
are the reddest sources among them. 

In the following, we discuss the potential contributions to our variability sample of these different 
types of stars. 

\textbf{WR, OB MS and Red supergiant}: 
Red supergiants, 
like IRS 7, are known for their variations with periods of 
years and amplitudes of about 1 mag~\citep{kis06}. On the other hand, 
variability of single WR and OB stars tends to have small amplitude 
($\sim$0.1 mag in the visual 
band, Figure 9d in~\citealt{lef09}).  
Therefore, variable WR and OB stars 
in our dataset are likely to be eclipsing binary 
systems (PPA=0.67$\pm$0.44 mag, mean and standard deviation; more 
discussion will be given in Dong et al. 2017c, in preparation). However, 
such massive stars could only explain a small fraction of the 
variables without spectroscopic identifications for several reasons: 
1) Red supergiants are very bright; for example, IRS 7 has  
$K$$\sim$ 7 mag~\citep{car00}. With 1 mag amplitude, the variability of such stars 
should easily have been detected in previous long-period monitoring programs 
conducted with $<$ 4m 
telescopes (see \S\ref{ss:previous}); 2) The sample of WR stars in the 
MWNSC, should have been roughly 
complete, except for dusty WC stars with red and 
featureless spectra~\citep{don12}; 3) Recent works~\citep{sto15,fel15} 
have found 90\% of 
the young massive stars within 0.5 pc (i.e. 12.5\arcsec ) of the GC. But  
this region only includes $\sim$3.5\% of the variables in our catalog. 

\textbf{CCEP}: The magnitude and colour ranges of CCEP well 
match with the range of variable stars having  
F153M $<$ 18 mag and F127M - F153M $<$ 4. 
\citet{mat11,mat15} found four CCEPs 
in the GC. According to the SFR given in~\citet{mat11} 
and~\citet{pfu11}, we estimate that less than 1 CCEP 
should be in our FoV. 

However, in \S\ref{ss:cross}, we found that 5 of the 13 
stars of age 50-500 Myr and in the $K$ magnitude range 
from 10.2 to 11.5 mag, as given in~\citet{nis16}, are in 
our variable catalog. Their light curves are given in 
Fig.~\ref{f:ccep} and we have marked their location in the top 
right panel in Fig.~\ref{f:cmd_demon_vari}. Except for ID 423, 
which has been only detected in the observations before 
Feb 28, 2014, the other four sources show clear variability 
among the exposures from Feb to Apr, 2014. 
Their location in the CMD (Fig. ~\ref{f:cmd_demon_vari}) could 
also been explained by the PL relationship of CCEP with $A_K$=2 to 3 mag. 
If we assume that these stars are in the MWNSC, we can 
use the extinction map of 
Nogueras-Lara et al. (2017, in preparation)\footnote{Nogueras et al. 
(2017, in preparation) produce an extinction map of the GC with a 
spatial resolution of $\sim$2\arcsec\ from 
the RC stars detected in VLT/HAWK-I survey of the central 100 pc 
at $J$, $H$ and $K$ bands.} 
to derive their absolute $K$ band magnitude. Then by using 
the Equation 11 of~\citet{mat13}, we can derive their periods, between 
14.5 d to 24.0 d, which are similar to those of the four CCEPs identified 
by Matsunaga et al. The reasons why SIRIUS survey did 
not identify these CCEP candidates are probably 1) they 
are relatively dimmer at the $K$ band due to the 
higher extinction and 2) SIRIUS observations suffer 
from strong confusion in the central few arc minutes. 
The only concern is their surface 
temperatures, between 3850 $K$ to 4670 $K$, which are near the lower 
limit of the instability strip of Cepheids~\citep{fio02}. However, 
considering the large temperature uncertainty given 
in~\citep{nis16}, $\sim$150 $K$, we cannot use their 
temperatures to conclude that they are not CCEPs. 

\textbf{Single Red Giant Stars}: 
Miras tend to be brighter 
 than 16 mag and are located in the top right part of the 
 CMD. However, sometimes such stars can have strong 
 dusty winds and thus account for most  
 variable stars with F127M-F153M$>$3.8, 51\% of which are variables (see 
 Fig.~\ref{f:var_frac}). Although OSARGs and SRVs are roughly 
 one mag dimmer than 
 Miras at the K band, but 
 still brighter than Cepheids (`G' and `F' sequences 
 for CCEPs and T2Cs in Figure 6 of~\citealt{ita04}). Therefore, they cannot 
 explain the majority of our variables with 
 F153M $>$ 18 mag. 
 
 \textbf{T2C}: As shown in Fig.~\ref{f:cmd_demon}, the colour 
 and magnitude ranges of T2Cs are consistent 
 with the majority of our variables. 
 \citet{sos11} also suggest that the Galactic bulge 
 contains a significant population 
 of T2Cs with a centrally concentrated distribution. 
  
 The problem with T2C stars is that they are metal poor ([M/H]$<$0)
 and are widely found in the low metallicity region, such as globular 
 clusters, thick disk, bulge and halo of the Galaxy~\citep{wal02}. 
\citet{fel17} 
claim that only 6\% (22\%) of sample stars ($<$700) towards the 
 GC have [M/H]$<$-0.5 (0) (see also~\citealt{do15}). 
 On the other hand,~\citet{mat13} identify 16  
 T2C in the GC, all within the range of 12.9$<$H$<$15.3. These stars may 
 represent the 
 bright end of the luminosity function of T2Cs towards the GC. From 
 the PL relationship,~\citet{mat13} suggest 
 that their distances lie between 4 and 9 kpc, with the mean of 7.4 kpc 
 and the standard deviation of 1.4 kpc. Therefore, 
 these 16 T2C stars, as well as the majority of our variables 
may lie not exactly in the MWNSC, but just in the inner Galactic bulge. 
 
Besides the single pulsating stars, the unclassified 
variables could also be eclipsing and ellipsoidal binaries. 
Due to the range of stellar parameters of their components, the 
potential eclipsing binaries could distribute widely in the CMD. 
Ellipsoidal binaries with a giant star component are proposed 
to explain the `E' sequence in the period-magnitude diagram 
of variable stars identified in the Large Magellanic Cloud 
~\citep[][and reference therein]{paw14}. Their periods span from 10 
d to nearly 300 d. The variability is caused by the 
deformed atmosphere of the giants, since they fill the Roche lobe. 
According to the CMD, the stars could not only be RGB stars, 
but also even dimmer RC stars~\citep{paw14}. Actually, as 
shown in the CMD of the bottom panel of Fig.~\ref{f:mag_cmd}, 
a significant number of our variables indeed fall in 
the region occupied by the RC and RGB stars in the GC. 

Confusion with nearby stars with high relative proper motion could 
also broad the light distribution and cause variability among years. 
This situation needs to satisfy three requirements: 1) two nearby stars 
are close by less than one WFC3/IR pixel, so that `DOLPHOT' package 
does not disentangle them as two sources; 2) they have similar brightness, 
for example, less than 2.5 mag difference; 3) they have large relative proper 
motion ($>$ 127 km s$^{-1}$, which cause at least one tenth WFC3/IR pixel shift at 
four-year baseline). We used the proper motion and $K$-band magnitude 
information given in~\citet{sch09} to examine the impact of confusion on 
our variable catalog.~\citet{sch09} use the \vlt/NACO dataset in the central 
$\sim$40\arcsec$\times$40\arcsec , which are deeper enough to provide the 
proper motion of stars 2.5 mag dimmer than the 50\% completeness limit of 
our \hst\ dataset in the same field. Fig.~\ref{f:con_frac} gives the fraction of stars, which 
have counterparts which satisfy the three requirements above. The fraction 
is close to 10\% near Sgr A*, but decreases to less than 2\% at 23\arcsec\ 
away. If we multiple the fraction with the real detected sources in our \hst/WFC3 
dataset, we expect $\sim$129 variables among the 3268 detected sources, while 
we find 403 variables. Considering that the confusion effect decreases with 
the galactocentric radii, we expect no more than 5\% of our variable catalog  
are due to confusion. 

 \section{Summary} \label{s:summary}
 We conducted a stellar variability study, based on a set of 
 \hst/WFC3 IR observations of the Milky Way 
 nuclear star cluster, which consist of 290 dithered exposures 
 taken from years 2010 to 2014. 
 This study bridges the similar works based on 
 large field-of-view surveys obtained with small ($<$ 4m) telescopes and 
 those on high-spatial-resolution adaptive-optics images from $\geq$ 8 m 
 telescopes. 
 The stable PSF and accurate photometric zeropoints of the 
\hst\ observations enable us to not only provide light curves with much 
 improved accuracy for such 
known variable stars as E60~\citep{pfu14}, but 
 also to identify many new variable sources. We have further compared our 
 detected sources with previous variability studies of the Galactic Centre 
 and have made tentative classifications for various sources, 
 individually or statistically, based on their spectroscopic and 
 magnitude/colour properties. Our results are 
 summarized below. 
 
 \begin{itemize}
\item In total, we detect $\sim$33,000 sources in the central $\sim$ 
2.3\arcmin$\times$2.3\arcmin\ field. 
3894 of them are detected as variable stars, 76 of which show evidence for 
intraday variability. 
We have calculated the periods of 36 of these 76 stars. 

\item Our detection recovers most of
 the bright variable stars, such as Miras and massive eclipsing binaries. Some variable stars, as identified by~\citet{ott99},~\citet{pee07} and~\citet{raf07} are missed by us 
 probably because 1) underestimated 
 measurement errors, which are most likely systematic, in these studies, and cause 
 mis-identification of non-variable stars; 2)  
 inadequate temporal sampling of the \hst\ observations; and  
 3) the poor resolution of \hst\ WFC3 IR camera, leading to severe source confusion in the central few arcseconds  field. 

\item 319 of our detected variables have spectroscopic counterparts, which include  
WR stars, OB stars, supergiants, Miras and giants. 

\item For the first time, we have discovered $\delta$ Scuti, RR Lyrae Type ab and BL Her 
stars toward the Galactic Centre. 

\item The majority of the remaining could be eclipsing/ellipsoidal binaries and Type II Cepheids.   

 \end{itemize}

These results demonstrate the power of using \hst/WFC3 near-IR observations to 
study variable stars in the Galactic Centre. However, the sampling of 
current observations is too poor to type stars with variability periods $>$ 2 d. 
Similar WFC3 observations with optimal cadences are highly desirable for us to accurate 
derive the periods, classify their types and derive the SFR of the MWNSC at different epoch periods. 
Such observations do not need to be deep, 
because the detection is largely limited by the source confusion. 

\section*{Acknowledgments}
We thank the anonymous referee for a thorough, detailed, and
constructive commentary on our manuscript. 
The research leading to these results has received funding from the European Research Council under the European Union's Seventh Framework Programme (FP7/2007-2013) / ERC grant agreement n° [614922]. The work is also supported by 
NASA via the grant AR-14589, 
provided by the Space Telescope Science Institute.
F N-L acknowledges financial support from a MECD predoctoral 
contract, code FPU14/01700. 
This work uses observations made 
with the NASA/ESA Hubble Space Telescope and 
the data archive at the Space Telescope Science Institute, which is 
operated by the Association of Universities for Research in Astronomy, 
Inc. under NASA contract NAS 5-26555.  
We are grateful to Zhiyuan Li, 
Francisco Najarro, Jon Mauerhan, Stephen Eikenberry and Farhad Yusef-Zadeh for many valuable
comments and discussion. We also want to thank Siro Benvenutto Gallego, the son of Eulalia 
Gallego-Cano to help us visually identify variable stars.

 \appendix
\section{Photometric Uncertainty}
Uncertainty ($\sigma_v$) is a key element in the least chi-square method in \S\ref{s:method} to identify 
variable stars. If we over- or underestimated 
the uncertainty, we would miss the real variable stars or detect too many variable stars. In 
this section, we will use the real dataset to judge the photometric uncertainty estimated 
from the artificial star tests. 

If we assume that most of detected stars are non-variable, the median value of the 
standard deviation of magnitudes from individual dithered exposures of stars in different 
magnitude and local surface brightness bins could be a good representative of $\sigma_v$. 
The top panel of Fig.~\ref{f:error_obs} shows the standard deviation for individual stars (black dots) 
and their median values as a function of the F153M magnitude (green diamonds). For each 
magnitude bin, the large scatting of the  standard deviation (black dots) is due to the variation of 
local surface brightness. 
The median value (green diamonds) is roughly 0.03 mag for stars with F153M$<$17 mag and 
then increases with the F153M magnitude. In the bottom panel, we compare the relationship 
between the median values of the standard deviation and the F153M magnitude for dithered exposures 
at different periods. The relationship at individual periods (dotted lines with different colours) are 
very similar, median standard deviation is $\sim$0.01 mag for F153M$<$17 mag and then increases 
with the magnitude. However, the median standard deviation at each magnitude bin increases with the 
duration of the dithered exposures; first all the observations in 2014, then those in 2010 and 2011, 2011 
and 2012, or 2012 and 2014. The median standard deviation derived from all the dithered exposures from 2010
to 2014 are the largest. The shift of the median standard deviation at different combination of dithered exposures seems to be independent to the magnitude 
and could be a systematic error. 

We want to understand the temporal evolution of the median standard deviation shown in the bottom panel of 
Fig.~\ref{f:error_obs}: 0.01 mag error for the single epoch and $\sim$0.025 mag for the whole 
dateset. Several factors could contribute the detected error: 1) Poisson statistic error due 
to the electrons. This error is very tiny for the bright stars and is 
independent on observation time. Therefore, Poisson statistic error is not the solution; 2) The temporal variation 
of the photometric zeropoint keyword (`PHOTFLAM'). This keyword is used to translate the image units 
(electron/s) into astrophysics units (such as Vega magnitude). The homepage of WFC3\footnote{http://www.stsci.edu/hst/wfc3/phot\_zp\_lbn} mentions $\sim$1\% uncertainty in PHOTFLAM from 
the calibration programs. The differences between PHOTFLAM among different dithered exposures could 
cause a systematic error in the observed magnitude. However, we do not expect a variation of PHOTFLAM 
in each period with duration as less than 0.3 hours (Group 4 in Table~\ref{t:obs}) to explain the 0.01 magnitude 
error of bright stars in the single epoch. We also try to derive the median magnitude correction 
of dithered exposures in 2010, 2011 and 2014, with respect to the 2012 observations. The correction is very small 
$<$0.01 mag. Even we correct this value, the median standard deviation for dataset with four-year 
baseline are only slightly reduced; 3) The uncertainty on the flat-fielding calibration files. The above result suggests that 
the photometry correction could change spatially. At different dithered exposures, the same source could fall into 
different locations of the instrument. The uncertainty on the instrument response (i.e. the flat-fielding calibration) 
could introduce a systematic error. \citet{dah13} suggest a 0.7$\%$ error for the flat-fielding uncertainty. This 
value could probably explain the 0.01 mag in the single epoch, but not the extra 0.02 mag difference between 
single epoch and multiple epochs; 4) WFC3/IR intra-pixel sensitivity. If the centroid of the PSF falls into the gaps 
between pixels, its total flux reduces, which is known as `intra-pixel sensitivity variation'. For an undersampling 
instrument, likes WFC3/IR, this effect is more significant in the short wavelength.~\citet{pav11} suggests a 0.44\% 
uncertainty due to the `intra-pixel sensitivity variation' in the F160W band (1.53 $\mu$m). 
Therefore, it still cannot completely solve 
our problem; 5) PSF fitting on the undersampling detector. The FWHM of the WFC3/IR PSF 
at 1.5 $\mu$m\footnote{http://www.stsci.edu/hst/wfc3/documents/handbooks/currentIHB/c07\_ir07.html} is only 1.136 
WFC3/IR pixels. All our observations use a coarse half-pixel dithering, making accurately determination of the 
stellar centroid very hard. Therefore, an error on the stellar centroid from the mosaic image could introduce a systematic 
uncertainty on the photometry determined from individual dithered exposures through the PSF fitting. For observations 
within short period, the relatively astrometry between dithered exposures could be determined accurately. Therefore, 
they suffered from the same systematic uncertainty introduced by the PSF fitting. Instead, observations at long period, 
with different pointing and rotation angles, they could suffer different systematic uncertainty, which in turn is 
transferred into the median standard deviation. In summary, the flat-fitting uncertainty and WFC3/IR intra-pixel 
sensitivity contribute 0.01 mag error in the single epoch, while the PSF fitting on the undersampling detector and 
the temporal variation of `PHOTFLAM' 
contribute an extra 0.015 mag error among multiple epochs. 

Figure.~\ref{f:fake_obs_compare} compares the median standard deviation from the observation and the artificial star test, but with 
different methods: 1) the photometric uncertainty, $\sigma_{F153M}$ (black dotted line), which is the square root of the quadratic sum of the systematic and the statistic parts, i.e. the median absolute value and 68\% percentiles of the difference between the output and the input magnitudes ; 2) 
standard deviation of output magnitudes of the dithered exposures ($\sigma_v$); 3) photometric uncertainty ($\sigma_{DOLPHOT}$) 
output by the `DOLPHOT' package, which only considered the Poisson uncertainty. We also add 1\% and 2.5\% systematic uncertainty into 
the median standard deviation derived from the second and third methods above. We can see that $\sigma_{F153M}$ seriously 
overestimate the observed median standard deviation for stars brighter than $>$16 mag 
in the F153M band. Instead, $\sigma_v$, plus 1\% and 2.5\% systematic error 
match the observed median standard deviation for single epoch and multiple epochs, very well to magnitudes $\sim$20.5 mag. 
Beyond that, $\sigma_{DOLPHOT}$ fits the observed one better. 

\section{Variability among Dithered Exposures}
We found that there is a correlation between the phenomenon, i.e., quick intensity variation among 
dithered exposures, and the `CHI' quality parameter. 
For each source, we defined the $\overline{CHI}$ as the mean 
`CHI$_{i}$' of dithered exposures in Groups 5 to 7 in Table~\ref{t:sou}, the dithered exposures in 
which are used 
to derive $\chi^2_d$ too. The top panel of Fig.~\ref{f:chi_short} shows the 
distribution of $\overline{CHI}$ for our sources. The peak of the distribution centered on 1.3 and 
68\% percentile is [0.73,2.66]. The bottom panel of Fig.~\ref{f:chi_short} gives the 
fraction of stars with $\chi^2_d>2.42$ (the upper 90\% percentile of the $\chi^2_d$ for all the sources 
are [0.58,2.42]) as a function of $\overline{CHI}$ (the bottom panel). We can see that the fraction of 
short variable stars increases, even the number of stars decreases. Considering that the majority 
of stars with large $\chi^2_d>2.42$ suffer from the artificial mentioned above, we believe that 
this artificial is  strong correlated with the `CHI' parameter. As a result of the bad PSF fitting, the 
uncertainty of the center position introduced by the dithering patterns artificially produced 
a regular variability among dithered exposures. 
 
\clearpage 

  \begin{figure*}[!thb]
  \centerline{    
       \epsfig{figure=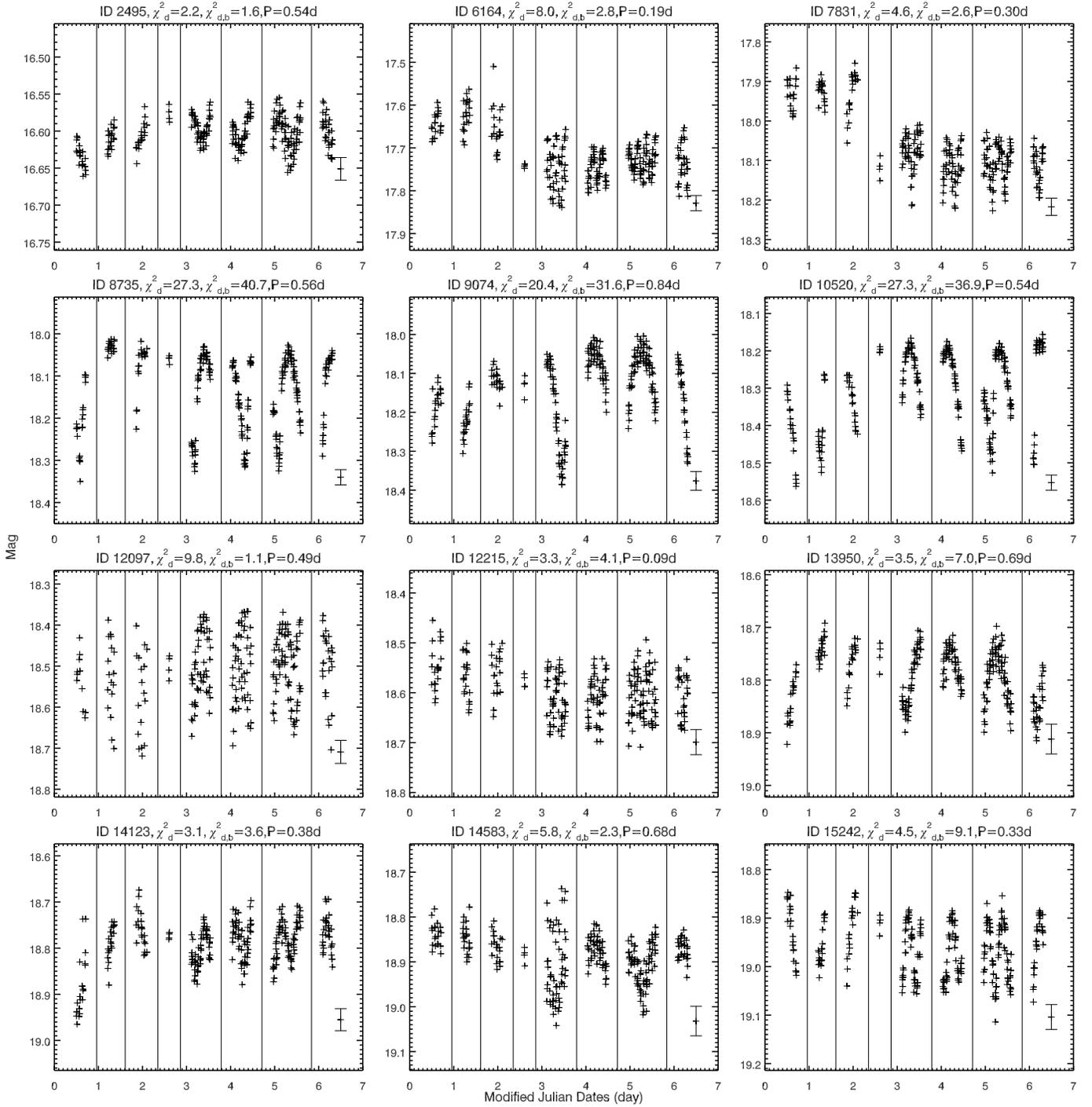,width=1.2\textwidth,angle=0}
       }
 \caption{The \hst/WFC3 F153M light curves of 28 sources with measured periods. 
 The vertical lines are the 
 same as those in Fig.~\ref{f:artifact}. The error bar in the lower right corner of 
 each panel represents the average uncertainty. In the title of each panel, we give the source ID, 
 $\chi^2_{d}$ and $\chi^2_{d,b}$, as defined in \S\ref{s:method}, as well as the periods 
 calculated in \S\ref{s:result}.}
\label{f:rrlyrae_1}
 \end{figure*}

 \begin{figure*}[!thb]
  \centerline{    
       \epsfig{figure=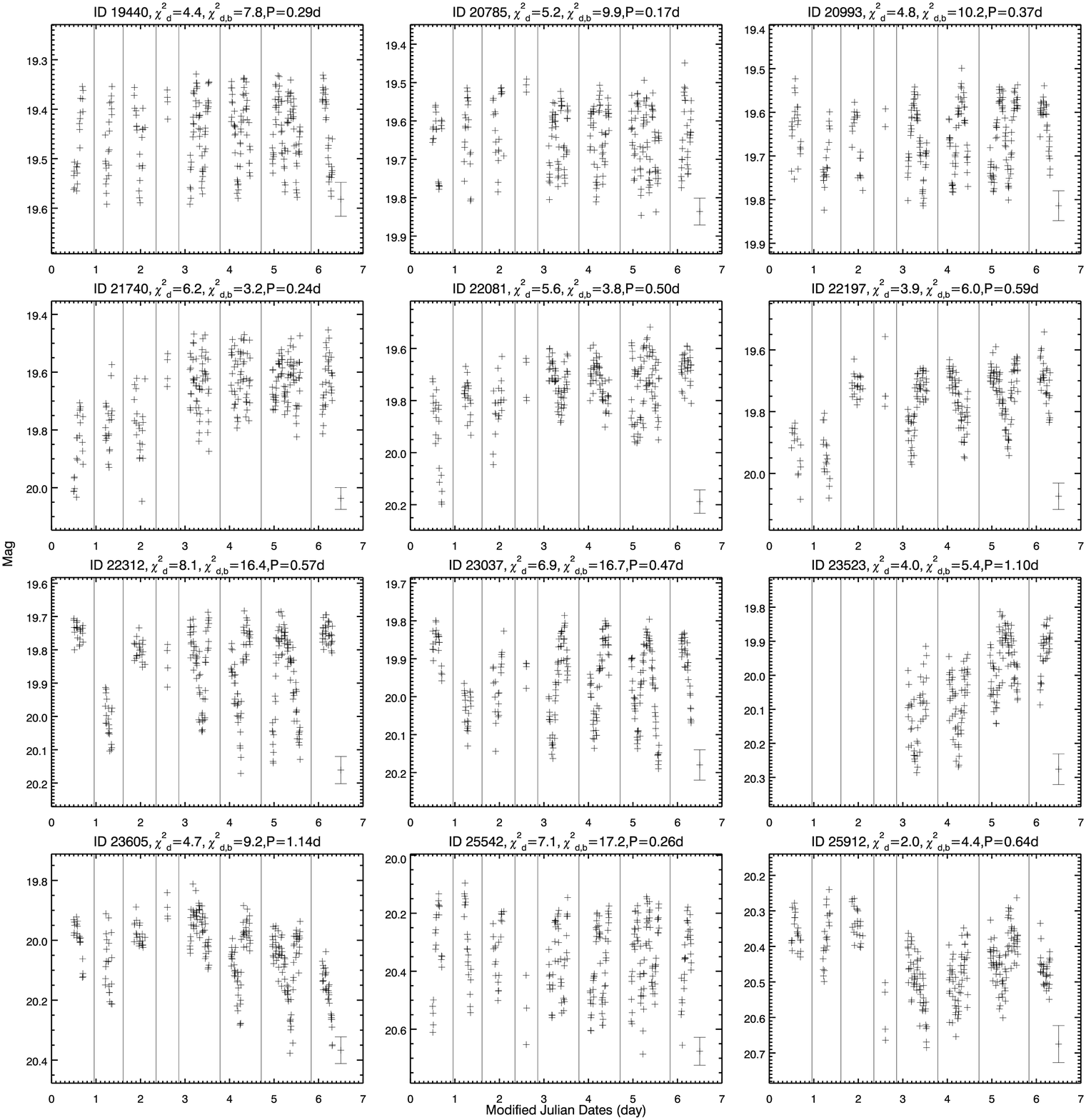,width=1.2\textwidth,angle=0}
       }
 \caption{Continued Fig.~\ref{f:rrlyrae_1}.}
\label{f:rrlyrae_2}
 \end{figure*}

 \begin{figure*}[!thb]
  \centerline{    
       \epsfig{figure=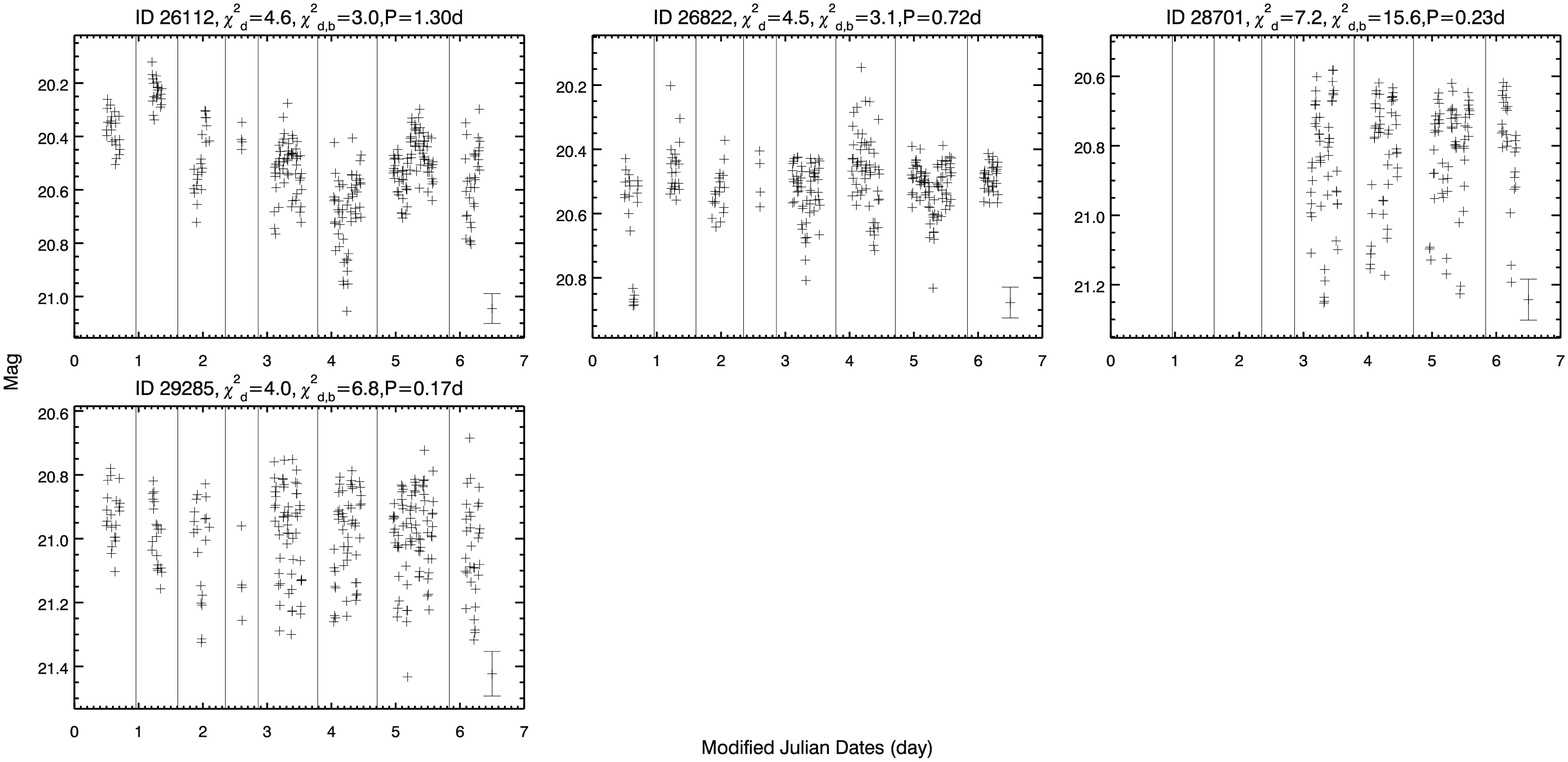,width=1.2\textwidth,angle=0}
       }
 \caption{Continued Fig.~\ref{f:rrlyrae_1}.}
\label{f:rrlyrae_3}
 \end{figure*}

 \begin{figure*}[!thb]
  \centerline{    
       \epsfig{figure=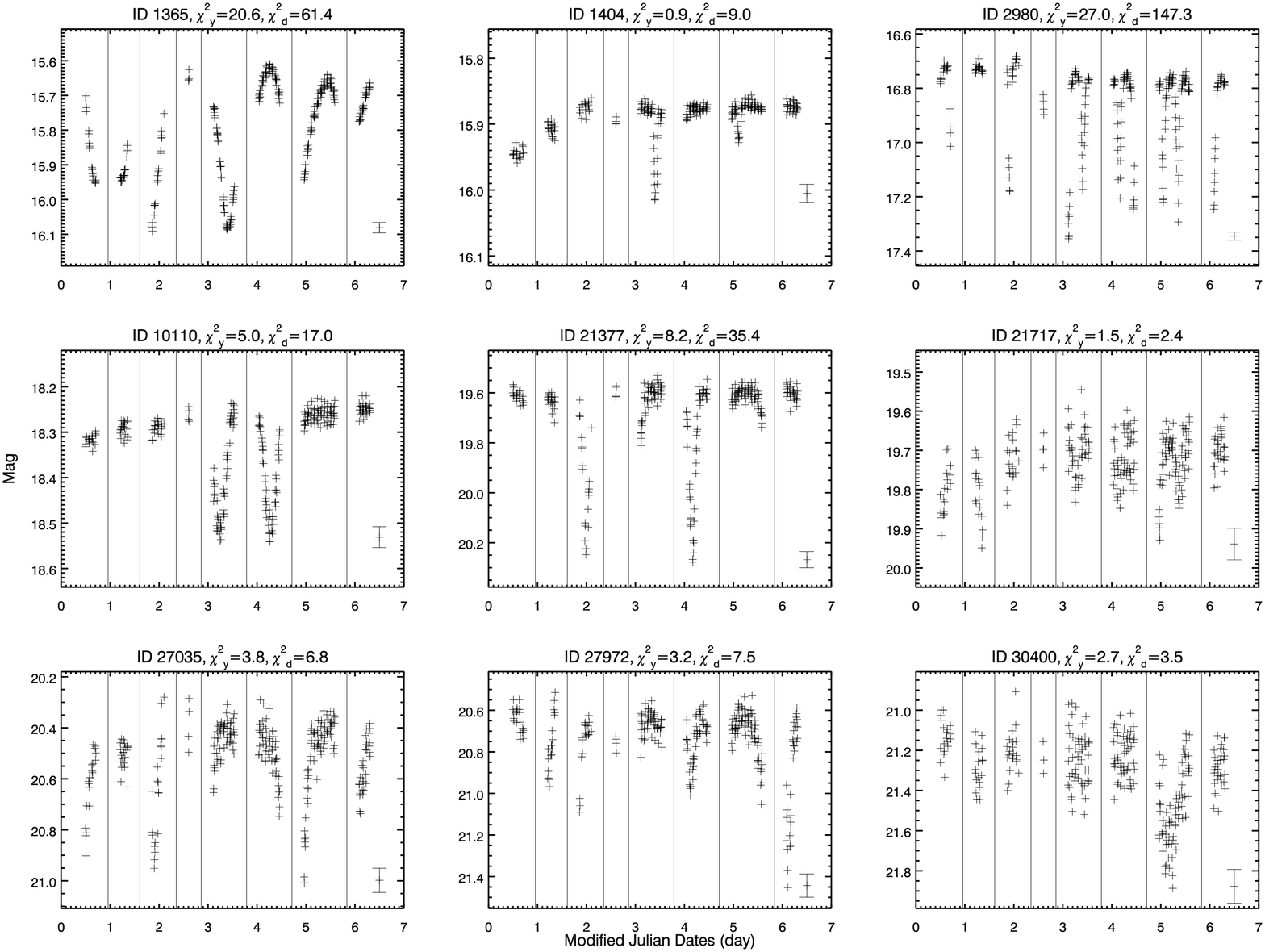,width=1.2\textwidth,angle=0}
       }
 \caption{The \hst/WFC3 F153M light curves of nine eclipsing binary candidates. 
 The vertical lines are the 
 same as those in Fig.~\ref{f:artifact}. The error bar in the lower right corner of 
 each panel represents the average uncertainty. In the title of each figure, 
 we give the source ID, 
 $\chi^2_{y}$ and  $\chi^2_{d}$, as defined in \S\ref{s:method}.}
\label{f:eclipse}
 \end{figure*}

\begin{figure*}[!thb]
  \centerline{    
       \epsfig{figure=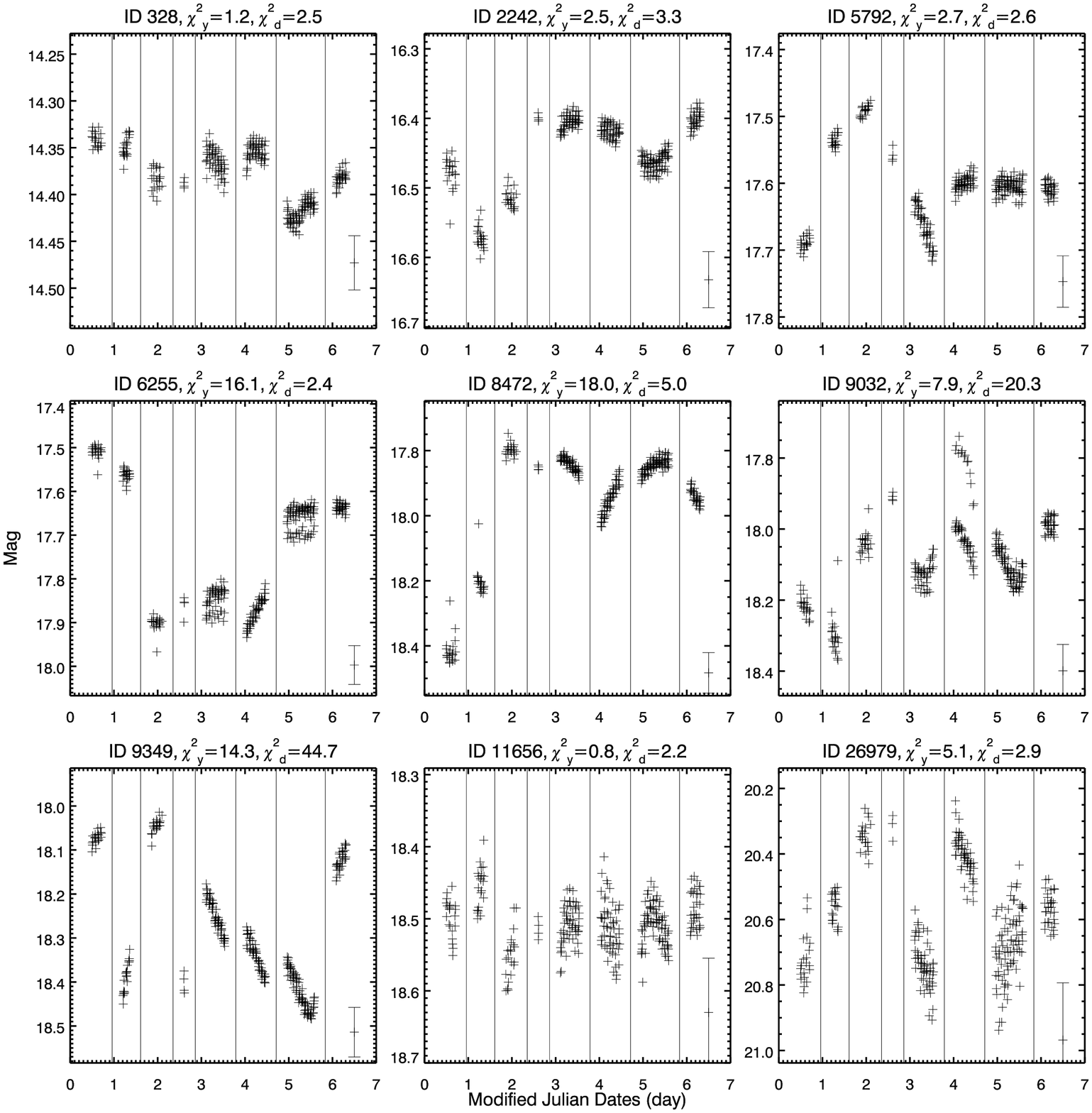,width=1.2\textwidth,angle=0}
       }
 \caption{The \hst/WFC3 F153M light curves of nine stars 
 with different magnitudes and intraday variability. The vertical lines are the 
 same as those in Fig.~\ref{f:artifact}. The error bar in the lower right corner of 
 each panel represents the average uncertainty. In the title of each figure, we give the source ID, 
 $\chi^2_{y}$ and  $\chi^2_{d}$, as defined in \S\ref{s:method}.}
\label{f:shortval}
 \end{figure*}

 \begin{figure*}[!thb]
  \centerline{    
       \epsfig{figure=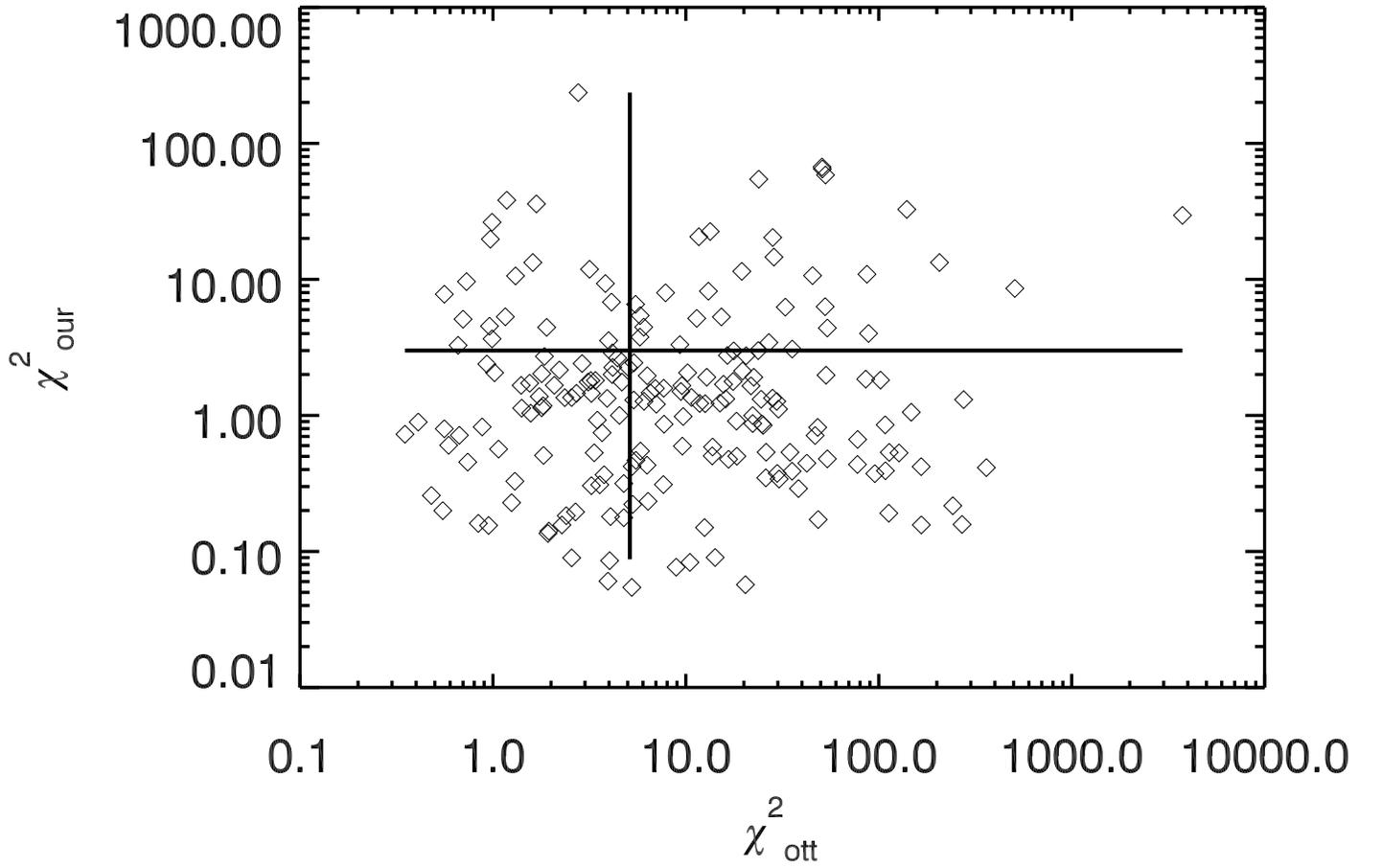,width=1.2\textwidth,angle=0}
       }
        \caption{Comparison of the reduced $\chi^2$ derived from~\citet{ott99} ($\chi^2_{ott}$) and 
our dataset ($\chi^2_{our}$). For each source, $\chi^2_{our}$ is the minimum of $\chi^2_y$ and $\chi^2_{y,b}$. 
The vertical and horizontal lines are the cut to separate the variable or non-variable stars: $\chi^2_{ott}$=5.13 
from~\citet{ott99} and $\chi^2_{our}$=3.}
\label{f:ott_compare}
 \end{figure*}
 
 \begin{figure*}[!thb]
  \centerline{    
       \epsfig{figure=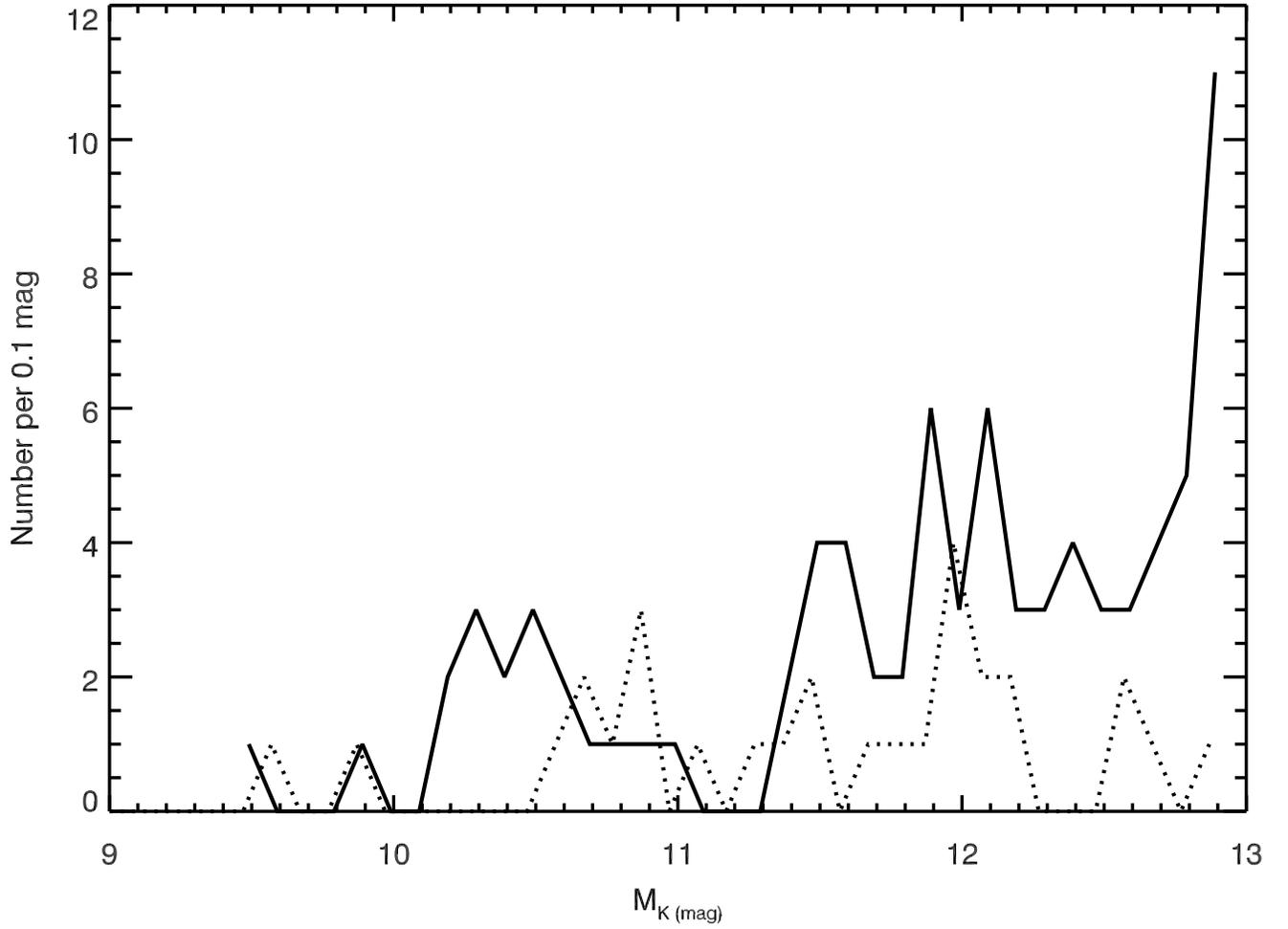,width=1.2\textwidth,angle=0}
       }
        \caption{The K-band magnitude distributions of variable stars defined in~\citet{ott99}, which 
        are classified as non-variable (solid line) and variable (dotted line) in our dataset.}
\label{f:ott_compare_dmk}
 \end{figure*}
 
  \begin{figure*}[!thb]
  \centerline{    
       \epsfig{figure=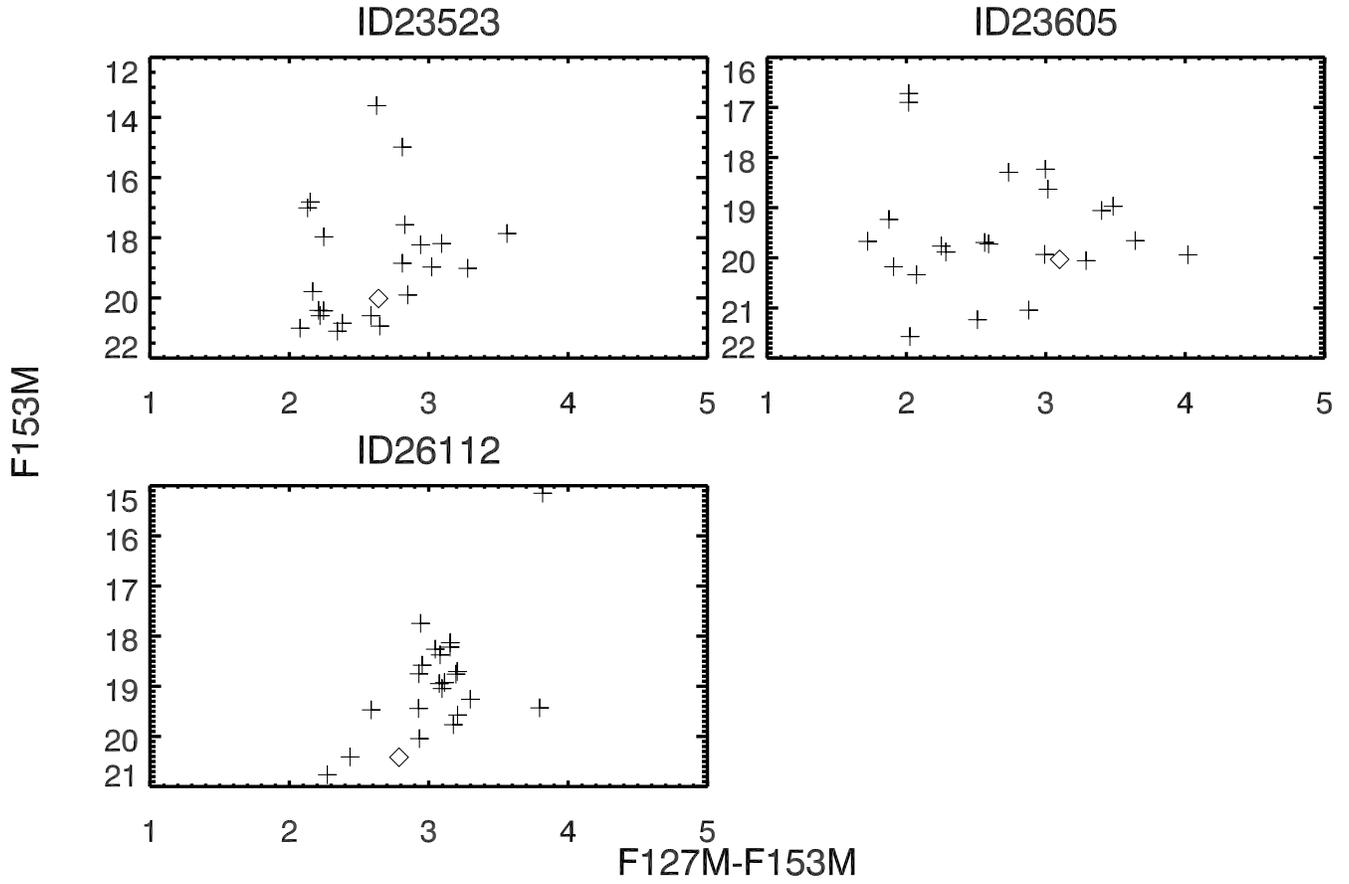,width=1.2\textwidth,angle=0}
       }
        \caption{The colour magnitude diagrams (F127M-F153M vs. F153M) of 
 the detected sources (pluses) within 2\arcsec\ of the three T2C candidates (diamonds), 
 the IDs of which are given in the titles.}
 \label{f:t2c_can_cmd}
 \end{figure*}
 
 \begin{figure*}[!thb]
  \centerline{   
       \epsfig{figure=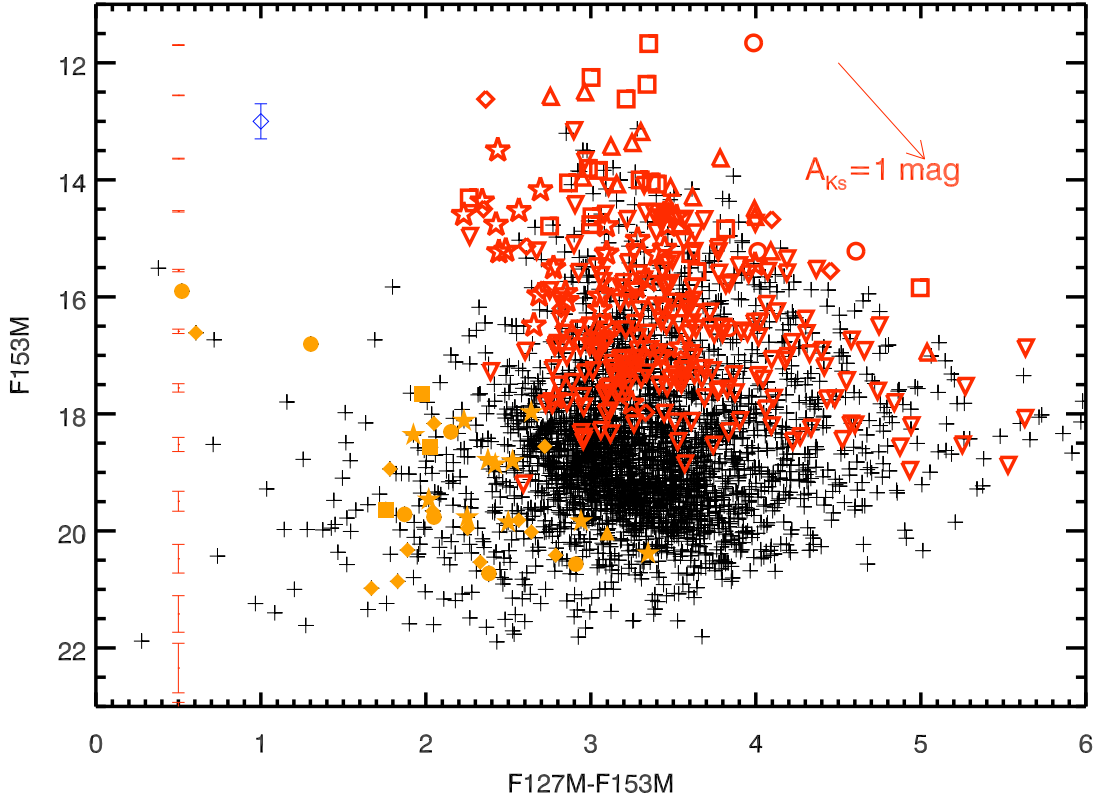,width=1.0\textwidth,angle=0}
       }
 \caption{Comparison of the CMD of variable stars with spectroscopic 
 identifications (red open symbols), 
 identifications determined from their light curves and periods 
 (yellow filled symbols) and those without available spectroscopic 
 identifications or variable type classification (black plus). 
Red diamonds represent WR stars, asterisks OB stars, 
squares Miras (LPVs), circles supergiants (`I' in~\citealt{blu03}), 
triangles giants (`III' in~\citealt{blu03}), and 
downward-pointing triangles other late-type stars. 
Yellow circles represent eclipsing binaries with 
well determined periods in~\S\ref{s:result}, squares 
3 $\delta$ Scuti, stars 11 RRLs, triangles T2Cs and filled diamonds  
11 candidates of eclipsing binaries. On the left, the red interval bars are 
 the $\sigma_{F153M}$ at different F153M magnitudes and the blue diamond with 
an error bar represents the photometric uncertainty introduced by a 1 kpc error in the 
distance modulus.}
\label{f:cmd_demon}
 \end{figure*}

\begin{figure*}[!thb]
  \centerline{    
       \epsfig{figure=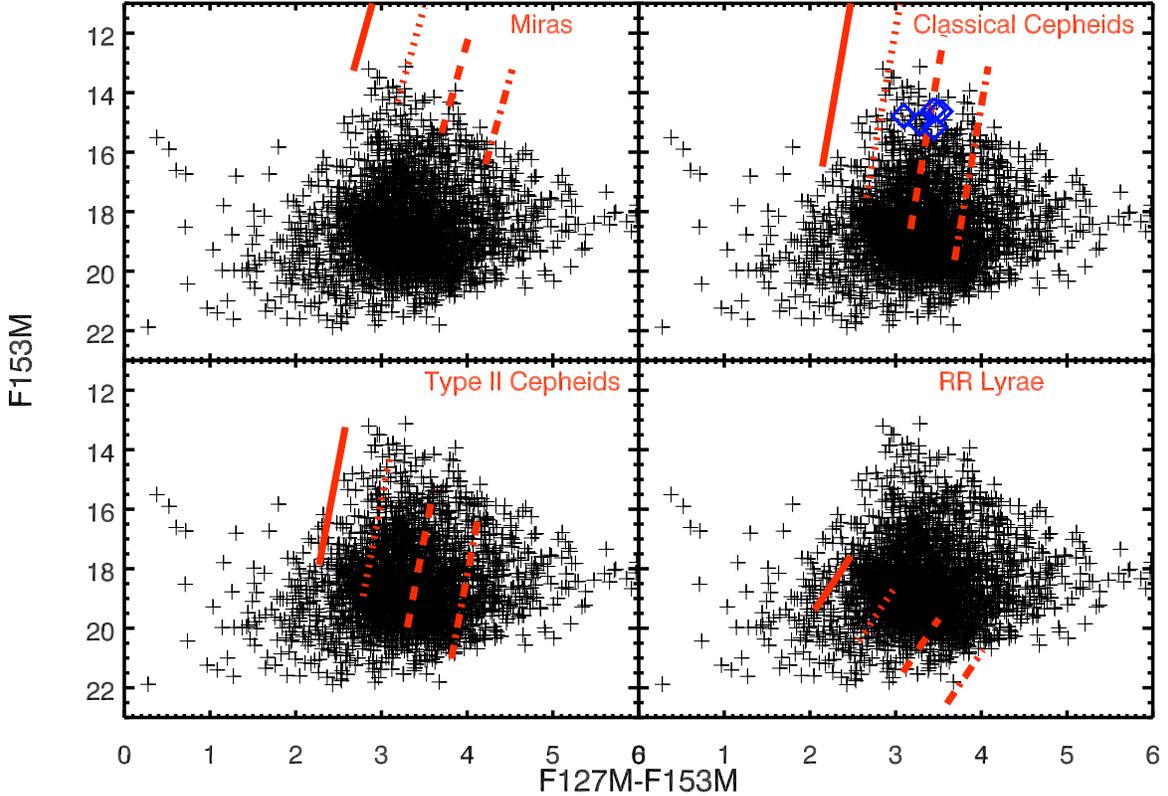,width=1.0\textwidth,angle=0}
       }
 \caption{Comparison of the CMD of variable stars with the ranges predicted 
 by the period-luminosity functions of variable stars with different types, assuming 
 that they are in the GC. 
We cannot calculate the period of 99\% of our sample in order 
to classify their variable type, so these ranges are plotted for 
qualitative purposes only. The 
 solid, dotted, dashed and dot-dashed red lines have been reddened with 
 $A_{Ks}$=2, 2.5, 3 and 3.5 mag. In the top right panel, the blue diamonds 
 around F127M-F153M$\sim$3.3 mag 
and F153M$\sim$ 15 mag are five CCEP candidates, which are 
 classified as intermediate-age stars (50 to 500 Myr old) 
 by~\citet{nis16} and show variability among years. 
  }
\label{f:cmd_demon_vari}
 \end{figure*}
 
 \begin{figure*}[!thb]
  \centerline{   
       \epsfig{figure=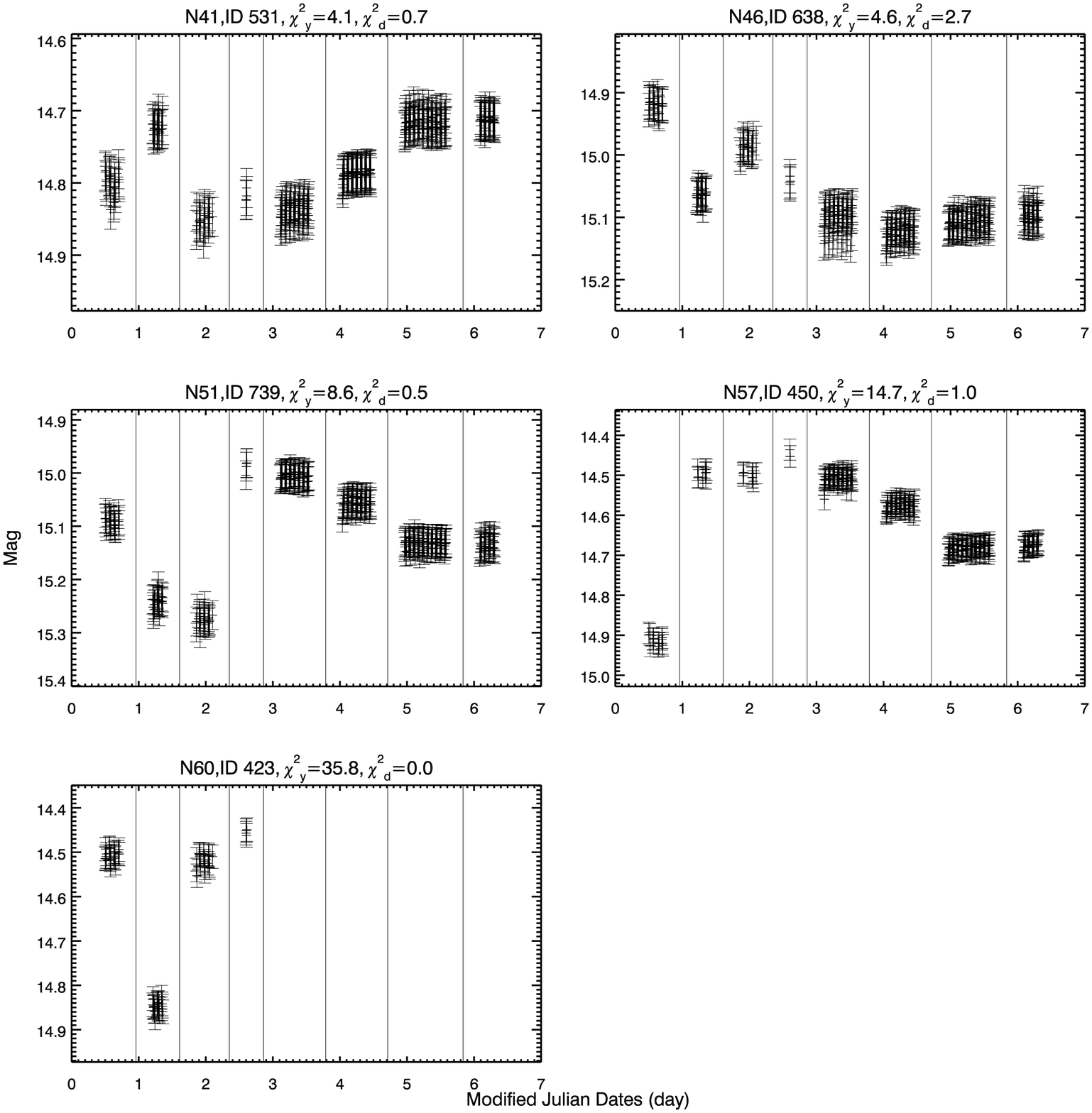,width=1.0\textwidth,angle=0}
       }
 \caption{The \hst/WFC3 F153M light curves of five CCEP candidates, which are 
 classified as intermediate-age stars (50 to 500 Myr old) by~\citet{nis16} and show 
 variability among years. The symbols 
 (`pluses' and vertical lines) are the 
 same as those in Fig.~\ref{f:artifact}. In the title of each figure, we give the ID~\citet{nis16}, which 
 begins with `N',  our source ID, 
 $\chi^2_{y}$ and  $\chi^2_{d}$, as defined in \S\ref{s:method}. ID 423 (`N60') is near the edge of 
 our FoV and was 
 only covered by the observations earlier than Feb 28, 2014. }
\label{f:ccep}
 \end{figure*}

\begin{figure*}[!thb]
  \centerline{    
       \epsfig{figure=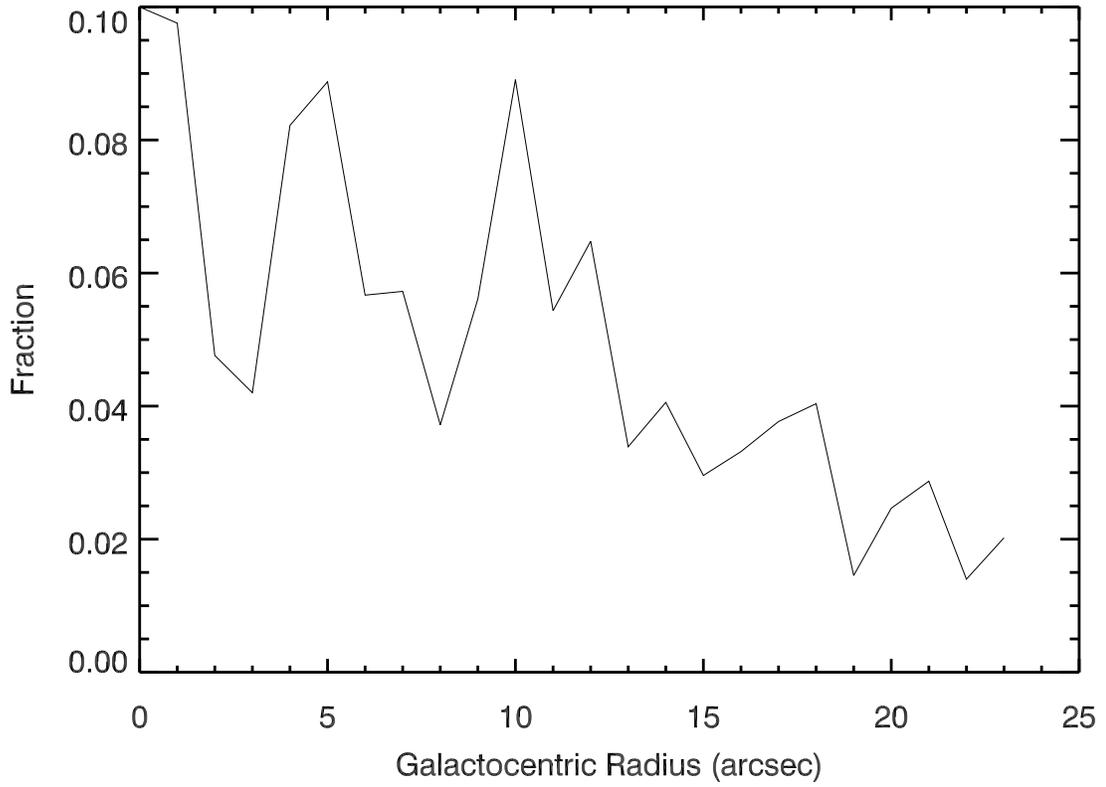,width=1.0\textwidth,angle=0}
       }
 \caption{The fraction of sources, which have counterparts less than 0.13\arcsec\ away, less than 
 2.5 mag difference at the $K$-band and relative proper motion $>$ 127 km s$^{-1}$ as 
 a function of galactocentric radius away from Sgr A*, 
 derived from the information given in~\citet{sch09}.}
\label{f:con_frac}
 \end{figure*}

  \begin{figure*}[!thb]
 \centerline{    
       \epsfig{figure=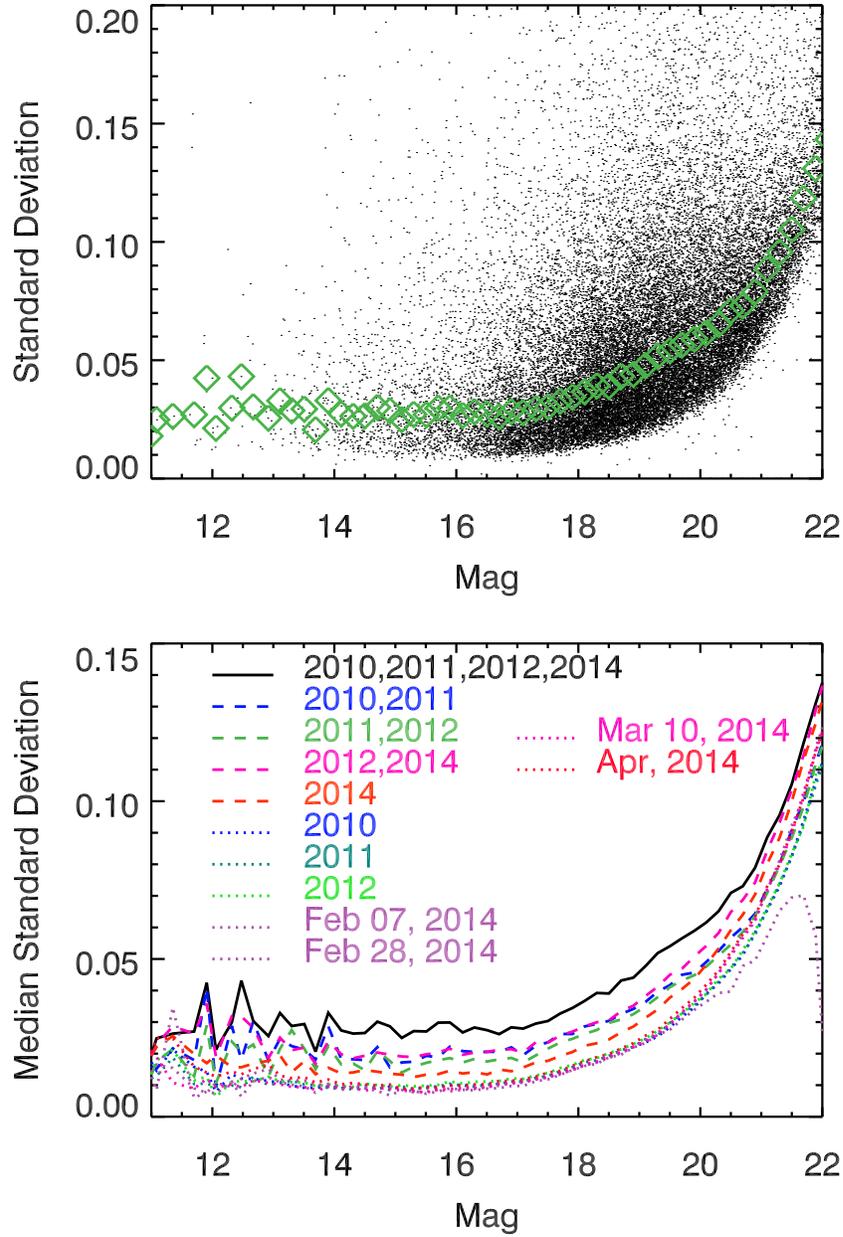,width=0.8\textwidth,angle=0}
       }
 \caption{Top panel: black dots represent 
 the standard deviation of magnitudes from 290 dithered exposures for individual sources as a function 
 of their F153M magnitude. For each 0.03 mag, we calculate the 
 median value of the standard deviation (green diamonds). Bottom panel, 
 the lines with different colours and styles show the median standard 
 deviations of magnitudes from dithered exposures at different periods as a function of F153M magnitude.}
\label{f:error_obs}
 \end{figure*}
 
\clearpage 
\begin{figure*}[!thb]
\centerline{    
     \epsfig{figure=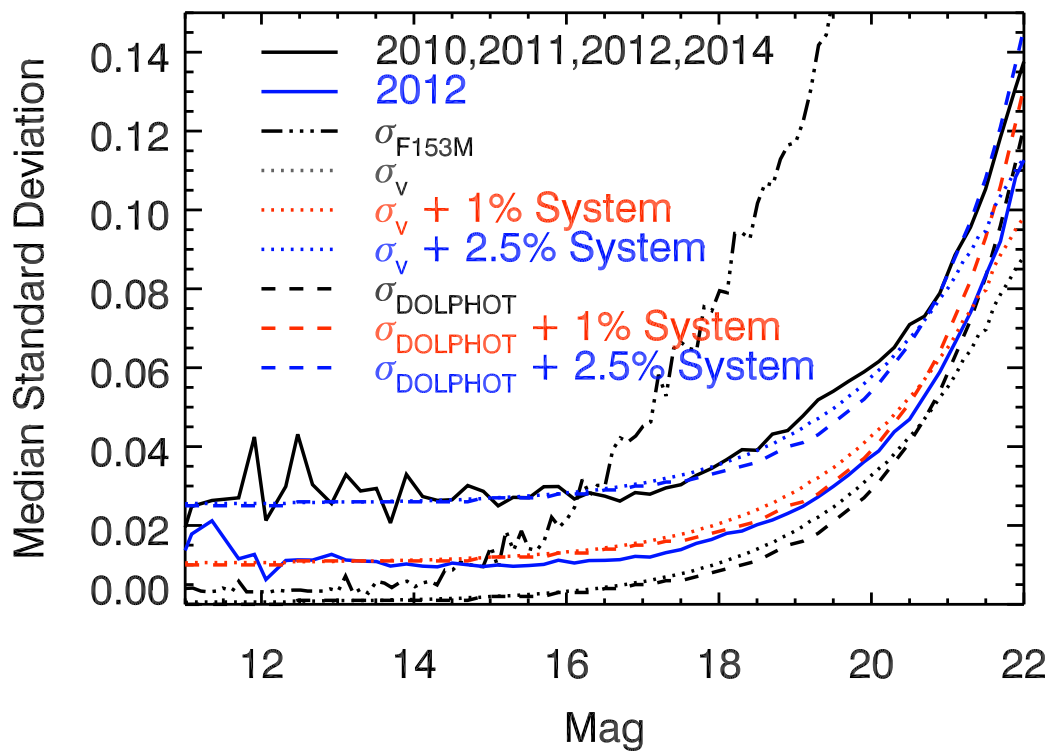,width=1.0\textwidth,angle=0}
     }
 \caption{Compare the median standard deviations of magnitudes from the real observations (solid lines) and those 
 from the artificial star tests (dotted lines and a dot dashed line) and the DOLPHOT (dashed lines).}
\label{f:fake_obs_compare}
 \end{figure*}

\begin{figure*}[!thb]
  \centerline{    
       \epsfig{figure=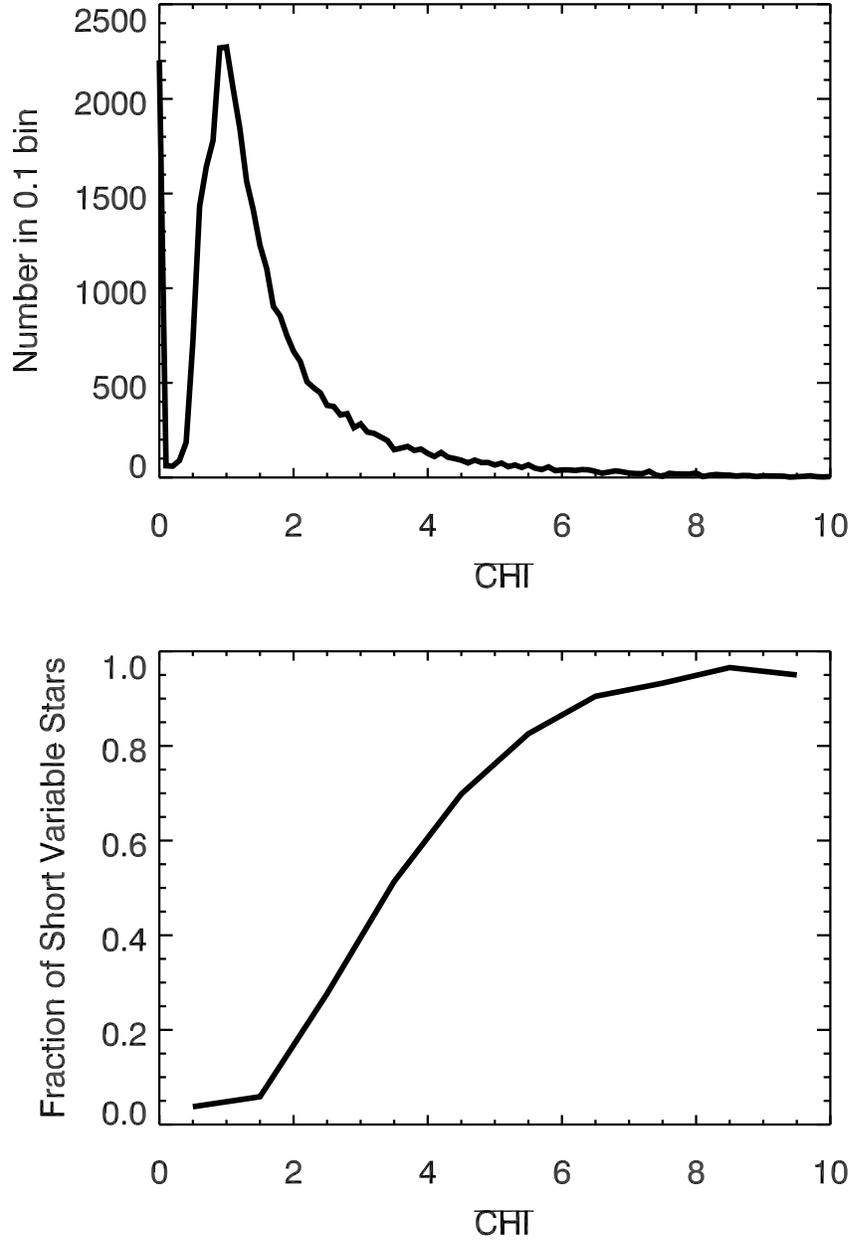,width=0.8\textwidth,angle=0}
       }
 \caption{Tope panel: the mean `CHI' ($\overline{CHI}$) distribution of dithered 
 exposures from Groups 5 to 7 in Table\ref{t:obs} 
 for individual sources. Bottom panel: the fraction of sources with $\chi^2_d>2.42$ as a function of
 $\overline{CHI}$. }
\label{f:chi_short}
 \end{figure*}

\input{table1}

\input{table2}
\input{table3}

\input{table4}

\input{table5}
\input{table6}

\input{table7}
\input{table8}

\input{table9}

\input{table10}

\end{document}

%% file: table1.tex
\begin{deluxetable}{ccccccc}
  \tabletypesize{\small}
 \tablecolumns{7}
  \tablecaption{\hst/WFC3 observations of the central 2.3\arcmin $\times$ 2.3\arcmin\ of the GC}
  \tablewidth{0pt}
  \tablehead{
    \colhead{Group} &
    \colhead{Observation Date} &
    \colhead{Program} & 
    \colhead{Duration} & 
    \colhead{Number of} &
    \colhead{Number of} &
    \colhead{Exposure } \\
    \colhead{ID} &
    \colhead{(GMT)} &
    \colhead{} &
    \colhead{(hours)} &
    \colhead{Pointings} &
    \colhead{Dithered Exposures} & 
    \colhead{Time (s)} 
}
\startdata
\multicolumn{7}{c}{F153M Observations}\\
\hline\\
1 & 2010-06-29 & GO-11671 & 5 & 1 & 21 & 7334 \\
2 & 2011-09-08 & GO-12318 & 3.5 & 1 & 21 & 7334 \\
3 & 2012-08-09 & GO-12667 & 6 & 1 & 21 & 7334 \\
4 & 2014-02-07 & GO-13049 & 0.3 & 1 & 4 & 997 \\
5 & 2014-02-28 & GO-13316/13403 & 10 & 14 & 56 & 10806 \\
6 & 2014-03-10 & GO-13316 & 10 & 14 & 55 & 16517 \\
7a$^1$ & 2014-04-02 & GO-13316 & 15 & 20 & 80 & 24035 \\
7b$^1$ & 2014-04-03 & GO-13316 & 5.5 & 8 & 32 & 9954 \\
\hline\\
\multicolumn{7}{c}{F127M Observations}\\
\hline\\
1 & 2010-08-17 & GO-11671 & 4 & 1 & 12 & 7191 \\
2 & 2011-05-20/21 & GO-12182 & 0.5 & 8 & 4 & 1797 \\
\enddata
\tablecomments{1) The first dithered exposure in the group 7b was
started 15.25 hours after the end of the last dithered exposure
in the group 7a. Therefore, we analyzed these two groups together
when 
we identified stars with short periods of days in \S\ref{s:result},
although we separated them when we plotted the light curves.}
\label{t:obs}
\end{deluxetable}

%% file: table2.tex
\begin{deluxetable}{cccccc}
  \tabletypesize{\small}
 \tablecolumns{6}
  \tablecaption{Detection Limit}
  \tablewidth{0pt}
  \tablehead{
    \colhead{Area} & 
    \colhead{Size} &
    \colhead{Surface Brightness} &
    \colhead{Input/Output} & 
    \colhead{90\% Detection} & 
    \colhead{50\% Detection} \\
    \colhead{ID$^a$} & 
    \colhead{arcsec$^2$} &
    \colhead{ergs s$^{-1}$ cm$^{-2}$ \AA$^{-1}$ arcsec$^{-2}$} &
    \colhead{Number$^b$ } & 
    \colhead{Limit} & 
    \colhead{Limit} 
}
\startdata
      1&         406&  5.1$\times$ $10^{-18}$&       63230/       36642&21.1&22.5\\
      2&        2396&  8.1$\times$ $10^{-18}$&      499053/      214436&19.8&21.5\\
      3&        4294& 12.7$\times$ $10^{-18}$&     1488127/      462869&19.0&20.7\\
      4&        5607& 20.1$\times$ $10^{-18}$&     2723407/      585244&18.3&20.0\\
      5&        3790& 31.8$\times$ $10^{-18}$&     2351949/      337882&17.5&19.2\\
      6&        1593& 50.1$\times$ $10^{-18}$&     1182158/      108677&16.7&18.5\\
      7&         529& 79.1$\times$ $10^{-18}$&      463717/       27890&15.6&17.9\\
      8&         148&124.9$\times$ $10^{-18}$&      136886/        4448&15.1&17.1\\
      9&          28&197.1$\times$ $10^{-18}$&       26316/         453&14.2&16.4\\
     10&           8&311.1$\times$ $10^{-18}$&        8549/          87&14.1&15.8\\
\enddata
\tablecomments{$^a$ The larger the `Area ID' is, the higher the surface brightness is;  
$^b$ `Input/Ouput Number' means the numbers of artificial stars inserted into each annulus 
and then recovered by the `DOLPHOT' package.}
\label{t:dl}
\end{deluxetable}

%% file: table3.tex
\begin{deluxetable}{cccccccccccc}
  \tabletypesize{\scriptsize}
 \tablecolumns{12}
  \tablecaption{Source Catalog}
  \tablewidth{0pt}
  \tablehead{
    \colhead{ID} & 
    \colhead{RA} &
    \colhead{Dec} & 
    \colhead{F153M$^a$} & 
    \colhead{$\sigma_{F153M}$$^b$} &
    \colhead{F127M$^a$} & 
    \colhead{$\sigma_{F127M}$$^b$}&
    \colhead{$\chi^2_y$}&
    \colhead{$\chi^2_{y,b}$}&
    \colhead{$\chi^2_d$}&
    \colhead{$\chi^2_{d,b}$}&
    \colhead{V?$^c$}
}
\startdata
       1&266.40831&-29.02624& 1.00&+0.004$\pm$0.001& 1.00&+9.900$\pm$9.900&  2.0&  0.3& 11.5&  0.6&0\\
      2&266.40634&-29.02650&10.38&+0.004$\pm$0.000&11.32&+0.003$\pm$0.000&  0.9&  0.1&  6.6&  0.2&0\\
      3&266.42049&-29.02490&10.51&+0.004$\pm$0.001&11.00&+0.003$\pm$0.000&  0.7&  0.1&  5.5&  0.2&0\\
      4&266.41602&-29.01495&10.66&+0.004$\pm$0.001&13.28&+0.003$\pm$0.001&  1.6&  0.7&  7.9&  0.1&0\\
      5&266.41749&-29.01606&10.98&+0.004$\pm$0.011&12.86&+0.003$\pm$0.000&  0.5&  0.1&  3.9&  0.2&0\\
      6&266.39496&-29.01075&11.09&+0.004$\pm$0.015&11.72&+0.003$\pm$0.001&  0.9&  0.2&  5.5&  0.2&0\\
      7&266.43080&-29.00743&11.34&+0.004$\pm$0.017&12.03&+0.003$\pm$0.001&  0.7&  0.1&  4.4&  0.1&0\\
      8&266.42854&-28.99368&11.37&+0.004$\pm$0.018&12.06&+0.003$\pm$0.001&  1.0&  0.2&  5.2&  0.0&0\\
      9&266.42271&-28.99041&11.62&+0.004$\pm$0.006&13.83&+0.003$\pm$0.001&  1.1&  0.5&  3.4&  0.1&0\\
     10&266.41685&-29.00629&11.66&+0.004$\pm$0.019&15.65&+0.004$\pm$0.002& 29.2& 22.5&  1.4&  0.1&1\\
     ...&...&...&...&...&...&...&...&...&...&...&...\\
\enddata
\tablecomments{The complete source list will be published online. The sources 
have been sorted according to their mean magnitudes. $^a$ If the source is 
saturated, its magnitude is 1; If the source is undetected, its magnitude is 99; $^b$ the first and the second values are
the systematic and statistic uncertainties determined from the artificial star tests (see \S\ref{ss:reduction}). Because of 
the confusion, the output magnitude is smaller than the input magnitude. Therefore, the systematic uncertainty is 
positive; $^c$ 
\textbf{The index which shows whether the source is variable or not:} “
`0', `1', `2', `3' means non-variable, stars varying on yearly time scales, 
stars with intraday variability and stars varying on both yearly and daily time scales, respectively.} 
\label{t:sou}
\end{deluxetable}

%% file: table4.tex
\begin{deluxetable}{cccccccc}
  \tabletypesize{\small}
 \tablecolumns{8}
  \tablecaption{Magnitudes in Individual Dithered Exposures}
  \tablewidth{0pt}
  \tablehead{
    \colhead{ID} & 
    \colhead{Julian Date(day)} &
    \colhead{Mag$^a$} & 
    \colhead{$\sigma_{v}$} & 
    \colhead{S/N} &
    \colhead{Sharp$^2$} &
    \colhead{Crowd} &
    \colhead{Flag}
}
\startdata
     ...&...&...&...&...&...&...&..\\
      3&55376.295&10.54&0.015&4142.3&0.0003&0.016&      4\\
      3&55376.300&10.54&0.015&4341.9&0.0002&0.014&      4\\
      3&55376.305&10.51&0.015&5536.4&0.0010&0.007&      0\\
      3&55376.310&10.53&0.015&4163.1&0.0011&0.014&      4\\
      3&55376.314&10.53&0.015&4416.6&0.0007&0.015&      4\\
      3&55376.357&10.53&0.015&4214.7&0.0003&0.017&      4\\
      3&55376.362&10.52&0.015&5531.3&0.0007&0.009&      0\\
      3&55376.367&10.51&0.015&5544.4&0.0007&0.008&      0\\
      3&55376.371&10.52&0.015&5524.5&0.0020&0.009&      0\\
      3&55376.376&10.52&0.015&5537.4&0.0025&0.010&      0\\
     ...&...&...&...&...&...&...&..\\
\enddata
\tablecomments{The complete source list will be published online. $^a$ If the source is 
saturated, its magnitude is 1; If the source is undetected, its magnitude is 99.}
\label{t:dithered}
\end{deluxetable}

%% file: table5.tex
\begin{deluxetable}{cccc}
  \tabletypesize{\small}
 \tablecolumns{4}
  \tablecaption{Statistic values of $\chi^2_y$ and $\chi^2_d$}
   \tablewidth{0pt}
  \tablehead{
    \colhead{} &
    \colhead{Median} &    
    \colhead{68\% percentile} &
    \colhead{90\% percentile} 
    }  
\startdata
$\chi^2_y$ & 0.94 & [0.41,3.04] & [0.23,8.21] \\
$\chi^2_{y,b}$ & 0.56 & [0.19,2.49] & [0.09,8.21] \\
$\chi^2_d$ & 1.00 & [0.58,2.42] & [0.42,6.44] \\
$\chi^2_{d,b}$ & 0.30 & [0.14,0.59] & [0.09,0.83] \\
\enddata
\label{t:sv}
\end{deluxetable}

%% file: table6.tex
\begin{deluxetable}{cccc}
  \tabletypesize{\small}
 \tablecolumns{4}
  \tablecaption{Periods}
   \tablewidth{0pt}
  \tablehead{
    \colhead{ID} &
    \colhead{Period} &
    \colhead{F153M} & 
    \colhead{F127M-F153M} 
}
\startdata
    2495&0.542&16.6&0.6\\
   6164&0.195&17.7&2.0\\
   7831&0.299&18.0&2.6\\
   8735&0.559&18.1&2.2\\
   9074&0.838&18.2&2.0\\
  10520&0.542&18.4&1.9\\
  12097&0.489&18.6&2.7\\
  12215&0.089&18.6&2.0\\
  13950&0.686&18.8&2.4\\
  14123&0.384&18.8&2.5\\
  14583&0.677&18.9&2.4\\
  15242&0.329&18.9&1.8\\
  19440&0.294&19.5&2.0\\
  20785&0.171&19.6&1.8\\
  20993&0.368&19.7&2.0\\
  21740&0.244&19.8&2.3\\
  22081&0.495&19.8&2.6\\
  22197&0.593&19.8&2.9\\
  22312&0.573&19.9&2.5\\
  23037&0.469&20.0&2.3\\
  23523&1.103&20.0&2.6\\
  23605&1.142&20.0&3.1\\
  25542&0.258&20.3&1.9\\
  25912&0.644&20.4&3.3\\
  26112&1.299&20.4&2.8\\
  26822&0.722&20.5&2.3\\
  28701&0.235&20.9&1.8\\
  29285&0.168&21.0&1.7\\
\hline
   1365&2.279&15.9&2.7\\
   1404&1.565&15.9&0.5\\
   2980&0.584&16.8&1.3\\
  10110&2.005&18.3&2.2\\
  21377&2.016&19.7&1.9\\
  21717&0.814&19.8&2.0\\
  27035&2.320&20.6&2.9\\
  27972&1.439&20.7&2.4\\ 
\enddata
\tablecomments{The lightcurves of the first 28 sources are shown in Figs.~\ref{f:rrlyrae_1} to~\ref{f:rrlyrae_3}, while the last 8 sources are given in Fig.~\ref{f:eclipse}.}
\label{t:period}
\end{deluxetable}

%% file: table7.tex
\begin{deluxetable}{cccccc}
  \tabletypesize{\scriptsize}
 \tablecolumns{6}
  \tablecaption{Previous Variability Studies}
   \tablewidth{0pt}
  \tablehead{
        \colhead{References$^a$} & 
        \colhead{Instruments} &
        \colhead{FoV} &
        \colhead{Time Range} &
        \colhead{Identified Variables$^b$} &
        \colhead{Recovered} 
        }
        \startdata
        (1) & Steward 1.54m & 5\arcmin$\times$5\arcmin\ & 1986-1988 & 12 without astrometry & \\
        (2) & IRTF 3m  & $\sim$23\arcsec$\times$23\arcsec\ & 1991-1993 & 6 & 6 \\
        (3) & CTIO 4m/OSIRIS & 2\arcmin$\times$2\arcmin\ & 1993, 1995 & 3 & 3\\
        (4) & NTT 3.6m/SHARP I & 20\arcsec$\times$20\arcsec\ & 1992-1998 & 108 (112) & 48 \\
        (5) & SAAO 0.75m/PANIC & 24\arcmin$\times$24\arcmin\ & 1994-1997 & 10 (409) & 7 \\
        (6) & CTIO Yalo 1m/ANDICAM & 112\arcsec$\times$112\arcsec\ &2000-2002 & 8(112) & 2 \\
        (7) & VLBA/VLA & & 1995-2006 & 12(15) & 11\\
        (8) & Keck 10m/NIRC & 5\arcsec$\times$5\arcsec\  &1995-2004 & 10(15) & 3\\
        (9) & IRSF 1.4m/SIRIUS & 20\arcmin$\times$20\arcmin\ & 2001-2008 & 17/(1364) & 15 \\
        (10) & VLT 8m/NAOC/SINFONI & 20\arcsec$\times$20\arcsec\  & 2003-2013 & 2 & 1\\
        \enddata
        \tablecomments{$^a$ (1)~\citet{hal89}; (2)~\citet{tam96}; (3)~\citet{blu96}; (4)~\citet{ott99}; (5)~\citet{gla01}; (6)~\citet{pee07}; (7)~\citet{rei07}; (8)~\citet{raf07}; (9)~\citet{mat09,mat11,mat13}; (10)~\citet{pfu14}. 
        $^b$ the first values are the numbers of variables, which have counterparts in our source catalog, while the values in the parentheses are the original numbers of variable stars, }
\label{t:references}
\end{deluxetable}

%% file: table8.tex
\begin{deluxetable}{cccccccccc}
  \tabletypesize{\tiny}
 \tablecolumns{10}
  \tablecaption{Cross Correlations With Previous Variability Study}
   \tablewidth{0pt}
  \tablehead{
    \colhead{ID} &
    \colhead{(1)} & 
    \colhead{(2)} & 
    \colhead{(3)} &
    \colhead{(4)} & 
    \colhead{(5)} & 
    \colhead{(6)} & 
    \colhead{(7)} &
    \colhead{(8)} & 
    \colhead{(9)} 
}
\startdata
           1&&&&&PSDJ174537.98-290134.4&&&&\\
           4&&&&&PSDJ174539.83-290053.9&&&&\\
          10&IRS7&IRS7&       1&&PSDJ174540.04-290022.7&IRS7&&&\\
          11&&&& 3-5   &PSDJ174542.72-285957.4&SiO6&&     680&\\
          14&&&       2&&PSDJ174540.25-290027.2&&IRS16NE  &&IRS16NE\\
          15&&&&&PSDJ174536.63-290015.4&&&&\\
          17&&&&&PSDJ174543.01-290011.9&&&&\\
          20&&&&&PSDJ174542.39-285950.7&&&&\\
          25&&&&&PSDJ174538.74-290012.7&&&&\\
          26&&&&&PSDJ174541.17-290046.9&&&&\\
...&...&...&...&...&...&...&...&...&..\\
\enddata
\tablecomments{References: (1)~\citet{tam96}; (2)~\citet{blu96}; (3)~\citet{ott99} (`ID' column in its Table 2); (4)~\citet{gla01} (`Iden.' column in its Table 2); (5)~\citet{pee07} (`ID' column in its Tables 1 and 2); (6)~\citet{rei07} (`Star' column in its Table 1); (7)~\citet{raf07} (`Star ID' column in its Tables 2 and 3); (8)~\citet{mat09} (`No.' column in its Table 6); (9)~\citet{pfu14}. The complete lists will be published online.}
\label{t:reference_var}
\end{deluxetable}

%% file: table9.tex
\begin{deluxetable}{ccccccccccc}
  \tabletypesize{\tiny}
 \tablecolumns{11}
  \tablecaption{Cross Correlations With Previous Spectroscopy Study}
   \tablewidth{0pt}
  \tablehead{
    \colhead{ID} &
    \colhead{Name} &
    \colhead{(1)} & 
    \colhead{(2)} & 
    \colhead{(3)} &
    \colhead{(4)} & 
    \colhead{(5)} & 
    \colhead{(6)} & 
    \colhead{(7)} &
    \colhead{(8)} & 
    \colhead{(9)} 
}
\startdata
          10&IRS7      &66(I)&&&&&&&&25352(Late)\\
          11&IRS24     &128(LPV)&&&&&&&&\\
          27&IRS17&105(LPV?)&&&&&&&&25351(Late)\\
          31&IRS23     &136(LPV?)&&&&&&&&\\
          38&&11(III)&&&&&&&&\\
          39&&140(III)&&&&&&&&\\
          42&IRS9      &91(LPV)&&&&&&&&25348(Late)\\
          43&IRS16SW           &71(24000)&&E23(Ofpe/WN9)&9(Early)&S1-34(l)&&97(Ofpe/WN9)&&\\
          70&IRS14SW   &60()&&&&&&&&1(Late)\\
          71&&126(III)&&&&&&&&\\
...&...&...&...&...&...&...&...&...&..\\
\enddata
\tablecomments{References: (1)~\citet{blu03}; (2)~\citet{man07}; (3)~\citet{pau06}; (4)~\citet{bar09}; (5)~\citet{do13}; (6)~\citet{sto15}; (7)~\citet{fel15}; (8)~\citet{nis16}; (9)~\citet{fel17}. The complete lists will be published online.}
\label{t:reference}
\end{deluxetable}

%% file: table10.tex
\begin{deluxetable}{ccccc}
  \tabletypesize{\small}
 \tablecolumns{5}
  \tablecaption{Type II Cepheid Candidates}
   \tablewidth{0pt}
  \tablehead{
    \colhead{ID} &
    \colhead{$A_{Ks,CCEPs}^a$} &
    \colhead{$Dis_{CCEPs}^a$} &
    \colhead{$A_{Ks,T2Cs}^b$} &
    \colhead{$Dis_{T2Cs}^b$} \\
}
\startdata
  23523&2.47$\pm$0.12&28.0$\pm$2.70&2.43$\pm$0.12&15.6$\pm$2.50\\
  23605&2.86$\pm$0.17&18.8$\pm$2.74&2.88$\pm$0.17&10.4$\pm$2.52\\
  26112&2.54$\pm$0.16&32.9$\pm$3.54&2.57$\pm$0.16&17.7$\pm$3.27\\
\enddata
\tablecomments{$^a$, \textbf{the values were determined} with the assumption 
of the PL relationship for CCEPs given in~\citet{mat13} and 
distances in units of kpc. 
$^b$, \textbf{the values were determined} with the assumption of the PL relationship for 
T2Cs  given in~\citet{mat13} and distances in units of kpc.}
\label{t:t2c}
\end{deluxetable}

%% file: gc_var_part1.bbl
\begin{thebibliography}{}
\bibitem[Alcock et al.(1999)]{alc99} Alcock, C., Allsman, R.~A., Alves, D.~R., et al.\ 1999, \aj, 117, 920 
\bibitem[Becker et al.(1977)]{bec77} Becker, S.~A., Iben, I., Jr., \& Tuggle, R.~S.\ 1977, \apj, 218, 633 
\bibitem[Bartko et al.(2009)]{bar09} Bartko, H., Martins, F., Fritz, T.~K., et al.\ 2009, \apj, 697, 1741 
\bibitem[Blum et al.(2003)]{blu03} Blum, R.~D., Ram{\'{\i}}rez, S.~V., Sellgren, K., \& Olsen, K.\ 2003, \apj, 597, 323 
\bibitem[Blum et al.(1996)]{blu96} Blum, R.~D., Sellgren, K., \& Depoy, D.~L.\ 1996, \apj, 470, 864 
\bibitem[Boehle et al.(2016)]{boe16} Boehle, A., Ghez, A.~M., Sch{\"o}del, R., et al.\ 2016, \apj, 830, 17 
\bibitem[Carr et al.(2000)]{car00} Carr, J.~S., Sellgren, K., \& Balachandran, S.~C.\ 2000, \apj, 530, 307 
\bibitem[Castelli \& Kurucz(2004)]{cas04} Castelli, F., \& Kurucz, R.~L.\ 2004, arXiv:astro-ph/0405087 
\bibitem[Catelan et al.(2004)]{cat04} Catelan, M., Pritzl, B.~J., \& Smith, H.~A.\ 2004, \apjs, 154, 633 
\bibitem[Catelan(2009)]{cat09} Catelan, M.\ 2009, \apss, 320, 261 
\bibitem[Chatzopoulos et al.(2015)]{cha15} Chatzopoulos, S., Fritz, T.~K., Gerhard, O., et al.\ 2015, \mnras, 447, 948 
\bibitem[Crowther(2007)]{cro07} Crowther, P.~A.\ 2007, \araa, 45, 177 
\bibitem[Dalcanton et al.(2012a)]{dal12a} Dalcanton, J.~J., 
Williams, B.~F., Lang, D., et al.\ 2012a, \apjs, 200, 18 
\bibitem[Dalcanton et al.(2012b)]{dal12b} Dalcanton, J.~J., Williams, B.~F., Melbourne, J.~L., et al.\ 2012b, \apjs, 198, 6 
\bibitem[Dahlen(2013)]{dah13} Dahlen, T.\ 2013, Space Telescope WFC Instrument Science Report,  
\bibitem[Deguchi et al.(2004)]{deg04} Deguchi, S., Imai, H., Fujii, T., et al.\ 2004, \pasj, 56, 261 
\bibitem[Do et al.(2009)]{do09} Do, T., Ghez, A.~M., Morris, M.~R., et al.\ 2009, \apj, 703, 1323 
\bibitem[Do et al.(2013)]{do13} Do, T., Lu, J.~R., Ghez, A.~M., et al.\ 2013, \apj, 764, 154 
\bibitem[Do et al.(2015)]{do15} Do, T., Kerzendorf, W., Winsor, N., et al.\ 2015, \apj, 809, 143 
\bibitem[Dolphin(2000)]{dol00} Dolphin, A.~E.\ 2000, \pasp, 
112, 1383 
\bibitem[Dong et al.(2011)]{don11} Dong, H., Wang, Q.~D., 
Cotera, A., et al.\ 2011, \mnras, 417, 114 
\bibitem[Dong et al.(2012)]{don12} Dong, H., Wang, Q.~D., 
\& Morris, M.~R.\ 2012, \mnras, 425, 884 
\bibitem[Dong et al.(2017b)]{don17} Dong, H., Schodel, R., Williams, B.~F., et al.\ 2017, arXiv:1706.03299 
\bibitem[Feast et al.(1989)]{fea89} Feast, M.~W., Glass, I.~S., Whitelock, P.~A., \& Catchpole, R.~M.\ 1989, \mnras, 241, 375 
\bibitem[Feldmeier-Krause et al.(2015)]{fel15} Feldmeier-Krause, A., Neumayer, N., Sch{\"o}del, R., et al.\ 2015, \aap, 584, A2 
\bibitem[Feldmeier-Krause et al.(2017)]{fel17} Feldmeier-Krause, A., Kerzendorf, W., Neumayer, N., et al.\ 2017, \mnras, 464, 194 
\bibitem[Figer et al.(1999)]{fig99} Figer, D.~F., McLean, 
I.~S., \& Morris, M.\ 1999, \apj, 514, 202
\bibitem[Figer et al.(2002)]{fig02} Figer, D.~F., Najarro, 
F., Gilmore, D., et al.\ 2002, \apj, 581, 258 
\bibitem[Figer et al.(2004)]{fig04} Figer, D.~F., Rich, 
R.~M., Kim, S.~S., Morris, M., \& Serabyn, E.\ 2004, \apj, 601, 319 
\bibitem[Fiorentino et al.(2002)]{fio02} Fiorentino, G., Caputo, F., Marconi, M., \& Musella, I.\ 2002, \apj, 576, 402 
\bibitem[Genzel et al.(2003)]{gen03} Genzel, R., Sch{\"o}del, 
R., Ott, T., et al.\ 2003, \apj, 594, 812 
\bibitem[Ghez et al.(2008)]{ghe08} Ghez, A.~M., Salim, S., Weinberg, N.~N., et al.\ 2008, \apj, 689, 1044-1062 
\bibitem[Gillessen et al.(2009)]{gil09} Gillessen, S., Eisenhauer, F., Fritz, T.~K., et al.\ 2009, \apjl, 707, L114 
\bibitem[Gillessen et al.(2012)]{gil12} Gillessen, S., Genzel, R., Fritz, T.~K., et al.\ 2012, \nat, 481, 51 
\bibitem[Glass \& Lloyd Evans(1981)]{gla81} Glass, I.~S., \& Lloyd Evans, T.\ 1981, \nat, 291, 303 
\bibitem[Glass et al.(2001)]{gla01} Glass, I.~S., Matsumoto, S., Carter, B.~S., \& Sekiguchi, K.\ 2001, \mnras, 321, 77 
\bibitem[Gray et al.(2009)]{gra09} Gray, M.~D., Wittkowski, M., Scholz, M., et al.\ 2009, \mnras, 394, 51 
\bibitem[Haller \& Rieke(1989)]{hal89} Haller, J.~W., \& Rieke, M.~J.\ 1989, The Center of the Galaxy, 136, 487 
\bibitem[Hornstein et al.(2002)]{hor02} Hornstein, S.~D., Ghez, A.~M., Tanner, A., et al.\ 2002, \apjl, 577, L9 
\bibitem[Hosek et al.(2015)]{hos15} Hosek, M.~W., Jr., Lu, J.~R., Anderson, J., et al.\ 2015, \apj, 813, 27 
\bibitem[Ita et al.(2004)]{ita04} Ita, Y., Tanab{\'e}, T., Matsunaga, N., et al.\ 2004, \mnras, 347, 720 
\bibitem[Kiss et al.(2006)]{kis06} Kiss, L.~L., Szab{\'o}, G.~M., \& Bedding, T.~R.\ 2006, \mnras, 372, 1721 
\bibitem[Klagyivik \& Szabados(2009)]{kla09} Klagyivik, P., \& Szabados, L.\ 2009, \aap, 504, 959 
\bibitem[Lan{\c c}on \& Wood(2000)]{lan00} Lan{\c c}on, A., \& Wood, P.~R.\ 2000, \aaps, 146, 217 
\bibitem[Lee(1992)]{lee92} Lee, Y.-W.\ 1992, \aj, 104, 1780 
\bibitem[Lef{\`e}vre et al.(2009)]{lef09} Lef{\`e}vre, L., Marchenko, S.~V., Moffat, A.~F.~J., \& Acker, A.\ 2009, \aap, 507, 1141 
\bibitem[Lomb(1976)]{lom76} Lomb, N.~R.\ 1976, \apss, 39, 447 
\bibitem[Lu et al.(2013)]{lu13} Lu, J.~R., Do, T., Ghez, A.~M., et al.\ 2013, \apj, 764, 155 
\bibitem[McNamara et al.(2007)]{mac07} McNamara, D.~H., Clementini, G., \& Marconi, M.\ 2007, \aj, 133, 2752 
\bibitem[Marconi et al.(2015)]{mar15} Marconi, M., Coppola, G., Bono, G., et al.\ 2015, \apj, 808, 50 
\bibitem[Maness et al.(2007)]{man07} Maness, H., Martins, F., Trippe, S., et al.\ 2007, \apj, 669, 1024 
\bibitem[Matsunaga et al.(2006)]{mat06} Matsunaga, N., Fukushi, H., Nakada, Y., et al.\ 2006, \mnras, 370, 1979 
\bibitem[Matsunaga et al.(2009)]{mat09} Matsunaga, N., Kawadu, T., Nishiyama, S., et al.\ 2009, \mnras, 399, 1709 
\bibitem[Matsunaga et al.(2011)]{mat11} Matsunaga, N., Kawadu, T., Nishiyama, S., et al.\ 2011, \nat, 477, 188 
\bibitem[Matsunaga et al.(2013)]{mat13} Matsunaga, N., Feast, M.~W., Kawadu, T., et al.\ 2013, \mnras, 429, 385 
\bibitem[Matsunaga et al.(2015)]{mat15} Matsunaga, N., Fukue, K., Yamamoto, R., et al.\ 2015, \apj, 799, 46 
\bibitem[Mossoux et al.(2016)]{mos16} Mossoux, E., Grosso, N., Bushouse, H., et al.\ 2016, \aap, 589, A116
\bibitem[Nagayama et al.(2003)]{nag03} Nagayama, T., 
Nagashima, C., Nakajima, Y., et al.\ 2003, \procspie, 4841, 459
\bibitem[Nishiyama et al.(2009)]{nis09} Nishiyama, S., Tamura, M., Hatano, H., et al.\ 2009, \apj, 696, 1407 \bibitem[Nishiyama et al.(2016)]{nis16} Nishiyama, S., Sch{\"o}del, R., Yoshikawa, T., et al.\ 2016, \aap, 588, A49 
\bibitem[Ott et al.(1999)]{ott99} Ott, T., Eckart, A., \& Genzel, R.\ 1999, \apj, 523, 248 
\bibitem[Paumard et al.(2006)]{pau06} Paumard, T., Genzel, R., Martins, F., et al.\ 2006, \apj, 643, 1011 
\bibitem[Pavlovsky et al.(2011)]{pav11} Pavlovsky, C., McCullough, P., \& Baggett, S.\ 2011, Space Telescope WFC Instrument Science Report,  
\bibitem[Pawlak et al.(2014)]{paw14} Pawlak, M., Soszy{\'n}ski, I., Pietrukowicz, P., et al.\ 2014, \actaa, 64, 293 
\bibitem[Peeples et al.(2007)]{pee07} Peeples, M.~S., Stanek, K.~Z., \& Depoy, D.~L.\ 2007, \actaa, 57, 173 
\bibitem[Reid et al.(2007)]{rei07} Reid, M.~J., Menten, K.~M., Trippe, S., Ott, T., \& Genzel, R.\ 2007, \apj, 659, 378 
\bibitem[Pfuhl et al.(2011)]{pfu11} Pfuhl, O., Fritz, T.~K., Zilka, M., et al.\ 2011, \apj, 741, 108 
\bibitem[Pfuhl et al.(2014)]{pfu14} Pfuhl, O., Alexander, T., Gillessen, S., et al.\ 2014, \apj, 782, 101 
\bibitem[Poleski et al.(2010)]{pol10} Poleski, R., Soszy{\'n}ski, I., Udalski, A., et al.\ 2010, \actaa, 60, 1 
\bibitem[Rafelski et al.(2007)]{raf07} Rafelski, M., Ghez, A.~M., Hornstein, S.~D., Lu, J.~R., \& Morris, M.\ 2007, \apj, 659, 1241 
\bibitem[Sandage \& Tammann(2006)]{san06} Sandage, A., \& Tammann, G.~A.\ 2006, \araa, 44, 93 
\bibitem[Scargle(1982)]{sca82} Scargle, J.~D.\ 1982, \apj, 263, 835 
\bibitem[Sch{\"o}del et al.(2007)]{sch07} Sch{\"o}del, R., Eckart, A., Alexander, T., et al.\ 2007, \aap, 469, 125 
\bibitem[Sch{\"o}del et al.(2009)]{sch09} Sch{\"o}del, R., Merritt, D., \& Eckart, A.\ 2009, \aap, 502, 91 
\bibitem[Sch{\"o}del et al.(2010)]{sch10} Sch{\"o}del, R., Najarro, F., Muzic, K., \& Eckart, A.\ 2010, \aap, 511, A18 
\bibitem[Sch{\"o}del(2010)]{sch10a} Sch{\"o}del, R.\ 2010, \aap, 509, A58 
\bibitem[Scoville et al.(2003)]{sco03} Scoville, N.~Z., Stolovy, S.~R., Rieke, M., Christopher, M., \& Yusef-Zadeh, F.\ 2003, \apj, 594, 294 
\bibitem[Soszynski et al.(2004)]{sos04} Soszynski, I., Udalski, A., Kubiak, M., et al.\ 2004, \actaa, 54, 129 
\bibitem[Soszy{\'n}ski et al.(2008)]{sos08} Soszy{\'n}ski, I., Udalski, A., Szyma{\'n}ski, M.~K., et al.\ 2008, \actaa, 58, 293 
\bibitem[Soszy{\'n}ski et al.(2011)]{sos11} Soszy{\'n}ski, I., Udalski, A., Pietrukowicz, P., et al.\ 2011, \actaa, 61, 285 
\bibitem[Soszy{\'n}ski et al.(2013)]{sos13} Soszy{\'n}ski, I., Wood, P.~R., \& Udalski, A.\ 2013, \apj, 779, 167 
\bibitem[Soszy{\'n}ski \& Udalski(2014)]{sos14} Soszy{\'n}ski, I., \& Udalski, A.\ 2014, \apj, 788, 13 
\bibitem[St{\o}stad et al.(2015)]{sto15} St{\o}stad, M., Do, T., Murray, N., et al.\ 2015, \apj, 808, 106 
\bibitem[Tamura et al.(1996)]{tam96} Tamura, M., Werner, M.~W., Becklin, E.~E., \& Phinney, E.~S.\ 1996, \apj, 467, 645 
\bibitem[Turner(1996)]{tur96} Turner, D.~G.\ 1996, \jrasc, 90, 82 
\bibitem[Verheyen et al.(2012)]{ver12} Verheyen, L., Messineo, M., \& Menten, K.~M.\ 2012, \aap, 541, A36 
\bibitem[Wallerstein(2002)]{wal02} Wallerstein, G.\ 2002, \pasp, 114, 689 
\bibitem[Walker(1989)]{wal89} Walker, A.~R.\ 1989, \pasp, 101, 570 
\bibitem[Walker \& Terndrup(1991)]{wal91} Walker, A.~R., \& Terndrup, D.~M.\ 1991, \apj, 378, 119 
\bibitem[Whitelock et al.(2008)]{whi08} Whitelock, P.~A., Feast, M.~W., \& van Leeuwen, F.\ 2008, \mnras, 386, 313 
\bibitem[Williams et al.(2014)]{wil14} Williams, B.~F., Lang, 
D., Dalcanton, J.~J., et al.\ 2014, \apjs, 215, 9 
\bibitem[Wood et al.(1999)]{woo99} Wood, P.~R., Alcock, C., Allsman, R.~A., et al.\ 1999, Asymptotic Giant Branch Stars, 191, 151 
\bibitem[Wood(2000)]{woo00} Wood, P.~R.\ 2000, \pasa, 17, 18 

\end{thebibliography}
